\documentclass{article}

\usepackage[dandb, final]{neurips_2025}

\usepackage[utf8]{inputenc} 
\usepackage[T1]{fontenc}    
\usepackage{hyperref}       
\usepackage{url}            
\usepackage{booktabs}       
\usepackage{amsfonts}       
\usepackage{nicefrac}       
\usepackage{microtype}      
\usepackage{xcolor}         

\usepackage{microtype}
\usepackage{multirow}
\usepackage{graphicx}
\usepackage{subfig}
\usepackage{booktabs} 
\usepackage[normalem]{ulem}

\usepackage{hyperref}

\usepackage{enumitem}
\usepackage[most]{tcolorbox}

\usepackage{booktabs}       
\usepackage{amsfonts}       
\usepackage{nicefrac}       
\usepackage{microtype}      
\usepackage{xcolor}         

\usepackage{tabularray}
\UseTblrLibrary{siunitx,counter}

\usepackage[frozencache,cachedir=minted-cache]{minted}

\usepackage[htt]{hyphenat}

\usepackage{amsmath}
\usepackage{amsfonts}
\usepackage{amssymb}
\usepackage{amsthm}
\usepackage{graphicx}
\usepackage{color}
\usepackage{url}
\usepackage{stackrel}
\usepackage[skip=6pt]{caption}
\usepackage{xspace}
\usepackage{tabularx}
\usepackage{balance}
\usepackage{wrapfig}
\usepackage{booktabs}
\usepackage{tabu}
\usepackage[linesnumbered,ruled,noend]{algorithm2e}
\usepackage{subfig}
\usepackage{graphicx}
\usepackage{float}

\theoremstyle{remark}

\newcommand{\squishlist}{
	\begin{list}{$\bullet$}{
		\setlength{\itemsep}{0pt}
		\setlength{\parsep}{3pt}
		\setlength{\topsep}{3pt}
		\setlength{\partopsep}{0pt}
		\setlength{\leftmargin}{1.0em}
		\setlength{\labelwidth}{1em}
		\setlength{\labelsep}{0.5em}
   }
}

\newcommand{\squishenum}{
	
	\begin{list}{\usecounter{scount}}{
		\setlength{\itemsep}{0pt}
		\setlength{\parsep}{3pt}
		\setlength{\topsep}{3pt}
		\setlength{\partopsep}{0pt}
		\setlength{\leftmargin}{1.2em}
		\setlength{\labelwidth}{1em}
		\setlength{\labelsep}{0.5em}
	}
}

\newcommand{\squishend}{
	\end{list}
}

\newenvironment{dialogue}{%
    \begin{tcolorbox}[
        enhanced,
        colback=white,
        arc=0mm,
        boxrule=0.5pt,
        left=1.5mm,
        right=1.5mm,
        top=1mm,
        bottom=1mm,
        boxsep=1mm,
        before skip=10pt,
        after skip=10pt
    ]
    \normalsize 
    \scriptsize 
    \begin{description}[
        align=left,
        labelwidth=!, 
        leftmargin=0em, 
        font=\normalfont\bfseries,
        noitemsep,
        parsep=0pt
    ]
}{%
    \end{description}
    \end{tcolorbox}
}

\newcommand{\user}[2][black]{\item[{\color{#1} User:}] {\color{#1}\texttt{#2}}}
\newcommand{\system}[2][black]{\item[{\color{#1} System:}] {\color{#1}\texttt{#2}}}
\newcommand{\assistant}[2][black]{\item[{\color{#1} Assistant:}] {\color{#1}\texttt{#2}}}

\usepackage{twemojis}

\newcommand{\textitt}[1]{\textit{\texttt{#1}}}

\usepackage{amsmath}
\usepackage{amssymb}
\usepackage{mathtools}
\usepackage{amsthm}

\usepackage[capitalize,noabbrev]{cleveref}

\theoremstyle{definition}
\usepackage[textsize=tiny]{todonotes}

\usepackage{amsmath,amsfonts,bm}

\makeatletter
\def\widebreve{\mathpalette\wide@breve}
\def\wide@breve#1#2{\sbox\z@{$#1#2$}%
     \mathop{\vbox{\m@th\ialign{##\crcr
\kern0.08em\brevefill#1{0.6\wd\z@}\crcr\noalign{\nointerlineskip}%
                    $\hss#1#2\hss$\crcr}}}\limits}
\def\brevefill#1#2{$\m@th\sbox\tw@{$#1($}%
  \hss\resizebox{#2}{\wd\tw@}{\rotatebox[origin=c]{90}{\upshape(}}\hss$}
\makeatletter

\def\1{\bm{1}}

\DeclareMathAlphabet{\mathsfit}{\encodingdefault}{\sfdefault}{m}{sl}
\SetMathAlphabet{\mathsfit}{bold}{\encodingdefault}{\sfdefault}{bx}{n}

\usepackage{wrapfig}
\usepackage{titletoc}

\newcommand\DoToC{%
  \startcontents
  \printcontents{}{1}{\textbf{Table of Contents}\vskip3pt\hrule\vskip5pt}
  \vskip3pt\hrule\vskip5pt
}

\newcommand{\oursys}{\texttt{Data-Juicer}\xspace}
\newcommand{\oursysI}{\texttt{Data-Juicer} 1.0\xspace}
\newcommand{\oursysII}{\texttt{Data-Juicer} 2.0\xspace}

\title{\oursysII: Cloud-Scale Adaptive Data Processing for and with Foundation Models}

\author{%
Daoyuan Chen$^*$,  
Yilun Huang$^*$, 
Xuchen Pan$^\dagger$,
Nana Jiang$^\dagger$, 
Haibin Wang$^\dagger$ \\
\textbf{
Yilei Zhang$^\dagger$,
Ce Ge$^\dagger$, 
Yushuo Chen,
Wenhao Zhang,
Zhijian Ma}  \\
\textbf{Jun Huang,
Wei Lin,
Yaliang Li$^\ddagger$,
Bolin Ding$^\ddagger$,
Jingren Zhou} \\ 
  Alibaba Group
}

\begin{document}

\maketitle

\begin{abstract}
  Foundation models demand advanced data processing for their vast, multimodal datasets.
However, traditional frameworks struggle with the unique complexities of multimodal data.
In response, we present \oursysII, a data processing system backed by 100+ data processing operators spanning text, image, video, and audio modalities, supporting more critical tasks including data analysis, synthesis, annotation, and foundation model post-training.
With seamless compatibility and dedicated optimization for popular dataset hubs like Hugging Face and computing engines like Ray, it improves upon its predecessor in terms of usability, efficiency, and programmability.
It features an easily accessible user interface layer that supports decoupled Python interactions, RESTful APIs, and conversational commands. 
Its new runtime layer offers adaptive execution across diverse scales and environments, abstracting away system complexities.
Extensive empirical evaluations demonstrate \oursysII's remarkable performance and scalability, highlighting its capability to efficiently process TB-level data with 10k+ CPU cores. 
The system is publicly available and has been widely adopted in diverse research fields and real-world products such as Alibaba Cloud PAI. 
We actively maintain the system and share practical insights to foster research and applications of next-generation foundation models.
\end{abstract}

\renewcommand*{\thefootnote}{\fnsymbol{footnote}}
\footnotetext[1]{Co-first authors.} 
\footnotetext[2]{Equal contribution.} 
\footnotetext[3]{Corresponding authors, email addresses: \{yaliang.li, bolin.ding\}@alibaba-inc.com}
\renewcommand*{\thefootnote}{\arabic{footnote}}

\section{Introduction}
\label{sec:intro}

\textbf{Data Processing Challenges for Foundation Models.}
Foundation models require sophisticated pipelines for multimodal data across evolving paradigms in pre-training and post-training. 
While existing frameworks address specific aspects of text processing \cite{together2023redpajama,DolmaToolkit} or traditional big data workloads \cite{apache-spark}, they lack essential capabilities for contemporary multimodal workflows. Three critical gaps emerge:

\textit{Multimodal Processing Limitations:} Current tools provide inadequate support for cross-modal alignment and semantic-aware transformations crucial for vision-language models \cite{li2024mgm}. The transition from text-only systems like the inaugural version, \oursysI. \footnote{In this paper, references to \oursysI pertain to the SIGMOD publication \cite{chen2024dj} and the open-source codes versioned with the \href{https://github.com/modelscope/data-juicer/tree/v0.1.2}{v0.1.2.}}
to multimodal processing introduces architectural challenges in handling heterogeneous data types and inter-modal relationships.

\textit{Efficiency-Scalability Tradeoffs:} Traditional big data engines \cite{dask} and Python-based solutions \cite{lhoest-etal-2021-datasets} struggle with foundation models' unique computational patterns - simple per-sample operations applied at petabyte scales. This creates reliability risks in large-scale processing scenarios where late-stage errors can invalidate terabytes of computation.

\textit{Ecosystem Fragmentation:} Disjoint APIs across popular frameworks force practitioners into suboptimal workflow choices. The lack of unified abstractions hampers portable optimization and cross-platform execution.

\begin{figure}[h!]
    \centering
    \includegraphics[width=0.9\linewidth]{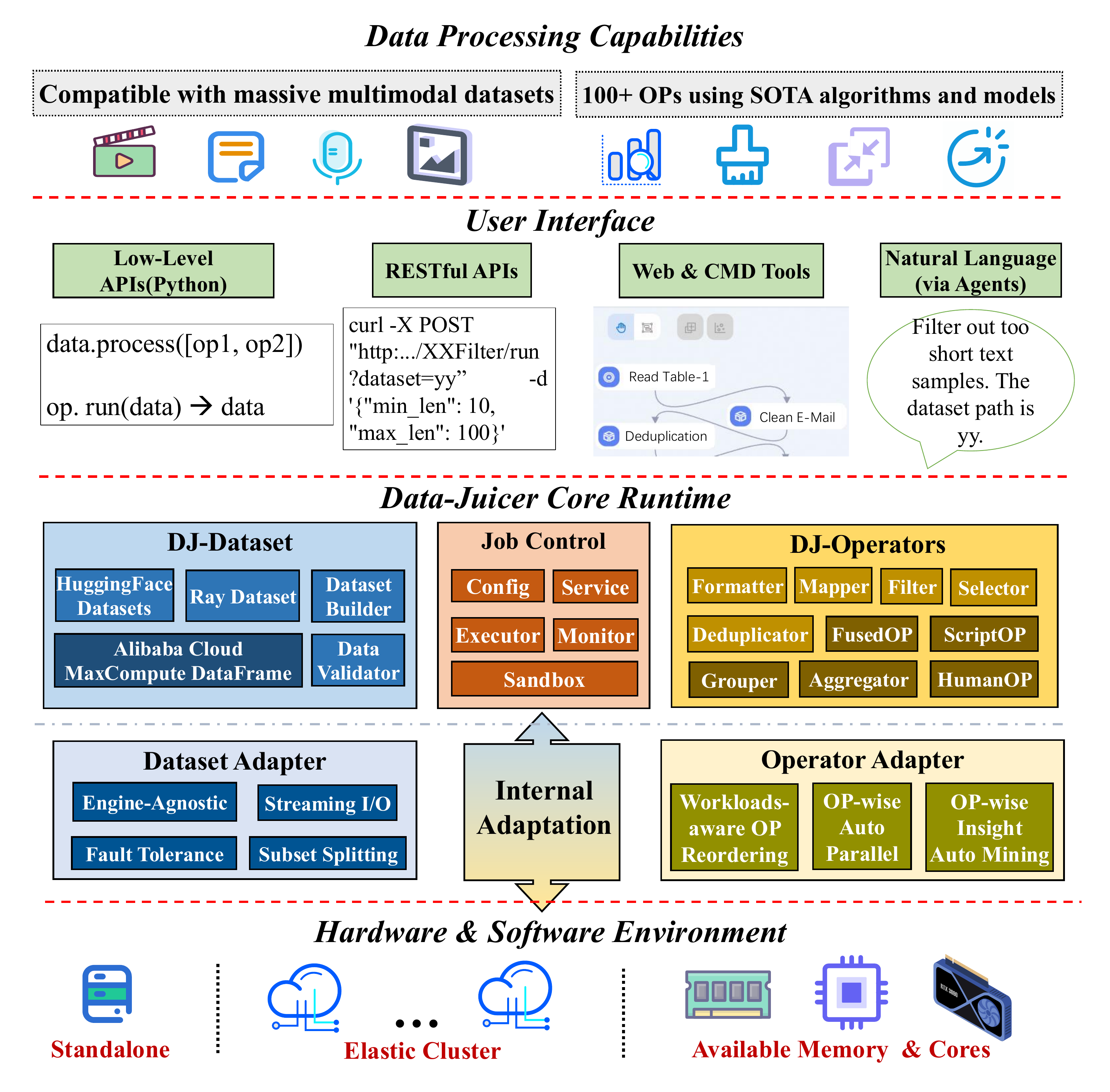}
        \vspace{-0.1in}
    \caption{The overview of \oursysII.}
        \vspace{-0.1in}

    \label{fig:overview}
\end{figure}

\textbf{Architecture Overview.}
\oursysII addresses these challenges with a layered architecture optimized for foundation models (Fig. \ref{fig:overview}):

\textit{Capability Layer:} Extends \oursysI's 50 text-only operators (OPs) for pre-training to 150+ multimodal OPs supporting text/image/video/audio processing and more post-training tasks. New OPs integrate foundation models (e.g., Tongyi-Qwen \cite{qwen3}, SDXL \cite{podell2023sdxl}) for semantic-aware filtering and cross-modal synthesis while maintaining compatibility with mainstream dataset hubs \cite{hf-dataset-hub,ms-dataset-hub}.

\textit{Interface Layer:} Provides multi-level APIs balancing flexibility and accessibility. Low-level Python APIs enable custom pipelines, while RESTful endpoints and a visual editor support rapid prototyping. Novel agent integration allows natural language specification of pipelines.

\textit{Runtime Layer:} Introduces four key enhancements:
(1) A unified \texttt{\oursys-Dataset} abstraction spanning local to cloud-scale execution environments; (2) Decoupled OPs with automatic optimization through adaptive batching and resource allocation; (3) Control plane supporting fault-tolerant execution and data-model co-development \cite{chen2024djsandbox} for insight mining; and (4) \texttt{Adapter} components enabling automatic hardware-software configuration and parallelism across diverse deployments.

\textbf{Contributions.} Our contributions are summarized below:

\textit{Enhanced System and Adaptive Techniques.} 
Incorporating feedback from numerous users and cutting-edge applications built upon its predecessor, \oursysII emerges as a new open-source system with enhanced multimodal processing capabilities. It emphasizes dedicated multi-layered adaptability and system optimization, integrating cloud-scale practices and diverse computation engines to dynamically and efficiently meet data processing demands.

\textit{Extensive Evaluation and Usages.}
We conduct thorough experimental analyses to assess the system's data processing capabilities under varied workloads, such as filtering-intensive, model-based, and semantically editable multimodal scenarios, using datasets ranging from millions to tens of billions of samples. We provide actionable insights and performance trade-offs across different use cases and resources, involving Ray and Alibaba MaxCompute \cite{maxcompute2024} with up to 100 nodes and 12,800 cores.

\textit{Community and Applications.} 
We have open-sourced this new system at \url{https://github.com/modelscope/data-juicer}, fostering sustained maintenance and engagement through practical events  \cite{qin2024synergy,dj-competition}, such as tutorials, surveys, data competitions, and co-optimization with community like Apache Arrow, Ray, and NVIDIA-Nemo-Curator teams. 
\oursysII also facilitates many foundation model researches such as those from Alibaba Tongyi \cite{ge2024data,qin2024federated,bai2024federated,jiao2024img,ling2025diversity,xu2025mindgym,zhou2024humanvbench,jiao2025detailmastertexttoimagemodelhandle}, and serves as the operator base for multiple Alibaba Cloud products such as PAI Designer \cite{paidesigner2024} and DLC-Ray \cite{paidlcray}, indirectly benefiting hundreds of internal and external customers across various real-world AI businesses. One of them has been running stably for over five months, processing data at a scale exceeding terabytes.

\section{Preliminaries and Design Rationale} 
\label{sec:preliminary}

\subsection{Related Work \& Core Challenges} 
\label{sec:related-work}
While existing systems address big data \cite{apache-spark} or text-centric model data \cite{together2023redpajama,DolmaToolkit,Jennings_NeMo-Curator_a_toolkit}, a dedicated, open-source framework for multimodal foundation models is lacking. This gap stems from three core challenges rooted in the unique demands of modern AI workflows \cite{qin2024synergy,bai2024survey}:
\textbf{1) Functionality:} Processing requires deep semantic understanding and cross-modal alignment (e.g., for video, image, text, audio) \cite{liu2023llava,li2024mgm}, moving beyond the structured data focus of traditional systems.
\textbf{2) Efficiency:} Workloads are dominated by simple, per-sample operations (mappers, filters) at massive scale, often involving costly model inference, which contrasts with the complex aggregation queries optimized by conventional engines \cite{apache-spark,dask}.
\textbf{3) Usability:} The practitioner ecosystem is centered on Python, demanding native, intuitive interfaces that abstract away backend complexities \cite{Paszke-NeurIPS-2019-Pytorch,lhoest-etal-2021-datasets}.
These challenges require new system designs, motivating the specific goals of \oursysII.

\subsection{System Design Goals in \oursysII}
\label{sec:sys-design-goal}

\textbf{Comprehensive Multimodal Processing.}
To address functionality gaps, extensive operators are required for collecting, cleansing, annotating, and synthesizing data across modalities like video, image, text, and audio, integrating both perceptual and cognitive information \cite{videoworldsimulators2024,podell2023sdxl}.

\textbf{Efficient and Optimized Data Flow.}
To tackle efficiency issues, we aim to accelerate high-frequency basic operators and efficiently manage high-cost semantic operations, while minimizing I/O and data transfer overhead.

\textbf{Intuitive and Engine-Agnostic Interface.}
To enhance usability, we aim to protect users from the complexities of underlying execution engines, with easy-to-use Python APIs, graphical interface, and natural language interaction.

\textbf{Adaptive and Scalable Execution.}
The system must adapt to diverse computational environments and workloads. It is designed to intelligently orchestrate and optimize data processing across various backends—from local execution to distributed computation on large-scale clusters.

\subsection{Key Differences from Prior Systems}
While its predecessor, \oursysI, laid a foundation for text data processing, it faced limitations in multimodal support, programmability, and fragmented workflows across different backends. \oursysII is architected to overcome these issues. Unlike modality-specific toolkits \cite{Jennings_NeMo-Curator_a_toolkit,penedo2024datatrove} or general big data systems \cite{apache-spark}, \oursysII introduces a composable operator system that generalizes across data types and training tasks while retaining scalability. It distinctively integrates deep learning models as first-class citizens in the data pipeline, emphasizing both computational efficiency and semantic richness. Table~\ref{tab:comparison} summarizes the key advancements from \oursysI to \oursysII.

\begin{table*}[t]
\centering
\small
\caption{Comparison of Data-Juicer 1.0 and 2.0.}
\label{tab:comparison}
\begin{tabularx}{\textwidth}{l|l|X}
\toprule
\textbf{Feature} & \textbf{Data-Juicer 1.0} & \textbf{New in Data-Juicer 2.0} \\
\midrule
Data Modality & Text-only (\textasciitilde50 OPs) & + \textasciitilde100 Image, video, audio OPs \\
\midrule
Major OP Types & Formatter, Filter, Mapper & + Grouper, Aggregator, FusedOP, HumanOP \\
\midrule
Deduplicator & Standalone-only & + Ray-based distributed deduplication \\
\midrule
Interaction & CLI, Low-level APIs & + RESTful APIs, Web UI, NL Interface \\
\midrule
Execution Engines & HF Datasets, Beam & + Tighter integration with Alibaba PAI, Ray \\
\midrule
Optimization & Greedy OP fusion & + Advanced OP reordering, auto resource allocation, batched/parallel execution, etc. \\
\midrule
Compute Scale & 1,000+ cores & 10,000+ cores \\
\midrule
Data Scale & \textasciitilde70M samples (TB-level) & \textasciitilde70B samples (PB-level) \\
\midrule
Code Contribution & 26 PRs, \textasciitilde34k LoC & + 100+ new PRs, \textasciitilde40k added, \textasciitilde8k deleted \\
\bottomrule
\end{tabularx}
\end{table*}

\section{Processing Capabilities Beyond Text \& Pretraining}
\label{sec:mm_abilities}

\oursysII extends its predecessor to support the processing of multimodal datasets on more tasks for a wide range of foundation models with about 100 new OPs. 
Images, videos, and audios are all valid inputs of \oursysII, and it's able to process datasets for post-training tasks, including supervised fine-tuning (SFT), Reinforcement fine-tuning (RFT), and so on. 
Detailed list of these OPs is in Appendix~\ref{sec:appendix:list_of_new_ops}. 
For clarity and ease of reference, the OPs are categorized from the following aspects, and their distributions are visually displayed in Fig.~\ref{fig:op_num_dist}:

\begin{figure}[h!]
\vspace{-0.1in}
    \centering
    \subfloat[Operator type.\label{fig:op_num_dist:op}]{\includegraphics[width=0.24\linewidth]{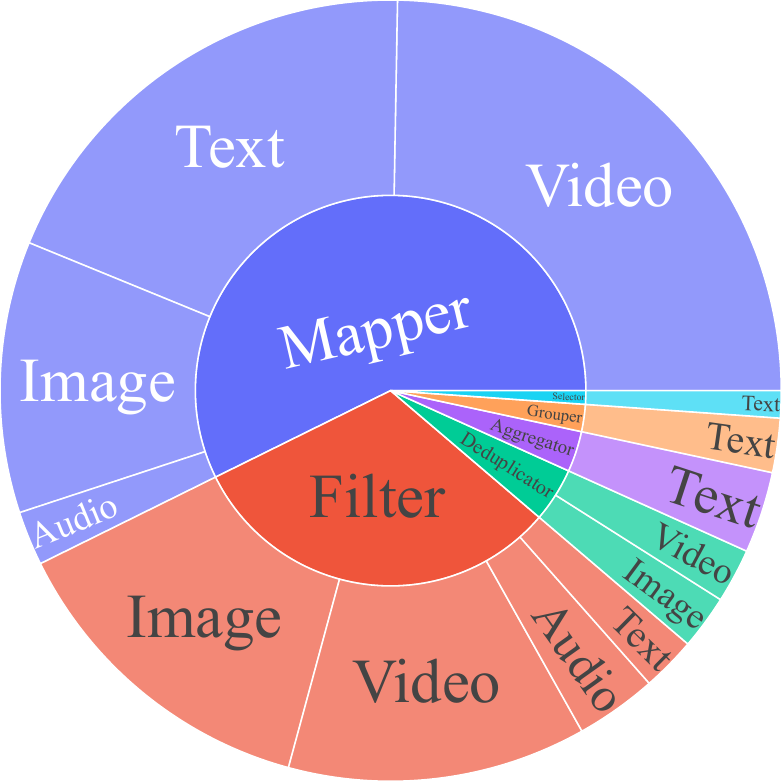}}\hspace{1pt}
    \subfloat[Modality type.\label{fig:op_num_dist:modality}]{\includegraphics[width=0.24\linewidth]{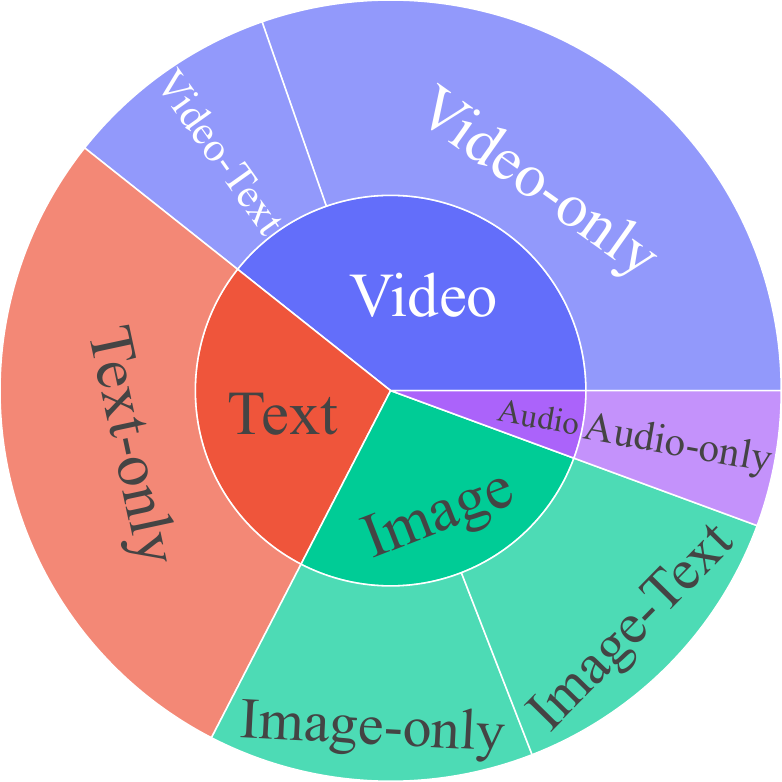}}\hspace{1pt}
    \subfloat[Function type.\label{fig:op_num_dist:func}]{\includegraphics[width=0.24\linewidth]{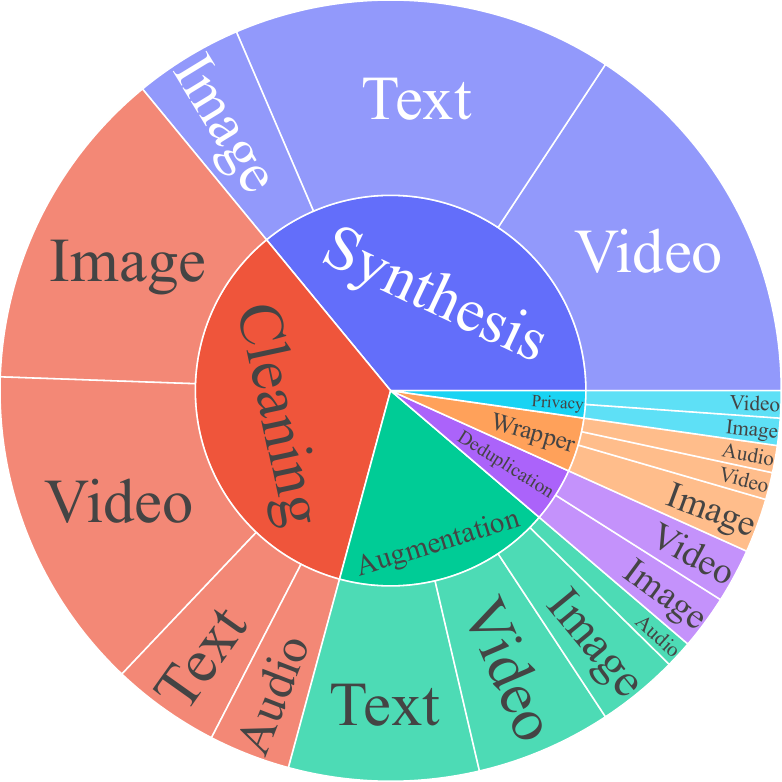}}\hspace{1pt}
    \subfloat[Implementation type.\label{fig:op_num_dist:impl}]{\includegraphics[width=0.24\linewidth]{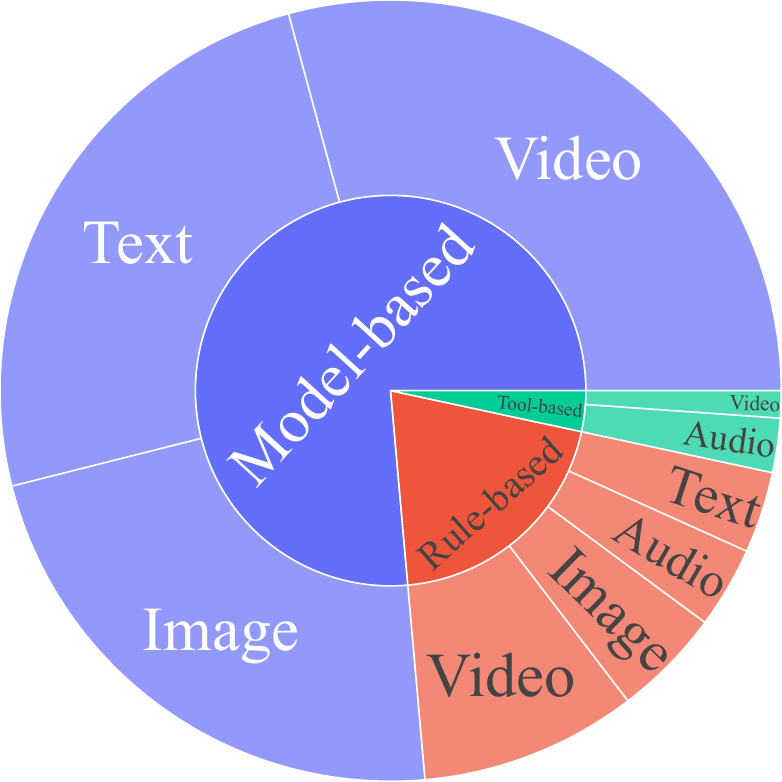}}
    \vspace{-0.05in}
    \caption{Distribution of new OPs across various dimensions. The high-resolution versions of these subfigures can be found in Appendix, Figure \ref{fig:op_num_dist:op_hd}, \ref{fig:op_num_dist:modality_hd}, \ref{fig:op_num_dist:func_hd}, and \ref{fig:op_num_dist:impl_hd}.}
    \vspace{-0.05in}
    \label{fig:op_num_dist}
\end{figure}

\textbf{Operator Types:} 
    These new OPs are built on both original types in \oursysI and types newly introduced by \oursysII. Among them, Formatters load datasets; Mappers edit samples; Filters compute data stats and remove samples accordingly; Deduplicators find redundant samples; Selectors sample data based on preferred ranks or rules; Groupers batch samples, and then Aggregators combine them into one. 
    Notably, \oursysII introduces a new type of OP named HumanOP. It's built upon Label Studio \cite{LabelStudio} and involves asynchronous human annotations and feedback during data processing and time-delayed downstream training tasks. This new OP helps to build SOTA human-in-the-loop procedures, such as reinforcement learning from human feedback (RLHF). These OP types will be detailed later in Sec.~\ref{sec:dj_op}. 
    Statistically, about 90\% of OPs are Mappers and Filters, reflecting trends of user needs in the foundation model domain. New variants in other types, such as Formatters and Deduplicators, remain stable due to the broad applicability of already supported formats (e.g., JSONL, Parquet, MP4) and classical algorithms like MinHash \cite{broder1997resemblance}.

\textbf{Modality Types:} The majority of new OPs are concentrated on video/image/audio/text-only processing, with about 20 OPs dedicated to cross-modal data processing.
Among these new OPs, almost 3/4 of them are aimed at multimodal data processing, covering images, videos, and audio. There are both single-modality OPs, such as \textitt{video\_motion\_score\_filter} that scores the dynamics of videos, and cross-modal OPs such as \textitt{phrase\_grounding\_recall\_filter} and \textitt{video\_captioning\_from\_summarizer\_mapper}, which measure the alignment between different modalities and generate contents from one modality to another, respectively. 
Detailed showcases are in Appendix \ref{sec:appendix:new_ops_showcase}.

    \textbf{Function Types:} Data cleaning and analysis operations constitute nearly 1/3 of the new OP suite. In particular, \oursysII introduces about 50 OPs for data synthesis and augmentation on cross-modal, post-training, and reinforcement learning scenarios, enabling any-to-any generation among 4 supported modalities and improving textual dialogs based on varied information and needs. Furthermore, new privacy protection OPs, such as removing not-safe-for-work (NSFW) contents or blurring human faces, and several wrapper OPs for established 3rd-party tools, such as FFmpeg \cite{ffmpeg}, have been integrated, enabling users to conveniently invoke existing professional commands.

\begin{wrapfigure}{r}{0.25\textwidth}
    \includegraphics[width=\linewidth]{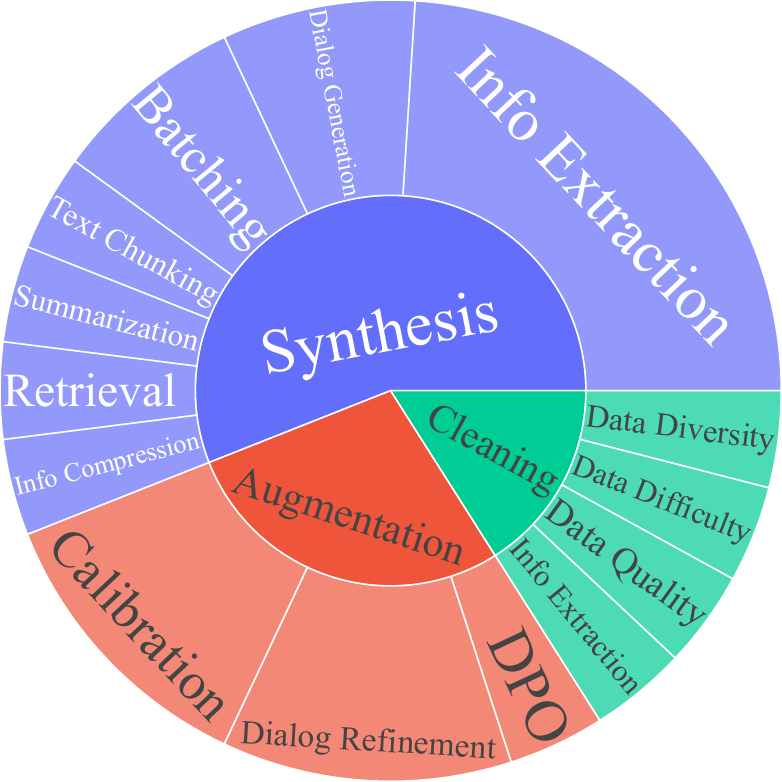}
    \caption{Distribution of fine-grained function types for text-only OPs. A high-resolution version can be found in Appendix, Figure \ref{fig:fg_func_type_hd}.}
\vspace{-0.18in}
    \label{fig:fg_func_type}
\end{wrapfigure}
Besides, 25 text-only OPs are introduced for post-training tasks. We tag them with fine-grained function types and summarize their distribution in Fig.~\ref{fig:fg_func_type}. Among them, there are information extraction and synthesis OPs for curating dialog samples, as well as calibration and refinement OPs for optimizing questions and responses in SFT and RFT tasks.
Beyond previous rule-based OPs to assess quality, difficulty, and diversity of datasets in \oursysI, more foundation-models-based OPs emerge to score the datasets from diverse dimensions and analyze rationale of given scores. 
This practice is flexibly and useful for vertical and cross-task scenarios. For example, rule-based methods usually struggle to analyze difficulties of math and financial problems, while this kind of OPs generally perform better with reduced customization and tuning efforts.

\textbf{Implementation Types:} 
The new OPs include both novel algorithms from \oursysII and implementations based on SOTA methodologies \cite{schuhmann2021laion,zhu2023multimodal,gadre2024datacomp} from the community. 
Some OPs offer multiple versions to accommodate varying computational resources, such as CPU-only (e.g., with OpenCV \cite{opencv_library}) or GPU-based configurations (e.g., with RAFT \cite{teed2020raft}).
In the distribution, model-based OPs dominate, as semantic-aware processing often requires advanced models, such as large models for general-purpose understanding and generation via SDXL, GPT, and Qwen. Data processing with foundation models is becoming more popular and helpful.

\section{Towards a More Accessible System}
\label{sec:interface}

\oursysII significantly improves accessibility over \oursysI with multiple new interfaces catering to both novice and expert users.

In terms of user interfaces, in addition to the way to process data with an all-in-one configuration file provided in the previous version, we introduce 4 more user interaction methods. (1) \textbf{Low-level APIs.} We make the underlying implementation more transparent, and expose many extensible interfaces of \texttt{\oursys-Dataset}, \texttt{\oursys-Operators}, such that users can conveniently integrate \oursys in their code. (2) \textbf{RESTful APIs.} For some web server usages, we provide one-click generation of high-performance web APIs capable of automatically discovering, registering, and adapting OP classes and tools. Users can start the server easily and trigger data processing across computing nodes. (3) \textbf{Web tools.} Based on RESTful APIs, \oursys's capabilities are integrated into Alibaba Cloud's visual modeling product, PAI-Designer \cite{paidesigner2024}, which provides a drag-and-drop UI for users to organize the data processing pipelines and allows users to make use of more products in the Alibaba Cloud ecosystem. (4) \textbf{Natural language interaction.} It's challenging to know when and how to use each of the 100+ diverse OPs. In response, we adopted AgentScope \cite{gao2024agentscope}, a multi-agent platform, for low-code integration, using prompt-optimized reasoning and acting (ReAct) \cite{Yao-arxiv-2022-ReAct} agents to align OP functionalities with our RESTful APIs. 
Users only need to tell the agents how they want to process the dataset, and the system will automatically handle the job analysis and execution. More details about these interfaces are in Appendix \ref{app:interface}.

As the 100+ OPs in \oursysII cover quite diverse multimodal data and training tasks, the complete capabilities require heavy dependencies and take the risk of runtime errors. This may slow down the system installation, initialization, and troubleshooting. 
We thus prepare a minimal set of requirements and split OPs' dependencies into subgroups based on their modalities and usage categories. Users only need to install the lite \oursys in a quick way, and the optional dependencies will be lazy-loaded when using specific OPs. Moreover, we maintain an automated unit and regression testing mechanism, ensuring over 85\% test coverage.
These actions make our system more user-friendly, helping both beginners and light users get started with it.
\vspace{-0.02in}
\section{Towards a More Flexible, Robust, \& Efficient Runtime}
\label{sec:core_runtime}
\vspace{-0.02in}
\subsection{\oursys-Dataset}
\label{sec:dj_dataset}

\textbf{Unified Execution Abstraction.} 
\oursysII introduces a \texttt{\oursys-Dataset} class that abstracts heterogeneous computational engines (Hugging Face Dataset, Ray Data, MaxFrame-DataFrame), while preserving native API compatibility through a Facade-pattern design \cite{gamma1995design}. 
It provides unified interfaces for standalone/distributed execution modes, enabling transitions between environments while hiding engine-specific complexities. The class supports chainable processing workflows via templated methods with multiple \texttt{process()} call manners. More implementation details about these modes are provided in Appendix~\ref{app:exec_modes} and Listing~\ref{listing:dj_dataset_interface}.

\textbf{Reliable Data Loading.}
Data loading is now more systematic, moving beyond simple path specifications.
The new \texttt{DatasetBuilder} class supports explicit dataset source specification (e.g., local, remote) and customizable configuration for loading datasets. To improve data loading reliability, a novel \texttt{DataValidator} module is introduced to check and validate data sources, schemas, and whether the dataset meets the processing goal. For example, post-tuning data should contain dialogs, and image paths should exist in image-captioning datasets.

\textbf{Token-Aligned Data Schema.} 
Our intermediate schema represents multimodal data through special tokens (e.g., \texttt{<\_\_dj\_\_image>}, where ``dj'' stands for \oursys) in text fields, with chunk-based alignment using \texttt{<|\_\_dj\_\_eoc|>} separators. This token-centric design supports both simple cross-modal pairs and complex interleaved datasets (MMC4-style \cite{zhu2023multimodal}), while preserving positional relationships and reducing redundant computing of shared media files. Core fields (``text'', ``query''/``response'') also align with the popular formats of language foundation models, supplemented by bi-directional conversion tools for training ecosystems like LLaMA-Factory \cite{zheng2024llamafactory} and ModelScope-SWIFT \cite{zhao2024swiftascalablelightweightinfrastructure} (implementation details are in Appendix~\ref{app:schema_impl}).

\textbf{Internal Adaptation with Industrial-Grade Optimization.}
The intermediate layer design enables us to adaptively take advantage of different underlying engines and systematically improve usability. We can optimize once and apply the enhancements with both execution modes. For example, our minhash-based Deduplicators achieve engine transparency and performance superiority, yielding 3.3x speedup over vanilla Ray with load-balanced union-find \cite{bts} and hash-based aggregation.

Moreover, large-scale data processing faces errors from data corruption or unreliable operators (e.g., malformed LLM outputs), which can halt entire pipelines. \oursysII enhances resilience at multiple levels. For operator-specific failures, it performs pre-flight validation on LLM responses and employs automatic retries with backoff. More systematically, to overcome the coarse-grained fault tolerance of underlying engines like Ray (which require full restarts), we introduce operator-level checkpointing and fine-grained recovery. This allows pipelines to resume from the last successful stage rather than from scratch, ensuring robust progress even with intermittent task failures. Schema-compatible empty samples are also used to maintain dataset integrity (e.g., see Fig. \ref{fig:demo_fault_tolerance} in Appendix~\ref{app:fault_tolerance}).
These optimizations are detailed in Appendix~\ref{app:dataset-optimization}.

\vspace{-0.02in}
\subsection{\oursys-Operators}
\label{sec:dj_op}

\textbf{Composable Processing Primitives.} 
We extend \oursysI's five atomic OPs (Formatter/Filter/Mapper/Deduplicator/Selector) with five compositional types: Grouper/Aggregator/FusedOP/ScriptOP/HumanOP. Using Strategy/Decorator patterns \cite{gamma1995design}, they enable flexible algorithm encapsulation and runtime behavior extension (OP taxonomy in Appendix~\ref{app:op_taxonomy}). FusedOP optimizes batch-wise OP fusion (Appendix~\ref{app:fusedop_example}), while ScriptOP integrates custom Python logic. 
The Grouper takes a \texttt{\oursys-Dataset} as input and groups data samples into batches, which can then be input into the Aggregator for subsequent aggregation.
All OPs follow a unified template method \texttt{run()} with automatic parallelism configuration, decoupling execution logic from runtime engines.

We design an abstract factory class that centralizes common functionalities (parameter preparation, serialization) while allowing for implementation of OP-specific execution logic through overridable methods (e.g., \texttt{compute\_stats()} for Filters). 
This eliminates executor dependencies and enables easy standalone customization/inspection/testing of individual OPs as their functionalists are constrained to be implemented in a self-contained manner. 

\textbf{OP-wise Optimization.}
We incorporate several automatic adaptation features for \texttt{\oursys-Operator}, aiming at balancing resource constraints and operational efficiency without requiring users to understand hardware specifics or implementation details.

We employ a dedicated \texttt{Adapter} class, which uses a \texttt{probe\_small\_batch()} method to systematically probe and analyze essential information by applying individual OPs on randomly sampled data in runtime.
As a result, we enhance \oursysI’s greedy OP fusion with adaptive reordering based on estimated OP speeds. Faster OPs precede slower ones within commutativity constraints, optimizing end-to-end latency (validation in Appendix \ref{exp:op-fusion}).
Moreover, using a uniform parallelism granularity across all OPs in a data pipeline can cause OOM issues for some and resource underutilization for others. In \oursysII, with auto-configuration, model-based OPs use GPU/quantization (e.g., vLLM), while I/O-bound OPs use hierarchical parallelism across batched processing, multiprocessing and multithreading, taking the concurrent opportunities between I/O and computation latencies.
The implementation details for OP \texttt{Adapter} are in Appendix~\ref{app:op-adapter}.

\textbf{OP Insight Mining.}
\label{sec:op_wise_insight}
The combined effects of sequential OPs are not always additive, as validated in \cite{chen2024djsandbox}. Existing tools like \oursysI \cite{chen2024dj} and Falcon \cite{refinedweb} focus on coarse-grained metrics (e.g., data volume changes via Sankey diagrams) but lack fine-grained analysis.
To formulate better data recipes, \oursysII tracks dataset statistics (e.g., perplexity) and semantic tags (e.g., image categories) after each OP execution. Built-in \texttt{Analyzer} leverages Filters and modality-specific tagging OPs to generate histograms of statistical distributions and semantic categories (an example is shown in Appendix, Fig.~\ref{app:fig:op-insight}).
\oursysII automates metric comparison between consecutive OPs, producing reports that highlight significant lineage-level variations. For example, a sudden text-length reduction after applying a BLIP-2 \cite{li2023blip2} image-text matching Filter could indicate noisy captions requiring adjustment. These insights help users systematically evaluate OP impacts, optimize data recipes, and identify unintended correlations.

\vspace{-0.02in}
\subsection{Processing Job Control}
\label{sec:dj_control}

\textbf{End-to-End Workflow Orchestration.} 
Our \texttt{Executor} module integrates configurable data pipelines with monitoring/checkpointing capabilities, codified into reusable data recipes (the red box in Fig.~\ref{fig:overview}). We also provide a sandbox suite that enables data-model co-development through template workflows connecting to model training/evaluation infrastructures \cite{chen2024djsandbox}, allowing cost-effective exploration of data-compute effects before full deployment.
More implementation details are in Appendix~\ref{app:job_detail}.
processing solution generation by foundation-model-based agents.

\textbf{Extensibility for Diverse Applications.}
Many effect-proven and illustrative workflows have been encapsulated in YAML recipes and maintained online \cite{dj-cookbook}, catering to various vertical domains, such as multimodal data synthesis and persona-oriented dialog processing \cite{jiao2024img, zhou2024humanvbench,role_play_recipe}. These facilitate interface exposure and reuse across different levels (Sec.~\ref{sec:interface}), simplifying recipe routing and tailored processing solution generation by foundation-model-based agents.
\vspace{-0.03in}

\section{Experiments \& Insights}
\label{sec:exps}
\label{sec:exp_efficiency_distributed}
\vspace{-0.02in}

\subsection{Experiment Setup}
\label{sec:exp_efficiency_dist:settting}
We evaluate \oursysII's efficacy across three data scales: small (560K-2.24M samples), medium (5.6M-56M samples), and large (56M-70B samples), covering both multimodal and text-only processing. Our test suite executes five representative data operations per recipe across three compute engines (Standalone, Ray, MaxCompute) using Alibaba Cloud resources (1-100 nodes, 64-12,800 CPU cores). All worker nodes maintain identical hardware configurations for fair comparison. More implementation details are in Appendix~\ref{app:exp-setup-details}.

\subsection{Overview of System Performance Gains}
\label{sec:overview_gains}

\textbf{Macro-level Scalability} We systematically evaluated end-to-end performance by scaling datasets from 1x to 12,500x. The results, presented later from \S\ref{sec:exp_scales_small} to \S\ref{exp:large-scale}, confirm the robust scalability of \oursys and provide actionable insights for choosing the optimal compute engine across different operational scales.

\textbf{Micro-level Optimizations} The strong end-to-end performance is underpinned by a suite of targeted optimizations. Below, we highlight the most impactful ones and point to where their effectiveness is validated (more details are in presented in Appendix \ref{sec:exp_efficiency_runtime}):
\begin{itemize}[leftmargin=*]
    \item \textbf{Resource Utilization \& Adaptive Splitting:} Our adaptive data splitting for Ray offers a 2x-3x acceleration on large datasets. The mechanism's impact on reducing network I/O and improving CPU consistency is visually analyzed in Fig.~\ref{fig:cloud_exps:text_large_resource} and discussed in \S\ref{exp:large-scale}.
    \item \textbf{Workload-aware OP Reordering:} For complex recipes, this strategy, along with OP fusion, can cut processing time by up to 70.22\%. The quantitative benefits are detailed in the ablation studies in Fig.~\ref{fig:probe_exps}.
    \item \textbf{Automatic GPU Resource Allocation:} Critical for multimodal workloads, this prevents OOM errors and can save up to 99\% of processing time. Its performance across various VRAMs is quantified in Table~\ref{tab:cuda_exp}.
    \item \textbf{Batched Data Processing:} Optimizing batch size and concurrency is key. As shown in Fig.~\ref{fig:hie_exps:batch}, this can reduce processing time by up to 84\%.
\end{itemize}
A key takeaway is that adaptive mechanisms for batching, resource allocation, and execution planning are crucial for mitigating hardware underutilization in modern data-centric AI pipelines.

\subsection{The Case of Small Scales}
\label{sec:exp_scales_small}

\textbf{Performance Profile.}
As shown in Fig.~\ref{fig:cloud_exps:mm_small}, when processing small-scale multimodal datasets, the standalone mode with Hugging Face Dataset is efficient and comparable to the Ray mode with a single node. Additional Ray nodes provide further but limited acceleration (speedup ratios between 138\% and 226\% with 4 nodes). For text-only datasets (Fig.~\ref{fig:cloud_exps:text_small}), the standalone mode remains efficient. However, for the Ray mode, 4x node increments yield smaller speedups (148\%) or even increased processing times due to dominated I/O and communication costs. 
Thus, \textit{with \oursys, processing datasets with hundreds of thousands of samples on a single machine is both efficient and cost-effective for most users}.

\textbf{Implications \& Typical Application Scenarios.}
In this scale, \oursys enables two critical capabilities for data-centric AI research: (1). \textit{Rapid recipe prototyping}. It simplifies data-model co-design as exemplified by extensive sandbox experiments \cite{chen2024djsandbox}, covering text-to-video generation, image-text pre-training, image-to-text generation for general image understanding, image captioning, and model prompt optimization.
(2). \textit{Structured insight mining}. It helps to support 5 open foundation model competitions on data filtering, augmentation, and synthetic data generation with 3,000+ teams \cite{dj-competition}. Key lessons learned reveal that standardized and systematic actions provided by \oursys (e.g., YAML recipes, probe sampling, visual analytics integration) accelerate data analysis and understanding compared to ad-hoc implementations.

\subsection{The Case of Medium Scales}
\textbf{Performance Profile.}
When datasets are scaled to 56M samples, processing times increase significantly with the standalone mode, thus, its performance is omitted in Fig. \ref{fig:cloud_exps:mm_medium} and Fig. \ref{fig:cloud_exps:text_medium}. Here, \textit{the Ray mode outperforms in all instances, demonstrating considerable speedups with increasing node counts, making it the recommended choice for medium-scale scenarios}. Moreover, compared to native nodes of Elastic Computing Service (ECS, green lines), the Ray mode on Deep Learning Containers (DLC, red lines) is faster, saving 24.8\% of processing time due to the Alibaba Cloud's optimization dedicated to cluster networking.

\textbf{Implications \& Typical Application Scenarios.}
In this scale, \oursys boosts data flywheels once we find high-quality data recipes.
There have been many synthesis-based and ready-to-use data recipes built upon \oursys like \cite{jiao2024img,zhou2024humanvbench,xu2025mindgym}, where more compute investment brings larger-size datasets. 
Several lessons were learned from them: 
(1) Although the loss functions of foundation models are relatively standardized, we can flexibly inject preferred inductive bias with dedicated data synthesis, such as contrastive learning \cite{jiao2024img} and data-difficulty based curriculum learning \cite{xu2025mindgym}.
(2) Foundation models emerge with expert-level knowledge, which can be used as proxy annotators, largely reducing manual labeling costs in benchmark construction \cite{zhou2024humanvbench}.

\subsection{The Case of Large Scales}
\label{exp:large-scale}
\begin{figure}[h!]
    \centering
    \subfloat[Small-scale text-only data.\label{fig:cloud_exps:text_small}]{\includegraphics[width=0.33\linewidth]{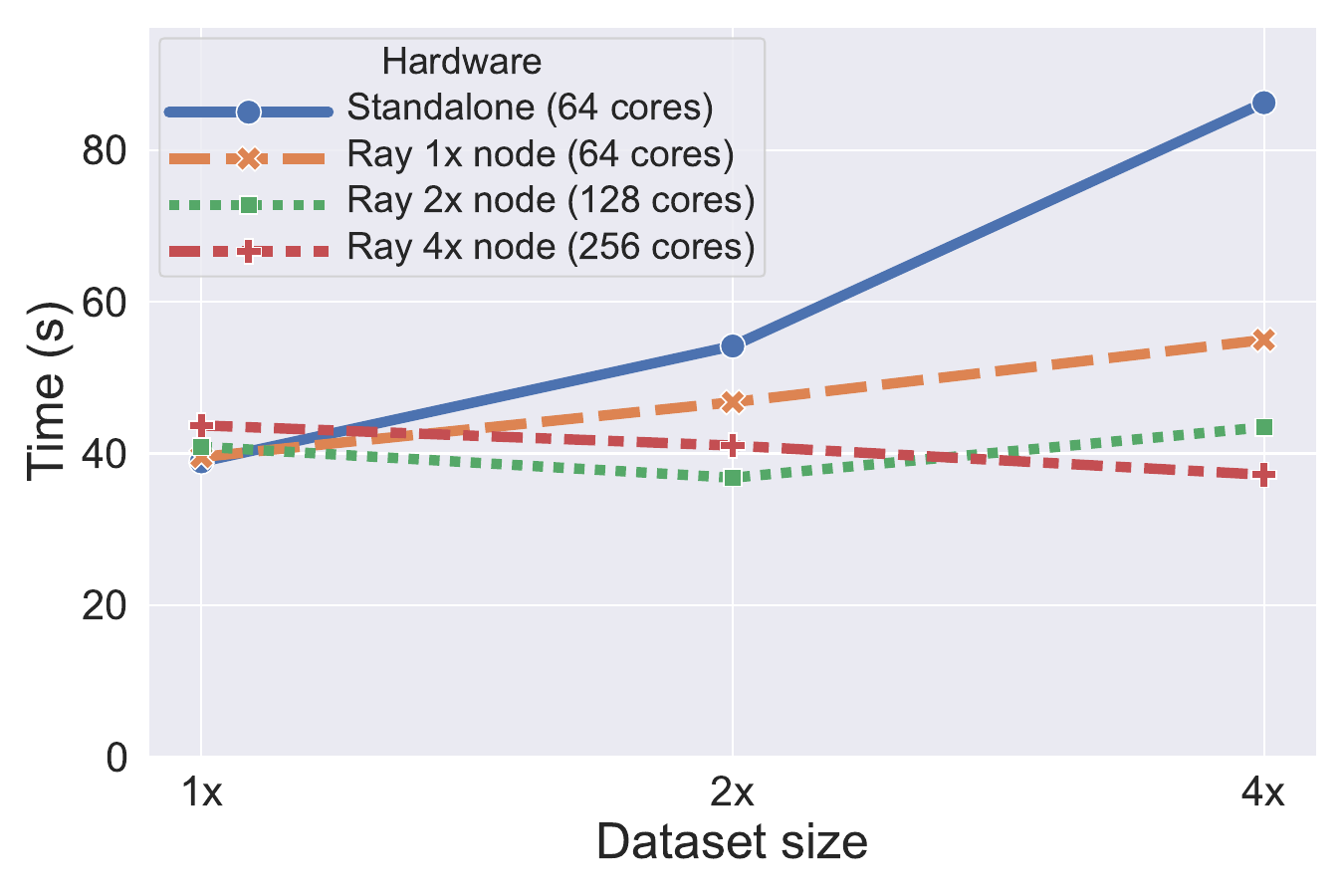}\vspace{-0.1in}}
    \subfloat[Medium-scale text-only data.\label{fig:cloud_exps:text_medium}]{\includegraphics[width=0.33\linewidth]{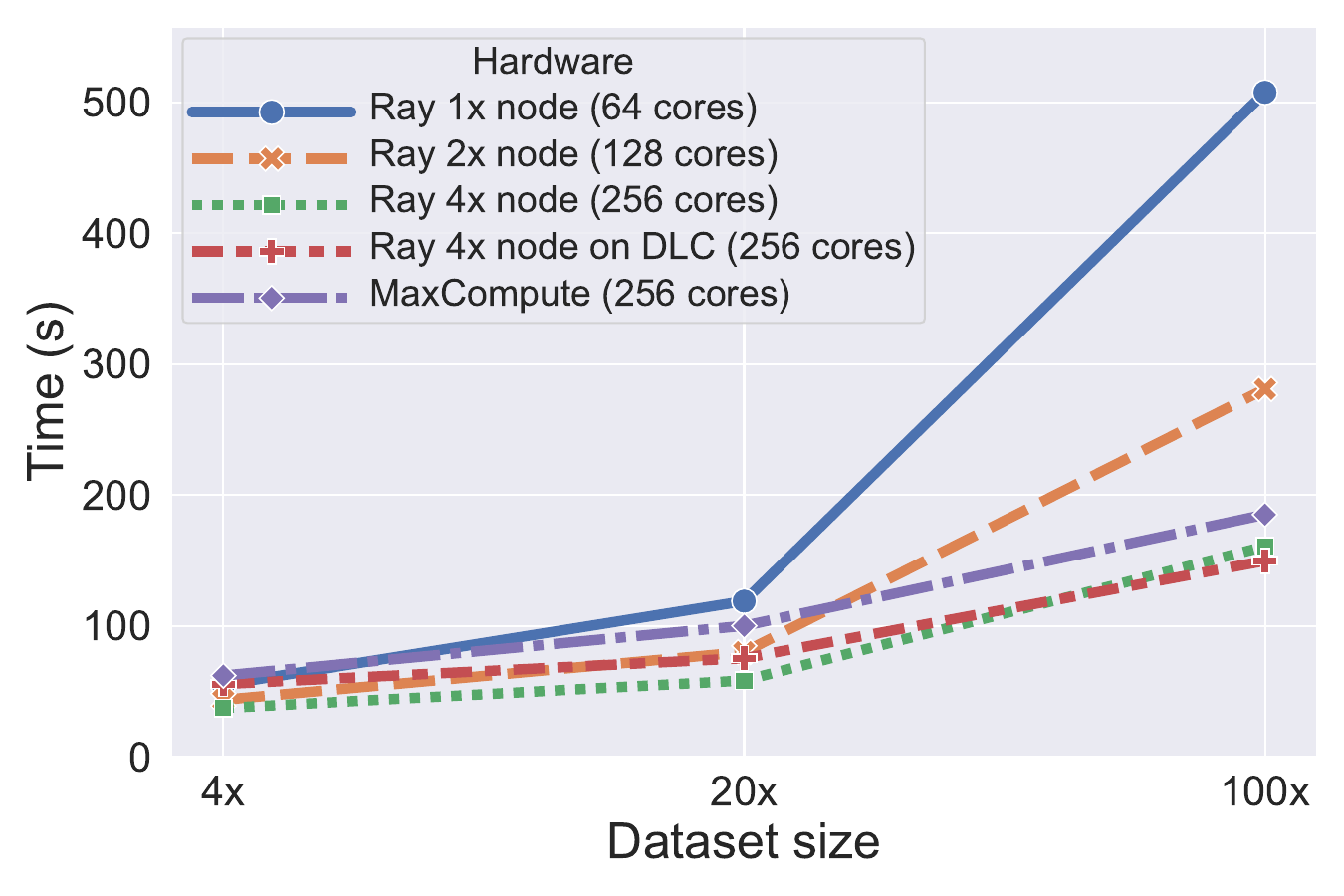}\vspace{-0.1in}}
    \subfloat[Large-scale text-only data.\label{fig:cloud_exps:text_large}]{\includegraphics[width=0.33\linewidth]{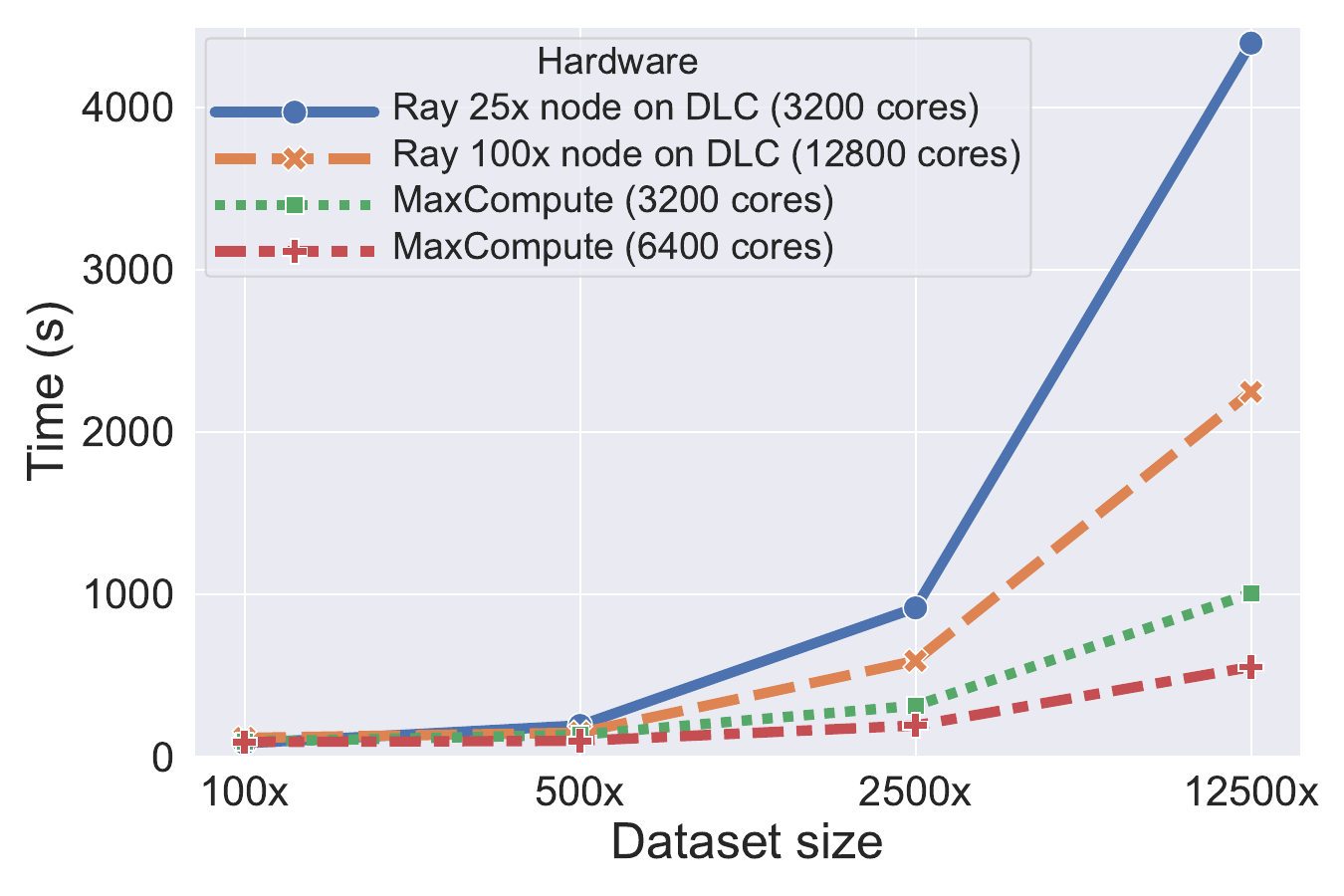}\vspace{-0.1in}}
    \\
    \vspace{-0.1in}
    \subfloat[Small-scale multimodal data.\label{fig:cloud_exps:mm_small}]{\includegraphics[width=0.33\linewidth]{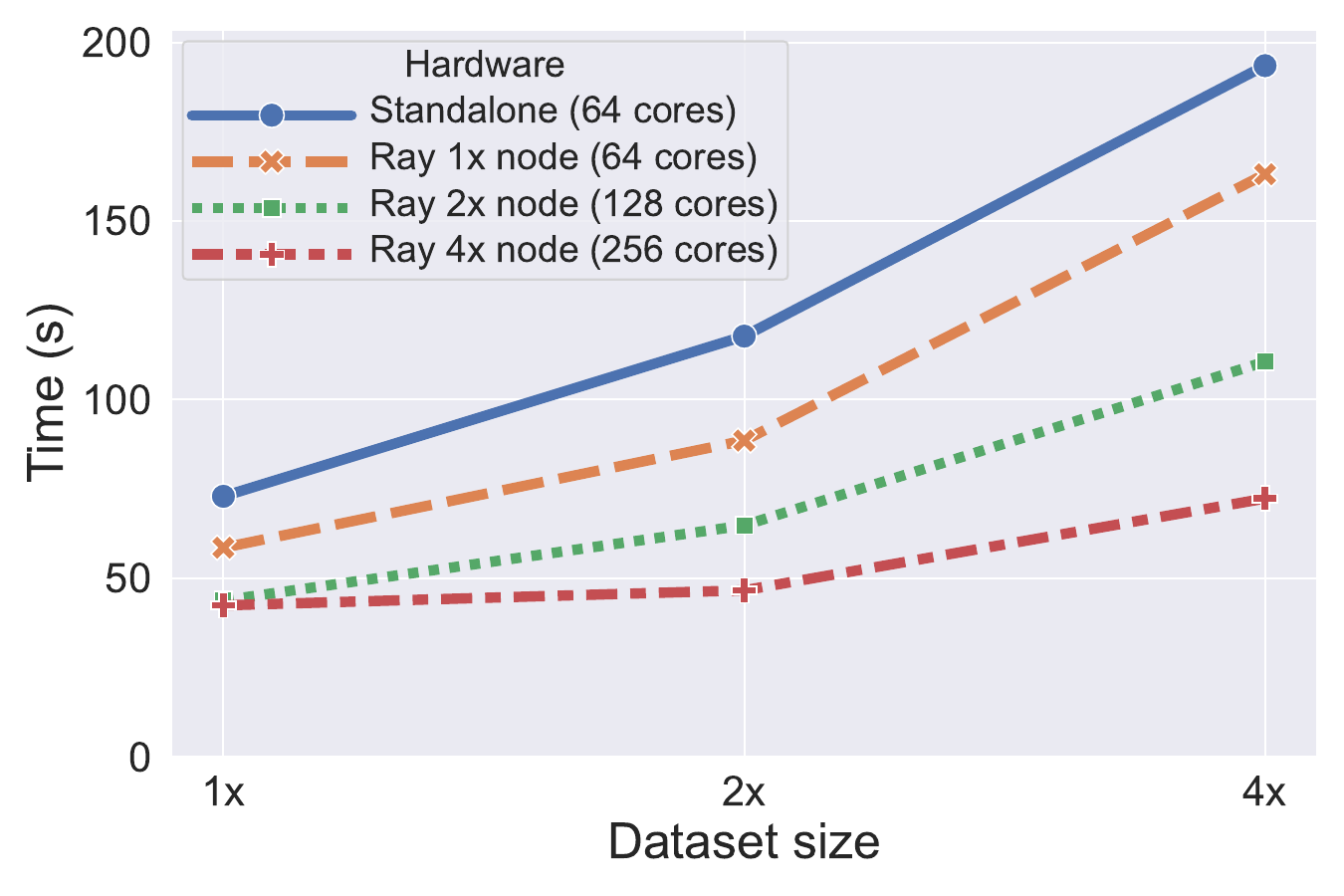} \vspace{-0.1in}}
    \subfloat[Medium-scale multimodal data.\label{fig:cloud_exps:mm_medium}]{\includegraphics[width=0.33\linewidth]{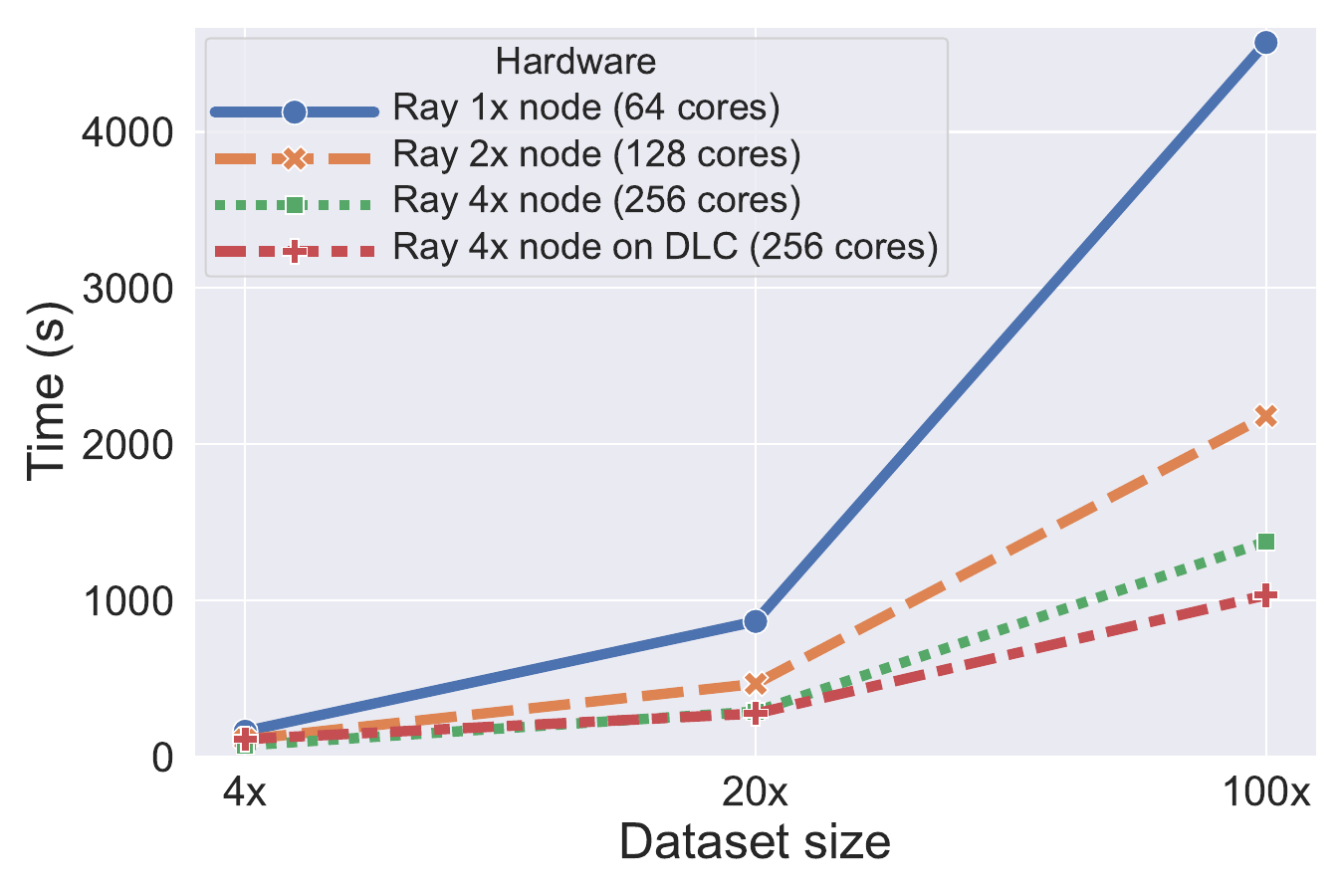}\vspace{-0.1in}}
    \subfloat[Resource utilization comparison.\label{fig:cloud_exps:text_large_resource}]{\includegraphics[width=0.33\linewidth]{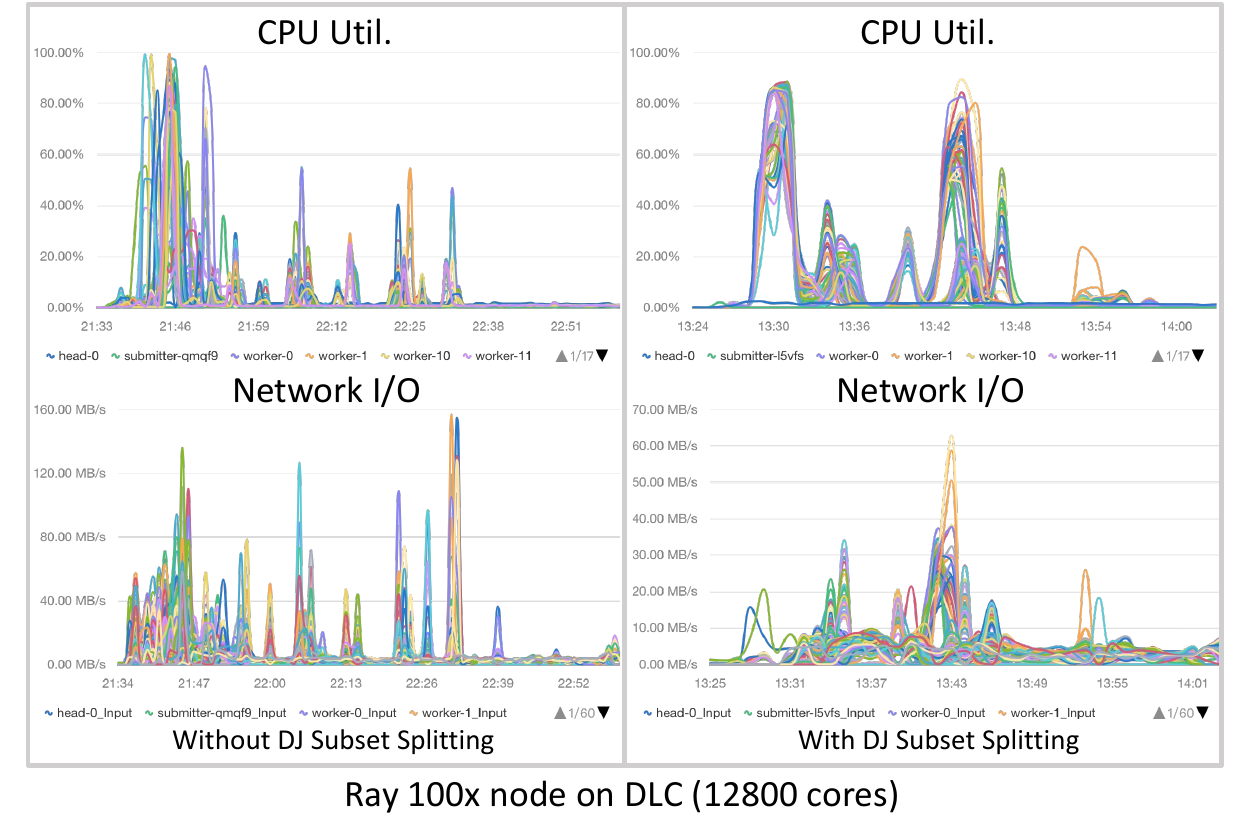}\vspace{-0.1in}}
    \caption{Processing time comparison across computing engines, dataset sizes, and workloads. A high-resolution version of subfigure (f) can be found in Appendix, Figure. \ref{fig:cloud_exps:text_large_resource_hd}}
        \vspace{-0.05in}
    \label{fig:cloud_exps}
\end{figure}

\textbf{Performance Profile.}
For datasets at the 70B-sample scale, professional cloud-scale distributed data processing products are advantageous. Users benefit from the vast cloud computing resources without dealing with the intricacies of setup and management. From our experiments, we recommend \textit{Ray-DLC for multimodal recipes and MaxCompute for text-only recipes at this scale}. 
(1) For multimodal recipes, using 3200 Ray-DLC cores process datasets in 1780.86s and 7083.5s for 500x and 2500x dataset sizes, respectively, indicating good scalability. On the other hand, the MaxCompute engine requires 1.5 times more processing time using the same resources, due to the challenges of loading large-size multimodal data. 
(2) For text-only recipes (Fig.~\ref{fig:cloud_exps:text_large}), although the Ray mode benefits from additional cores even at the scale of ten thousand, MaxCompute is the fastest, requiring about 1/4 of the time and using 1/2 fewer cores. This is attributed to MaxCompute’s co-optimization of distributed computing and storage, a feature not as advanced in Ray’s implementation.

\textbf{Impact of I/O and Data Splitting.}
I/O is critical for data processing workloads. For distributed processing, large-scale datasets need to be distributed to nodes, stored on shared disks like network attached storage (NAS), or on cloud storage services like Alibaba Cloud object storage service (OSS) and cloud parallel file storage (CPFS). All these introduce inevitable I/O and network communication costs. In our experiments shown in Fig.~\ref{fig:cloud_exps}, we use Alibaba Cloud standard CPFS, compared to which NAS and OSS increase time costs by 20-30\%. Furthermore, using another AI-oriented CPFS product with a 3x larger network bandwidth and optimized remote direct memory access to process the 12,500x dataset with 3200 cores in Ray mode takes 1,625s, 2.7x faster than the standard CPFS (4,396s). 
We then scale up the dataset to 125,000x, resulting in times of 14,617s and 7,611s with 3,200 and 6,400 cores, respectively, further demonstrating the efficiency and scalability of the system. 

Additionally, the adaptive subset splitting capability for Ray mode, introduced in Sec.~\ref{sec:dj_dataset}, offers 2x$\sim$3x acceleration in our experiments. 
For example, with 100 Ray-DLC nodes and 12,800 cores, our subset splitting method reduces the processing time of the 12,500x dataset from over 5,000s to about 2,000s. 
Analysis of Ray's resource utilization shows that its automatic block-splitting strategy incurs high network communication costs and low CPU utilization when many nodes handle a few dataset files. Our strategy pre-splits datasets according to size and node count, optimizing alignment with Ray’s features and enhancing efficiency. As seen in Fig.~\ref{fig:cloud_exps:text_large_resource}, this strategy reduces peak network I/O from about 160MB/s to 60MB/s and achieves more consistent CPU utilization across all nodes.




\begin{wraptable}{r}{0.5\textwidth}
\small
\caption{Deduplication time (minutes) with Ray nodes across varying dataset sizes and CPU counts (160 cores per node). }
\label{tab:dedup}
\begin{tblr}{
  width = \linewidth,
  colspec = {Q[140]Q[270]Q[240]Q[240]},
  hline{1,4} = {-}{0.08em},
  hline{2} = {-}{},
}
\# CPU & 200GB Time & 1TB Time  & 5TB Time   \\
4*160  & 11.13 min  & 50.83 min & 285.43 min        \\
8*160  & 7.47 min   & 30.08 min & 168.10 min 
\end{tblr}
\end{wraptable}

\textbf{Large-scale Deduplication.}
We tested the MinHash-based \textitt{RayDeduplicator} with datasets from Redpajama \cite{together2023redpajama} and Common Crawl chunks \cite{commoncrawl}, sized at 200GB, 1TB, and 5TB, using CPU counts ranging from 640 to 1,280 cores. 
As shown in Table \ref{tab:dedup}, \textitt{RayDeduplicator} efficiently scales with increased data size and computing resources, indicating its capability to handle large-scale deduplication tasks effectively. When the data size increases by 5x, the processing time increases by 4.02x$\sim$5.62x. When the number of CPU cores doubles, the processing time decreases to 58.9\%$\sim$67.1\% of the original time. Notably, \textitt{RayDeduplicator} can process 5TB of data in 2.8 hours using 8 Ray nodes (8*160 CPU cores). As a comparison, NVIDIA's NeMo Curator, which leverages cuDF \cite{cuDF} and Dask \cite{dask}, takes 1.8 hours to process 1.1TB of Redpajama data using 64 A100 GPUs (64*6,192 CUDA cores), as shown in \cite{nemo-dedup-results}.

\textbf{Implications \& Typical Application Scenarios.}
The optimizations in \oursys benefit many emerging paradigms to meet the needs of large-scale scenarios. For example, it powers \textit{enterprise foundation model training} 
from Alibaba Tongyi and Alibaba Cloud's production deployments, especially for TB-token 
pre-training and costly video/image processing for spatial intelligence.
In addition, it helps to explore the data scaling law \cite{ge2024data} and reinforced fine-tuning \cite{Trinity-RFT}, where users need to efficiently process scalable feedback data for learning from experience with environment interactions. 
Moreover, the key lesson is that complex distributed system introduces non-linear scaling tradeoffs that depend on data modality and access patterns, and storage-compute-software co-design becomes critical beyond 10M samples.


\section{Conclusions, Limitations \& Future Works}
\label{sec:conclusion}
\oursysII emerges as a versatile scaffold in the evolution of foundation models, providing efficient and adaptive data processing solutions for handling the diversity and scale of modern datasets. Our re-envisioned architecture leverages multi-layered adaptability to coordinate different modules, operators, and runtime environments. Extensive evaluations reveal its high performance across diverse cloud-scale workloads. 
By open-sourcing this system, we aim to foster a vibrant community of contributors and users, encouraging collaborative development and driving innovation to underpin the next generation of foundation models.

There are several limitations in \oursysII. From the perspective of the computing system, future work includes further enhancing processing adaptability through model-driven agents \cite{gao2024agentscope}, scalability by optimizing the single-node transmission bottleneck in Ray's head node \cite{ray-head-worker}, and efficiency by supporting GPU backend engines like NeMo Curator \cite{Jennings_NeMo-Curator_a_toolkit}. 
Looking ahead, one key direction for future development is enhancing the framework to cover multilingualism, more scenarios (e.g., AI4science, self-driving, embodied intelligence), and a wider range of data governance and safety considerations, such as ensuring data privacy across enterprise-level security protocols.
Besides, scaling to the next order of magnitude of processed data is important and requires forward-thinking architectural design like more advanced pipeline optimization, as we discussed in Appendix \ref{app:pipeline-opt}.

\bibliographystyle{plain}
\bibliography{main}

\begin{thebibliography}{10}

\bibitem{apache-arrow}
Apache Arrow.
\newblock \url{https://arrow.apache.org/}, 2023.

\bibitem{bai2024federated}
Jiamu Bai, Daoyuan Chen, Bingchen Qian, Liuyi Yao, and Yaliang Li.
\newblock Federated fine-tuning of large language models under heterogeneous language tasks and client resources.
\newblock In {\em NeurIPS}, 2024.

\bibitem{bai2024survey}
Tianyi Bai, Hao Liang, Binwang Wan, Ling Yang, Bozhou Li, Yifan Wang, Bin Cui, Conghui He, Binhang Yuan, and Wentao Zhang.
\newblock A survey of multimodal large language model from a data-centric perspective.
\newblock {\em arXiv preprint arXiv:2405.16640}, 2024.

\bibitem{opencv_library}
G.~Bradski.
\newblock {The OpenCV Library}.
\newblock {\em Dr. Dobb's Journal of Software Tools}, 2000.

\bibitem{broder1997resemblance}
Andrei~Z Broder.
\newblock On the resemblance and containment of documents.
\newblock In {\em Proceedings. Compression and Complexity of SEQUENCES 1997 (Cat. No. 97TB100171)}, pages 21--29. IEEE, 1997.

\bibitem{BRODER2000630-minhash}
Andrei~Z Broder, Moses Charikar, Alan~M Frieze, and Michael Mitzenmacher.
\newblock Min-wise independent permutations.
\newblock {\em Journal of Computer and System Sciences}, 60(3):630--659, 2000.

\bibitem{videoworldsimulators2024}
Tim Brooks, Bill Peebles, Connor Holmes, Will DePue, Yufei Guo, Li~Jing, David Schnurr, Joe Taylor, Troy Luhman, Eric Luhman, et~al.
\newblock Video generation models as world simulators.
\newblock 2024.

\bibitem{chen2024dj}
Daoyuan Chen, Yilun Huang, Zhijian Ma, Hesen Chen, Xuchen Pan, Ce~Ge, Dawei Gao, Yuexiang Xie, Zhaoyang Liu, Jinyang Gao, et~al.
\newblock Data-juicer: A one-stop data processing system for large language models.
\newblock In {\em Companion of the 2024 International Conference on Management of Data}, pages 120--134, 2024.

\bibitem{chen2024djsandbox}
Daoyuan Chen, Haibin Wang, Yilun Huang, Ce~Ge, Yaliang Li, Bolin Ding, and Jingren Zhou.
\newblock Data-juicer sandbox: A feedback-driven suite for multimodal data-model co-development.
\newblock {\em ICML}, 2025.

\bibitem{chen2015microsoft}
Xinlei Chen, Hao Fang, Tsung-Yi Lin, Ramakrishna Vedantam, Saurabh Gupta, Piotr Doll{\'a}r, and C~Lawrence Zitnick.
\newblock Microsoft coco captions: Data collection and evaluation server.
\newblock {\em arXiv preprint arXiv:1504.00325}, 2015.

\bibitem{cpfs2024}
Alibaba Cloud.
\newblock Cloud parallel file system (cpfs).
\newblock \url{https://www.alibabacloud.com/en/product/cpfs?_p_lc=1}, 2024.
\newblock Accessed: 2024-10-25.

\bibitem{pai-dlc2024}
Alibaba Cloud.
\newblock Containerized deep compute clusters (pai-dlc).
\newblock \url{https://www.alibabacloud.com/help/en/pai/user-guide/container-training/}, 2024.
\newblock Accessed: 2024-10-25.

\bibitem{maxcompute2024}
Alibaba Cloud.
\newblock Create and manage maxcompute nodes.
\newblock \url{https://www.alibabacloud.com/tech-news/a/maxcompute/gsxwf5i9fs-create-and-manage-maxcompute-nodes}, 2024.
\newblock Accessed: 2024-10-25.

\bibitem{paidesigner2024}
Alibaba Cloud.
\newblock Machine learning designer overview.
\newblock \url{https://www.alibabacloud.com/help/en/pai/user-guide/machine-learning-designer-overview}, 2024.
\newblock Accessed: 2024-10-25.

\bibitem{nas2024}
Alibaba Cloud.
\newblock Nas file system.
\newblock \url{https://www.alibabacloud.com/en/product/nas?_p_lc=1}, 2024.
\newblock Accessed: 2024-10-25.

\bibitem{oss2024}
Alibaba Cloud.
\newblock Object storage service (oss).
\newblock \url{https://www.alibabacloud.com/help/en/pai/user-guide/container-training/}, 2024.
\newblock Accessed: 2024-10-25.

\bibitem{paidlcray}
Alibaba Cloud.
\newblock Pai-dlc: Quickly submit a datajuicer task.
\newblock \url{https://www.alibabacloud.com/help/en/pai/user-guide/quickly-submit-a-datajuicer-task}, 2024.
\newblock Accessed: 2025-10-22.

\bibitem{together2023redpajama}
Together Computer.
\newblock Redpajama: An open source recipe to reproduce llama training dataset, April 2023.

\bibitem{commoncrawl}
Common Crawl.
\newblock \url{https://commoncrawl.org/}, 2023.

\bibitem{cuDF}
cuDF Team.
\newblock cudf - gpu dataframes.
\newblock \url{https://github.com/rapidsai/cudf}, 2024.

\bibitem{ffmpeg}
FFmpeg Developers.
\newblock ffmpeg tool.
\newblock \url{http://ffmpeg.org/}.

\bibitem{gadre2024datacomp}
Samir~Yitzhak Gadre, Gabriel Ilharco, Alex Fang, Jonathan Hayase, Georgios Smyrnis, Thao Nguyen, Ryan Marten, Mitchell Wortsman, Dhruba Ghosh, Jieyu Zhang, et~al.
\newblock Datacomp: In search of the next generation of multimodal datasets.
\newblock {\em Advances in Neural Information Processing Systems}, 36, 2024.

\bibitem{gamma1995design}
Erich Gamma.
\newblock Design patterns: elements of reusable object-oriented software.
\newblock {\em Person Education Inc}, 1995.

\bibitem{gao2024agentscope}
Dawei Gao, Zitao Li, Xuchen Pan, Weirui Kuang, Zhijian Ma, Bingchen Qian, Fei Wei, Wenhao Zhang, Yuexiang Xie, Daoyuan Chen, et~al.
\newblock Agentscope: A flexible yet robust multi-agent platform, 2024.

\bibitem{ge2024data}
Ce~Ge, Zhijian Ma, Daoyuan Chen, Yaliang Li, and Bolin Ding.
\newblock Data mixing made efficient: A bivariate scaling law for language model pretraining.
\newblock {\em arXiv preprint arXiv:2405.14908}, 2024.

\bibitem{harris2020array}
Charles~R. Harris, K.~Jarrod Millman, St{\'{e}}fan~J. van~der Walt, Ralf Gommers, Pauli Virtanen, David Cournapeau, Eric Wieser, Julian Taylor, Sebastian Berg, Nathaniel~J. Smith, et~al.
\newblock Array programming with {NumPy}.
\newblock {\em Nature}, 585(7825):357--362, September 2020.

\bibitem{Jennings_NeMo-Curator_a_toolkit}
Joseph Jennings, Mostofa Patwary, Sandeep Subramanian, Shrimai Prabhumoye, Ayush Dattagupta, Vibhu Jawa, Jiwei Liu, Ryan Wolf, Sarah Yurick, and Varun Singh.
\newblock {NeMo-Curator: a toolkit for data curation}.

\bibitem{jiao2024img}
Qirui Jiao, Daoyuan Chen, Yilun Huang, Yaliang Li, and Ying Shen.
\newblock Img-diff: Contrastive data synthesis for multimodal large language models.
\newblock In {\em CVPR}, 2024.

\bibitem{jiao2025detailmastertexttoimagemodelhandle}
Qirui Jiao, Daoyuan Chen, Yilun Huang, Xika Lin, Ying Shen, and Yaliang Li.
\newblock Detailmaster: Can your text-to-image model handle long prompts?, 2025.

\bibitem{bts}
Chaeeun Kim, Changhun Han, and Ha{-}Myung Park.
\newblock {BTS:} load-balanced distributed union-find for finding connected components with balanced tree structures.
\newblock In {\em {ICDE}}, pages 1090--1102. {IEEE}, 2024.

\bibitem{kwon2023efficient}
Woosuk Kwon, Zhuohan Li, Siyuan Zhuang, Ying Sheng, Lianmin Zheng, Cody~Hao Yu, Joseph~E. Gonzalez, Hao Zhang, and Ion Stoica.
\newblock Efficient memory management for large language model serving with pagedattention.
\newblock In {\em Proceedings of the ACM SIGOPS 29th Symposium on Operating Systems Principles}, 2023.

\bibitem{lhoest-etal-2021-datasets}
Quentin Lhoest, Albert Villanova~del Moral, Yacine Jernite, Abhishek Thakur, Patrick von Platen, Suraj Patil, Julien Chaumond, Mariama Drame, Julien Plu, Lewis Tunstall, et~al.
\newblock Datasets: A community library for natural language processing.
\newblock In {\em Proceedings of the 2021 Conference on Empirical Methods in Natural Language Processing: System Demonstrations}, pages 175--184, 2021.

\bibitem{li2023blip2}
Junnan Li, Dongxu Li, Silvio Savarese, and Steven Hoi.
\newblock {BLIP-2:} bootstrapping language-image pre-training with frozen image encoders and large language models.
\newblock In {\em ICML}, 2023.

\bibitem{li2024mgm}
Yanwei Li, Yuechen Zhang, Chengyao Wang, Zhisheng Zhong, Yixin Chen, Ruihang Chu, Shaoteng Liu, and Jiaya Jia.
\newblock Mini-gemini: Mining the potential of multi-modality vision language models.
\newblock {\em arXiv:2403.18814}, 2023.

\bibitem{ling2025diversity}
Zhenqing Ling, Daoyuan Chen, Liuyi Yao, Qianli Shen, Yaliang Li, and Ying Shen.
\newblock Diversity as a reward: Fine-tuning llms on a mixture of domain-undetermined data.
\newblock In {\em NeurIPS}, 2025.

\bibitem{liu2023llava}
Haotian Liu, Chunyuan Li, Qingyang Wu, and Yong~Jae Lee.
\newblock Visual instruction tuning, 2023.

\bibitem{minderer2022simple}
Matthias Minderer, Alexey Gritsenko, Austin Stone, Maxim Neumann, Dirk Weissenborn, Alexey Dosovitskiy, Aravindh Mahendran, Anurag Arnab, Mostafa Dehghani, Zhuoran Shen, et~al.
\newblock Simple open-vocabulary object detection.
\newblock In {\em European Conference on Computer Vision}, pages 728--755. Springer, 2022.

\bibitem{ray}
Philipp Moritz, Robert Nishihara, Stephanie Wang, Alexey Tumanov, Richard Liaw, Eric Liang, Melih Elibol, Zongheng Yang, William Paul, Michael~I. Jordan, et~al.
\newblock Ray: {A} distributed framework for emerging {AI} applications.
\newblock In {\em {OSDI}}, pages 561--577, 2018.

\bibitem{hf-dedup}
Chenghao Mou.
\newblock Large-scale near-deduplication behind bigcode, 2023.

\bibitem{Trinity-RFT}
Xuchen Pan, Yanxi Chen, Yushuo Chen, Yuchang Sun, Daoyuan Chen, Wenhao Zhang, Yuexiang Xie, Yilun Huang, Yilei Zhang, Dawei Gao, Yaliang Li, Bolin Ding, and Jingren Zhou.
\newblock Trinity-rft: A general-purpose and unified framework for reinforcement fine-tuning of large language models, 2025.

\bibitem{reback2020pandas}
The pandas~development team.
\newblock pandas-dev/pandas: Pandas, February 2020.

\bibitem{Paszke-NeurIPS-2019-Pytorch}
Adam Paszke, Sam Gross, Francisco Massa, Adam Lerer, James Bradbury, Gregory Chanan, Trevor Killeen, Zeming Lin, Natalia Gimelshein, Luca Antiga, et~al.
\newblock Pytorch: An imperative style, high-performance deep learning library.
\newblock In {\em {NIPS}}, pages 8024--8035, 2019.

\bibitem{penedo2024datatrove}
Guilherme Penedo, Hynek Kydlíček, Alessandro Cappelli, Mario Sasko, and Thomas Wolf.
\newblock Datatrove: large scale data processing, 2024.

\bibitem{refinedweb}
Guilherme Penedo, Quentin Malartic, Daniel Hesslow, Ruxandra Cojocaru, Alessandro Cappelli, Hamza Alobeidli, Baptiste Pannier, Ebtesam Almazrouei, and Julien Launay.
\newblock The {R}efined{W}eb dataset for {F}alcon {LLM}: outperforming curated corpora with web data, and web data only.
\newblock {\em arXiv preprint arXiv:2306.01116}, 2023.

\bibitem{podell2023sdxl}
Dustin Podell, Zion English, Kyle Lacey, Andreas Blattmann, Tim Dockhorn, Jonas M{\"u}ller, Joe Penna, and Robin Rombach.
\newblock Sdxl: Improving latent diffusion models for high-resolution image synthesis.
\newblock {\em arXiv preprint arXiv:2307.01952}, 2023.

\bibitem{qin2024federated}
Zhen Qin, Daoyuan Chen, Bingchen Qian, Bolin Ding, Yaliang Li, and Shuiguang Deng.
\newblock Federated full-parameter tuning of billion-sized language models with communication cost under 18 kilobytes.
\newblock In {\em ICML}, 2024.

\bibitem{qin2024synergy}
Zhen Qin, Daoyuan Chen, Wenhao Zhang, Liuyi Yao, Yilun Huang, Bolin Ding, Yaliang Li, and Shuiguang Deng.
\newblock The synergy between data and multi-modal large language models: A survey from co-development perspective.
\newblock {\em IEEE Transactions on Pattern Analysis and Machine Intelligence (TPAMI)}, 2025.

\bibitem{Rombach_2022_CVPR}
Robin Rombach, Andreas Blattmann, Dominik Lorenz, Patrick Esser, and Bj\"orn Ommer.
\newblock High-resolution image synthesis with latent diffusion models.
\newblock In {\em Proceedings of the IEEE/CVF Conference on Computer Vision and Pattern Recognition (CVPR)}, pages 10684--10695, June 2022.

\bibitem{schuhmann2021laion}
Christoph Schuhmann, Richard Vencu, Romain Beaumont, Robert Kaczmarczyk, Clayton Mullis, Aarush Katta, Theo Coombes, Jenia Jitsev, and Aran Komatsuzaki.
\newblock Laion-400m: Open dataset of clip-filtered 400 million image-text pairs.
\newblock {\em arXiv preprint arXiv:2111.02114}, 2021.

\bibitem{DolmaToolkit}
Luca Soldaini, Kyle Lo, Rodney Kinney, Aakanksha Naik, Abhilasha Ravichander, Akshita Bhagia, Dirk Groeneveld, Dustin Schwenk, Ian Magnusson, and Khyathi Chandu.
\newblock {The Dolma Toolkit}, 2023.
\newblock {Apache 2.0 License, Version \texttt{0.9.0}, \url{https://github.com/allenai/dolma}}.

\bibitem{qwen3}
Alibaba Tongyi~Qwen Team.
\newblock Qwen3 technical report, 2025.

\bibitem{dask}
Dask~Development Team.
\newblock Dask: flexible parallel computing library for analytics.
\newblock \url{https://github.com/dask/dask}, 2024.

\bibitem{dj-cookbook}
Data-Juicer Team.
\newblock \url{https://github.com/modelscope/data-juicer?tab=readme-ov-file#dj-cookbook}, 2025.

\bibitem{role_play_recipe}
Data-Juicer Team.
\newblock \url{https://github.com/modelscope/data-juicer/tree/main/demos/role_playing_system_prompt}, 2025.

\bibitem{nemo-dedup-results}
NVIDIA NeMO-Curator Team.
\newblock Nemo curator: Module ablation and compute performance, December 2024.

\bibitem{ray-head-worker}
Ray Team.
\newblock Ray doc: Configuring the head node, December 2024.

\bibitem{dj-competition}
The Data-Juicer Team.
\newblock Better synth: Multimodal large model data synthesis challenge.
\newblock \url{https://tianchi.aliyun.com/competition/entrance/532251?lang=en-us}, 2024.

\bibitem{hf-dataset-hub}
The~Huggingface Team.
\newblock The datasets hub from huggingface.
\newblock \url{https://huggingface.co/datasets/}, 2024.

\bibitem{ms-dataset-hub}
The~Modelscope Team.
\newblock The datasets hub from modelscope.
\newblock \url{https://modelscope.cn/datasets}, 2024.

\bibitem{teed2020raft}
Zachary Teed and Jia Deng.
\newblock Raft: Recurrent all-pairs field transforms for optical flow.
\newblock In {\em Computer Vision--ECCV 2020: 16th European Conference, Glasgow, UK, August 23--28, 2020, Proceedings, Part II 16}, pages 402--419. Springer, 2020.

\bibitem{LabelStudio}
Maxim Tkachenko, Mikhail Malyuk, Andrey Holmanyuk, and Nikolai Liubimov.
\newblock {Label Studio}: Data labeling software, 2020-2025.
\newblock Open source software available from https://github.com/HumanSignal/label-studio.

\bibitem{wang2024emu3}
Xinlong Wang, Xiaosong Zhang, Zhengxiong Luo, Quan Sun, Yufeng Cui, Jinsheng Wang, Fan Zhang, Yueze Wang, Zhen Li, Qiying Yu, et~al.
\newblock Emu3: Next-token prediction is all you need.
\newblock {\em arXiv preprint arXiv:2409.18869}, 2024.

\bibitem{xu2023youku}
Haiyang Xu, Qinghao Ye, Xuan Wu, Ming Yan, Yuan Miao, Jiabo Ye, Guohai Xu, Anwen Hu, Yaya Shi, Guangwei Xu, et~al.
\newblock Youku-mplug: A 10 million large-scale chinese video-language dataset for pre-training and benchmarks.
\newblock {\em arXiv preprint arXiv:2306.04362}, 2023.

\bibitem{xu2025mindgym}
Zhe Xu, Daoyuan Chen, Zhenqing Ling, Yaliang Li, and Ying Shen.
\newblock Mindgym: Enhancing vision-language models via synthetic self-challenging questions.
\newblock In {\em NeurIPS}, 2025.

\bibitem{Yao-arxiv-2022-ReAct}
Shunyu Yao, Jeffrey Zhao, Dian Yu, Nan Du, Izhak Shafran, Karthik Narasimhan, and Yuan Cao.
\newblock React: Synergizing reasoning and acting in language models.
\newblock {\em CoRR}, abs/2210.03629, 2022.

\bibitem{apache-spark}
Matei Zaharia, Mosharaf Chowdhury, Michael~J. Franklin, Scott Shenker, and Ion Stoica.
\newblock Spark: Cluster computing with working sets.
\newblock In {\em HotCloud}, page~10, 2010.

\bibitem{zhao2024swiftascalablelightweightinfrastructure}
Yuze Zhao, Jintao Huang, Jinghan Hu, Xingjun Wang, Yunlin Mao, Daoze Zhang, Zeyinzi Jiang, Zhikai Wu, Baole Ai, Ang Wang, et~al.
\newblock Swift:a scalable lightweight infrastructure for fine-tuning, 2024.

\bibitem{zheng2024llamafactory}
Yaowei Zheng, Richong Zhang, Junhao Zhang, Yanhan Ye, Zheyan Luo, Zhangchi Feng, and Yongqiang Ma.
\newblock Llamafactory: Unified efficient fine-tuning of 100+ language models.
\newblock In {\em Proceedings of the 62nd Annual Meeting of the Association for Computational Linguistics (Volume 3: System Demonstrations)}, Bangkok, Thailand, 2024. Association for Computational Linguistics.

\bibitem{zhou2024humanvbench}
Ting Zhou, Daoyuan Chen, Qirui Jiao, Bolin Ding, Yaliang Li, and Ying Shen.
\newblock Humanvbench: Exploring human-centric video understanding capabilities of mllms with synthetic benchmark data.
\newblock {\em arXiv preprint arXiv:2412.17574}, 2024.

\bibitem{zhu2023multimodal}
Wanrong Zhu, Jack Hessel, Anas Awadalla, Samir~Yitzhak Gadre, Jesse Dodge, Alex Fang, Youngjae Yu, Ludwig Schmidt, William~Yang Wang, and Yejin Choi.
\newblock {Multimodal C4}: An open, billion-scale corpus of images interleaved with text.
\newblock {\em arXiv preprint arXiv:2304.06939}, 2023.

\end{thebibliography}

\clearpage
\newpage
\appendix

\section*{Appendix}
\label{sec:appendix}

\DoToC

\newpage

\section{List of New Operators in \oursysII}
\label{sec:appendix:list_of_new_ops}
The full list of new OPs in \oursysII is shown in Table \ref{tab:new_ops}.

\clearpage
{
\tiny
\begin{longtblr}[
  caption = {List of new OPs in \oursysII. The full OP list can be found in our document in GitHub repository.},
  label = {tab:new_ops},
  entry = none,
]{
  width = \linewidth,
  colspec = {Q[30]Q[200]Q[400]Q[85]Q[80]},
  cell{2}{1} = {r=12}{},
  cell{14}{1} = {r=12}{},
  cell{26}{1} = {r=27}{},
  cell{53}{1} = {r=8}{},
  cell{61}{1} = {r=5}{},
  cell{66}{1} = {r=25}{},
}
\hline
Modality Type & OP Name (OP Type as the last suffix)                                    & OP
  Description                                                                                                                      & Function Type             & Implementation Type \\
  \hline
Image-only    & image\_blur\_mapper                         & Blur images                                                                                                                           & Augmentation              & Rule-based          \\
              & image\_remove\_background\_mapper                   & Remove background of images                                                                   & Augmentation                   & Model-based         \\
              & image\_face\_blur\_mapper                   & Blur faces detected in images                                                                                                         & Privacy                   & Model-based         \\
              & image\_aesthetics\_filter                   & Keeps samples containing images whose
  aesthetics scores are within the specified range                                              & Cleaning                  & Model-based         \\
              & image\_aspect\_ratio\_filter                & Keeps samples containing images with
  aspect ratios within the specified range                                                       & Cleaning                  & Rule-based          \\
              & image\_face\_ratio\_filter                  & Keeps samples containing images with face
  area ratios within the specified range                                                    & Cleaning                  & Model-based         \\
              & image\_nsfw\_filter                         & Keeps samples containing images with NSFW
  scores below the threshold                                                                & Cleaning                  & Model-based         \\
              & image\_shape\_filter                        & Keeps samples containing images with
  widths and heights within the specified range                                                  & Cleaning                  & Rule-based          \\
              & image\_size\_filter                         & Keeps samples containing images whose
  size in bytes are within the specified range                                                  & Cleaning                  & Rule-based          \\
              & image\_watermark\_filter                    & Keeps samples containing images with
  predicted watermark probabilities below the threshold                                          & Cleaning                  & Model-based         \\
              & image\_deduplicator                         & Deduplicates samples at document-level
  using exact matching of images between documents                                             & Deduplication             & Model-based         \\
              & ray\_image\_deduplicator                    & Deduplicates samples at document-level
  using exact matching of images between documents on ray                                      & Deduplication             & Model-based         \\
  \hline
Image-Text    & image\_captioning\_from\_gpt4v\_mapper      & Generate texts based on GPT-4V for images                                                                                             & Synthesis                 & Model-based         \\
              & image\_captioning\_mapper                   & Generate texts based on image-to-text
  models for images                                                                             & Synthesis                 & Model-based         \\
              & image\_diffusion\_mapper                    & Generate images based on text-to-image
  models for texts                                                                             & Synthesis                 & Model-based         \\
              & mllm\_mapper                                & Use multimodal large language models for
  visual question answering tasks                                                            & Wrapper of external tool~ & Model-based         \\
              & sdxl\_prompt2prompt\_mapper                 & Use the generative model SDXL and image
  editing technique Prompt-to-Prompt to generate pairs of similar images                      & Synthesis                 & Model-based         \\
              & image\_segment\_mapper                      & Perform segment-anything on images and
  return the bounding box values                                                               & Wrapper of external tool~ & Model-based         \\
              & sentence\_augmentation\_mapper              & Augment sentences using large language
  models                                                                                       & Augmentation              & Model-based         \\
              & image\_text\_matching\_filter               & Keeps samples with image-text
  classification matching score within the specified range based on a BLIP
  model                      & Cleaning                  & Model-based         \\
              & image\_text\_similarity\_filter             & Keeps samples with image-text feature
  cosine similarity within the specified range based on a CLIP model                            & Cleaning                  & Model-based         \\
              & phrase\_grounding\_recall\_filter           & Keeps samples whose locating recalls of
  phrases extracted from text in the images are within a specified range                      & Cleaning                  & Model-based         \\
              & image\_pair\_similarity\_filter             & Keeps image pairs whose cosine similarity
  of features is within a specified range, based on a CLIP model.                           & Cleaning                  & Model-based         \\
              & text\_pair\_similarity\_filter              & Keeps text pairs whose cosine similarity
  of features is within a specified range, based on a CLIP model                             & Cleaning                  & Model-based         \\
  \hline
Video-only    & video\_face\_blur\_mapper                   & Blur faces detected in videos                                                                                                         & Privacy                   & Model-based         \\
              & video\_ffmpeg\_wrapped\_mapper              & Wrapper to run a FFmpeg video filter                                                                                                  & Wrapper                   & Tool-based          \\
              & video\_remove\_watermark\_mapper            & Remove watermarks in videos                                                                                                           & Cleaning                  & Model-based         \\
              & video\_resize\_aspect\_ratio\_mapper        & Resize video aspect ratio to a specified
  range                                                                                      & Augmentation              & Rule-based          \\
              & video\_resize\_resolution\_mapper           & Map videos to ones with given resolution
  range                                                                                      & Augmentation              & Rule-based          \\
              & video\_split\_by\_duration\_mapper          & Split video by duration                                                                                                               & Augmentation              & Rule-based          \\
              & video\_spit\_by\_key\_frame\_mapper         & Split video by key frames                                                                                                             & Augmentation              & Rule-based          \\
              & video\_split\_by\_scene\_mapper             & Split videos into scene clips                                                                                                         & Augmentation              & Model-based         \\
              & video\_human\_tracks\_extraction\_mapper    & Extracts human tracks by linking face and
  body bounding boxes across frames                                                         & Synthesis                 & Model-based         \\
              & video\_human\_demographics\_mapper                 & determines demographic attributes
  (gender, age, race) for face tracks by aggregating frame-level detections.                        & Synthesis                 & Model-based         \\
              & video\_human\_description\_mapper           & generates individual-focused videos from
  body bounding box tracks and processes it for appearance description and
  simple actions. & Synthesis                 & Model-based         \\
              & video\_facial\_description\_mapper          & generates face-focused videos from face
  bounding box tracks and processes them for emotion and facial description.                  & Synthesis                 & Model-based         \\
              & video\_active\_speaker\_detection\_mapper   & detects active speaking by analyzing face
  track sequences alongside corresponding audio                                             & Synthesis                 & Model-based         \\
              & video\_ASR\_mapper                          & generate the automatic speech recognition
  result for video                                                                          & Synthesis                 & Model-based         \\
              & video\_speech\_emotion\_recognition\_mapper & Detect the emotion category of the speech
  in the video                                                                              & Synthesis                 & Model-based         \\
              & video\_voice\_demographics\_mapper          & analyzes voice demographics (such as
  gender, age) in video                                                                          & Synthesis                 & Model-based         \\
              & video\_aesthetics\_filter                   & Keeps samples whose specified frames have
  aesthetics scores within the specified range                                              & Cleaning                  & Model-based         \\
              & video\_aspect\_ratio\_filter                & Keeps samples containing videos with
  aspect ratios within the specified range                                                       & Cleaning                  & Rule-based          \\
              & video\_duration\_filter                     & Keep data samples whose videos' durations
  are within a specified range                                                              & Cleaning                  & Rule-based          \\
              & video\_face\_ratio\_filter                  & Keep samples whose frame ratio containing
  faces is greater than a certain threshold are retained.                                   & Cleaning                  & Model-based         \\
              & video\_motion\_score\_filter                & Keep samples with video motion scores
  within a specific range                                                                       & Cleaning                  & Rule-based          \\
              & video\_nsfw\_filter                         & Keeps samples containing videos with NSFW
  scores below the threshold                                                                & Cleaning                  & Model-based         \\
              & video\_ocr\_area\_ratio\_filter             & Keep data samples whose detected text
  area ratios for specified frames in the video are within a specified range                    & Cleaning                  & Model-based         \\
              & video\_resolution\_filter                   & Keeps samples containing videos with
  horizontal and vertical resolutions within the specified range                                 & Cleaning                  & Rule-based          \\
              & video\_watermark\_filter                    & Keeps samples containing videos with
  predicted watermark probabilities below the threshold                                          & Cleaning                  & Model-based         \\
              & video\_deduplicator                         & Deduplicates samples at document-level
  using exact matching of videos between documents                                             & Deduplication             & Model-based         \\
              & ray\_video\_deduplicator                    & Deduplicates samples at document-level
  using exact matching of videos between documents on ray                                      & Deduplication             & Model-based         \\
  \hline
Video-Text    & video\_captioning\_from\_audio\_mapper      & Generate texts for videos according to
  their audio streams based on audio LLMs                                                      & Synthesis                 & Model-based         \\
              & video\_captioning\_from\_frames\_mapper     & Generate texts for videos according to
  sampled frame images based on image-to-text models                                           & Synthesis                 & Model-based         \\
              & video\_captioning\_from\_video\_mapper      & Generate texts for videos based on
  video-to-text models                                                                             & Synthesis                 & Model-based         \\
              & video\_tagging\_from\_audio\_mapper         & Generate tags from audio streams
  extracted from the video.                                                                          & Synthesis                 & Model-based         \\
              & video\_tagging\_from\_frames\_mapper        & Generate video tags from frames extracted
  from the video.                                                                           & Synthesis                 & Model-based         \\
              & video\_captioning\_from\_summarizer\_mapper & Generate texts by summarizing several
  types of generated texts (from video/audio/frames, tags from audio/frames,
  ...)             & Synthesis                 & Model-based         \\
              & video\_frames\_text\_similarity\_filter     & Keep data samples whose similarities
  between sampled video frame images and text are within a specific range                        & Cleaning                  & Model-based         \\
              & video\_tagging\_from\_frames\_filter        & Keep samples containing videos with given
  tags                                                                                      & Cleaning                  & Model-based         \\
  \hline
Audio-only    & audio\_ffmpeg\_wrapped\_mapper              & Wrapper to run a FFmpeg audio filter                                                                                                  & Wrapper                   & Tool-based          \\
              & audio\_add\_gaussian\_noise\_mapper                     & Add gaussian noise to audio.                                                              & Augmentation                  & Tool-based          \\
              & audio\_duration\_filter                     & Keep data samples whose audios' durations
  are within a specified range                                                              & Cleaning                  & Rule-based          \\
              & audio\_nmf\_snr\_filter                     & Keep data samples whose audios'
  Signal-to-Noise Ratios are within a specified range                                                 & Cleaning                  & Rule-based          \\
              & audio\_size\_filter                         & Keep data samples whose audios' sizes are
  within a specified range                                                                  & Cleaning                  & Rule-based          \\
  \hline
Text-only     & calibrate\_qa\_mapper                       & Calibrate
  question-answer pairs based on reference text                                                                             & Augmentation                 & Model-based         \\
              & calibrate\_query\_mapper                    & Calibrate
  query in question-answer pairs based on reference text                                                                    & Augmentation                 & Model-based         \\
              & calibrate\_response\_mapper                 & Calibrate
  response in question-answer pairs based on reference text                                                                 & Augmentation                 & Model-based         \\
              & extract\_entity\_attribute\_mapper          & Extract
  attributes for given entities from the text.                                                                                & Synthesis                 & Model-based         \\
              & extract\_entity\_relation\_mapper           & Extract
  entities and relations in the text for knowledge graph.                                                                     & Synthesis                 & Model-based         \\
              & extract\_event\_mapper                      & Extract
  events and relevant characters in the text.                                                                                 & Synthesis                 & Model-based         \\
              & extract\_keyword\_mapper                    & Generate
  keywords for the text.                                                                                                     & Synthesis                 & Model-based         \\
              & extract\_nickname\_mapper                   & Extract
  nickname relationship in the text.                                                                                          & Synthesis                 & Model-based         \\
              & extract\_support\_text\_mapper              & Extract support sub text for a summary.                                                                                               & Synthesis                 & Model-based         \\
              & extract\_tables\_from\_html\_mapper              & Extract tables from HTML content.                                                                                               & Cleaning                 & Rule-based         \\
              & generate\_qa\_from\_examples\_mapper        & Generate
  question and answer pairs based on examples.                                                                               & Synthesis                 & Model-based         \\
              & generate\_qa\_from\_text\_mapper            & Generate
  question and answer pairs from text.                                                                                       & Synthesis                 & Model-based         \\
              & optimize\_qa\_mapper                        & Optimize both the query and response in question-answering samples.                                                                   & Augmentation                 & Model-based         \\
              & optimize\_query\_mapper                     & Optimize the query in question-answering samples.                                                                                     & Augmentation                 & Model-based         \\
              & optimize\_response\_mapper                  & Optimize the response in question-answering samples.                                                                                  & Augmentation                 & Model-based         \\
              & pair\_preference\_mapper                    & Construct paired preference samples.                                                                                                  & Augmentation                 & Model-based         \\
              & text\_chunk\_mapper                         & Split input text to chunks.                                                                                                           & Synthesis                 & Model-based         \\
              & naive\_grouper                              & Group all samples to one batched sample.                                                                                              & Synthesis                 & Rule-based          \\
              & key\_value\_grouper                         & Group samples to batched samples according
  values in given keys.                                                                    & Synthesis                 & Rule-based          \\
              & entity\_attribute\_aggregator               & Return conclusion of the given entity's
  attribute from some docs.                                                                   & Synthesis                 & Model-based         \\
              & most\_relavant\_entities\_aggregator        & Return most relevant entities with the given
  entity from some docs.                                                                 & Synthesis                 & Model-based         \\
              & nested\_aggregator                          & Considering
  the limitation of input length, nested aggregate contents for each given
  number of samples.                           & Synthesis                 & Model-based         \\
  & llm\_quality\_score\_filter                          & Keep sample with high quality score estimated by LLM.                           & Cleaning                 & Model-based         \\
  & llm\_difficulty\_score\_filter                          & Keep sample with high difficulty score estimated by LLM.                           & Cleaning                 & Model-based         \\
  & domain\_diversity\_selector                          & Select samples based on the data's domain diversity.                           & Cleaning                 & Model-based         \\
  \hline
\end{longtblr}
}
\clearpage

\section{Showcase of Typical Multimodal Operators}
\label{sec:appendix:new_ops_showcase}

The new OPs include some unique contributions to \oursysII, and others partly inspired by SOTA data processing methodologies for foundation models \cite{schuhmann2021laion,zhu2023multimodal,gadre2024datacomp}. Below are three representative examples that showcase the diverse computational operations and requirements of these OPs.

\begin{figure}[h!]
    \centering
    \subfloat[Example of \textitt{phrase\_grounding\_recall\_filter}.\label{fig:2_op_examples:phrase_grounding}]{\includegraphics[width=\linewidth]{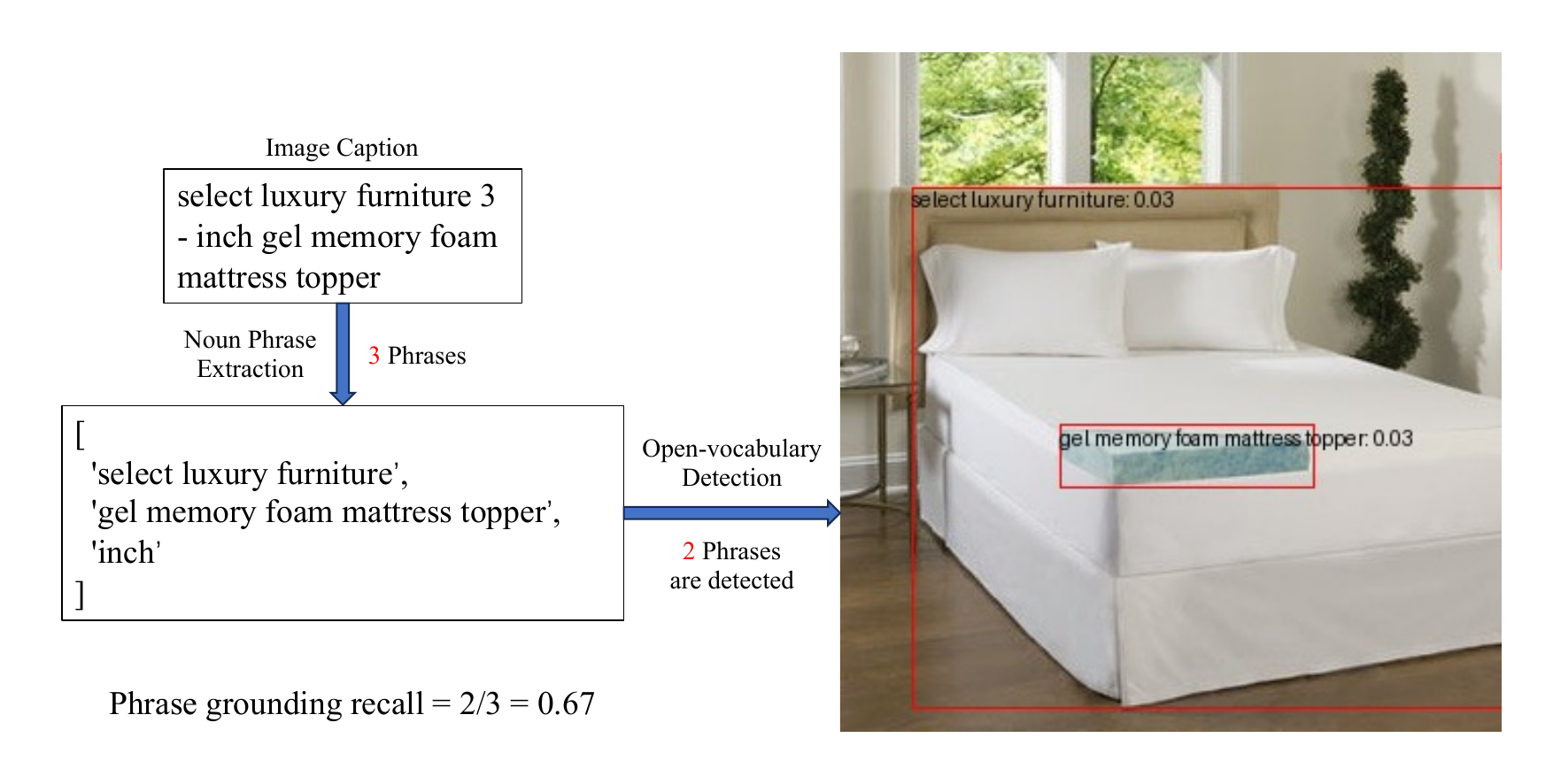}       
    }\\

    \subfloat[Example of \textitt{video\_motion\_score\_filter}.\label{fig:2_op_examples:motion_score}]{\includegraphics[width=\linewidth]{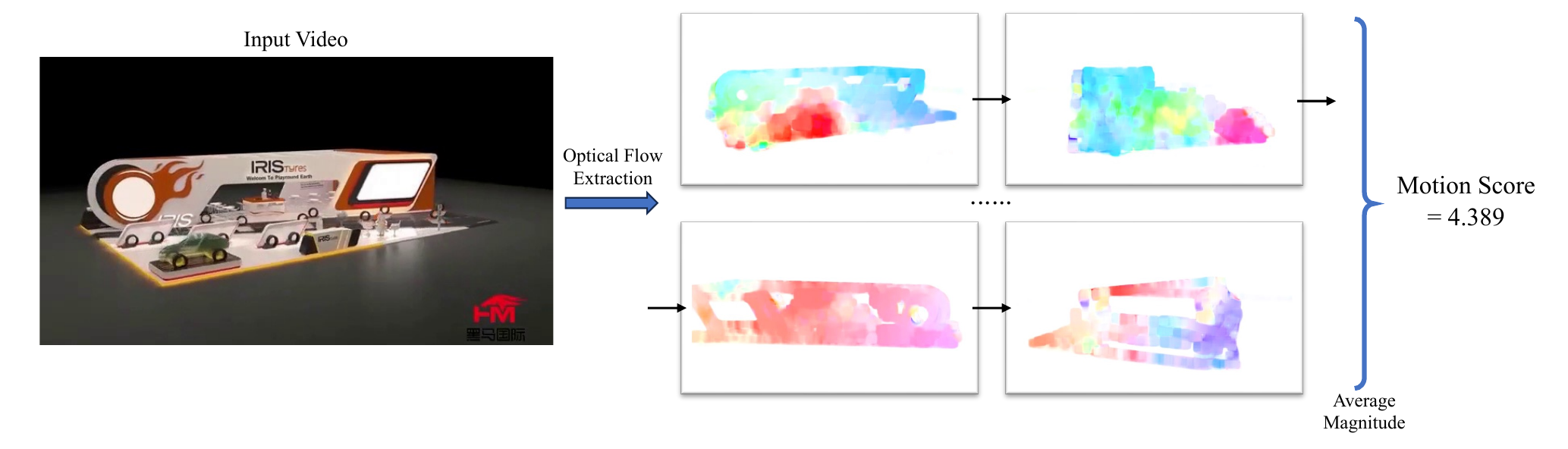}
        
    }\\

    \subfloat[Example of \textitt{video\_captioning\_from\_summarizer\_mapper}.\label{fig:2_op_examples:video_caption_from_summarizer}]{\includegraphics[width=\linewidth]{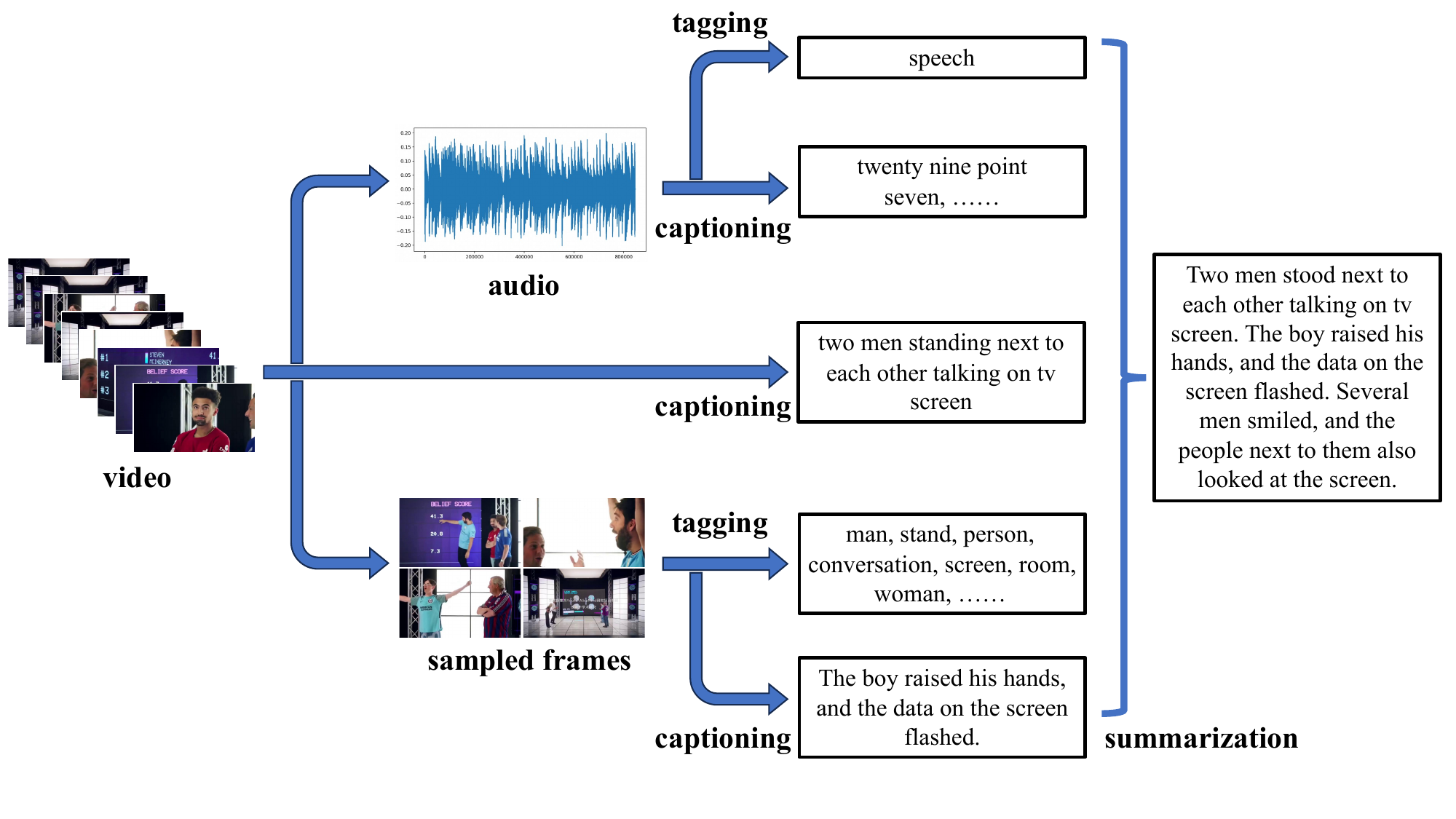}     
}
    \caption{Illustrative examples of new operators.}
    
    \label{fig:2_op_examples}
\end{figure}

\textbf{\textitt{phrase\_grounding\_recall\_filter}:} This OP, newly developed by \oursysII, assesses alignment and consistency between images and textual descriptions. As illustrated in Fig.~\ref{fig:2_op_examples:phrase_grounding}, it identifies noun phrases in the text that refer to key entities. Subsequently, an open-vocabulary object detection model, such as Owl-ViT \cite{minderer2022simple}, attempts to detect corresponding entities within the image. The OP then calculates the recall of detected phrases to evaluate consistency, where a higher recall indicates better image-text coherence.

\textbf{\textitt{video\_motion\_score\_filter}:} This OP quantifies video dynamics. As shown in Fig.~\ref{fig:2_op_examples:motion_score}, it samples multiple frames at a specified frames-per-second (FPS) rate and computes optical flows. The average magnitude of these flows determines the motion score, with higher scores indicating more dynamic content. To accommodate various computational resources, both a GPU-based RAFT version \cite{teed2020raft} and a CPU-based OpenCV version \cite{opencv_library} are available.

\textbf{\textitt{video\_captioning\_from\_summarizer\_mapper}:} This OP generates new video captions by combining tagging and captioning capabilities from various OPs. As illustrated in Fig.~\ref{fig:2_op_examples:video_caption_from_summarizer}, it uses six OPs from different angles: two for tagging and captioning the audio stream, two for static visual frame analysis, and two for dynamic video stream and information integration. A summarizer finally incorporates the three captions and two tag sets into a new caption. This composition yields more accurate and comprehensive captions by considering multi-dimensional and multi-perspective content.

\section{User Interfaces}
\label{app:interface}


\subsection{Low-level APIs}
\label{sec:interface:low-level-api}
The framework's core capabilities are exposed through Python-based programmatic interfaces, providing object-oriented logical encapsulation for both the fundamental \texttt{\oursys-Dataset}, \texttt{\oursys-Operators} and other runtime modules. This design delivers developers maximum flexibility and customizability. Processing workflows can be automatically chained by passing a series of OP instances to a loaded \texttt{\oursys-Dataset} object (e.g., \texttt{data.process([op1, op2])}), enabling various operations to be performed on the dataset in a single pass. Additionally, the framework supports applying an instantiated \texttt{\oursys-Operators} to a target dataset (e.g., \texttt{op.run(data)}), enhancing the reusability of OP instances. More details about the \texttt{\oursys-Dataset} and \texttt{\oursys-Operators} are introduced in Sec. \ref{sec:core_runtime} and Appendix \ref{app:core_runtime}. This dual approach---chained processing and individual OP application---optimizes both efficiency and modularity in data processing tasks, while leveraging the inherent advantages of Python's ecosystem.

\subsection{RESTful APIs}
\label{sec:interface:restful-api}
Utilizing standard Python type hints, we provide one-click generation of high-performance web APIs, capable of automatically discovering, registering, and adapting OP classes and related tools. Users can rapidly initiate a web server supporting the Asynchronous Server Gateway Interface by simply executing a service script, eliminating the need for manual code writing. The asynchronous concurrency mechanism enables options such as lightweight background tasks and mitigates potential bottlenecks for endpoints that may experience prolonged network I/O blocking. Each OP is accessible via POST requests, typically executing the OP's \texttt{run()} method as the endpoint. The target dataset path is passed through query parameters, with additional configurable OP parameters transmitted via JSON payload. Upon completion, the path to the processed dataset is returned. This invocation through Web APIs allows for a centralized host with distributed access, reducing deployment complexity. It also facilitates the separation of application logic from execution, potentially fostering the development and release of more applications built upon \oursys. Importantly, the extensive customization parameters available in the programming API can be seamlessly passed through the Web API, maintaining full functionality without compromising usability, and facilitating serviceful calling by higher-level interfaces as introduced later.

\subsection{Web \& CMD tools}
\label{sec:interface:web-cmd}
\begin{figure}[h!]
    \centering
    \includegraphics[width=0.95\linewidth]{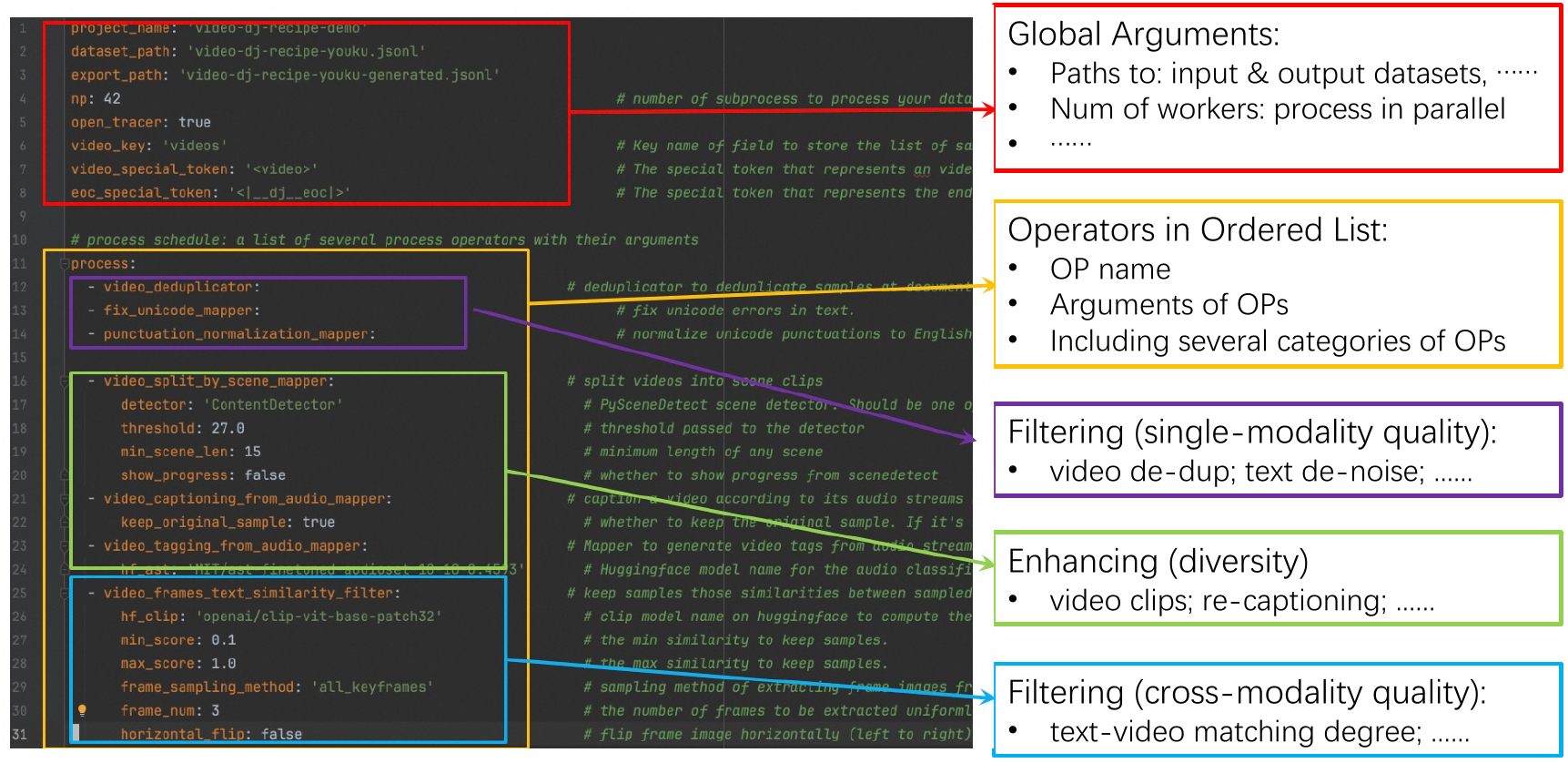}
    \caption{An example of config file for data recipe.}
    \label{fig:recipe_cfg}
\end{figure}

Thanks to the low-level and RESTful APIs, we further establish service-level capabilities across various scenarios in \oursysII, encapsulating scenario-specific processing solutions into built-in data recipes, described as YAML configuration files for end-to-end workflows (an example for video-data-synthesis is shown in Fig.~\ref{fig:recipe_cfg}). 
Centered around these data recipes, we provide user-friendly, highly encapsulated command-line interfaces as illustrated in Listing \ref{listing:dj_cmd_tools}, supporting automated lightweight installation and recipe-level invocation. These tools automatically gather and install the requisite of OPs within the recipe, perform automated OP fusion, and maximize adaptive utilization of computing resources while minimizing user cognitive load and operational cost.

\begin{listing}[!ht]
\setlength{\abovecaptionskip}{0pt}
\setlength{\belowcaptionskip}{0pt}
\begin{minted}[
    frame=lines,
    framesep=2mm,
    baselinestretch=1.2,
    fontsize=\scriptsize, 
    linenos,
    breaklines=true, 
    bgcolor=lightgray!10, 
    rulecolor=black, 
    numbersep=5pt, 
]{bash}
dj-install --config recipe.yaml  # install dependencies only for this recipe
dj-analyze --auto --dataset_path xx.jsonl # default OPs, or via specific recipe 
dj-process --config recipe.yaml # for both standalone or distributed
\end{minted}
\caption{Representative tools available in \oursysII.}%
\label{listing:dj_cmd_tools}
\end{listing}

Utilizing the aforementioned APIs and tools, \oursys's capabilities are seamlessly integrated into Alibaba Cloud's visual modeling product, PAI-Designer \cite{paidesigner2024}. As shown in Fig. \ref{fig:overview}, \texttt{\oursys-Operators} are encapsulated as visual components within Designer, presented in a directory tree structure for easy access. Users can manage data by intuitive drag-and-drop manners, and import built-in or customized recipes in YAML format using graphical canvas interface. 
With Designer as entry point, users can activate MaxCompute cluster nodes \cite{maxcompute2024} or containerized deep compute clusters (PAI-DLC) \cite{pai-dlc2024} directly for efficient cloud-native distributed data processing, leveraging Alibaba Cloud's data storage solutions (e.g., Object Storage OSS \cite{oss2024}, Cloud Parallel File System CPFS \cite{cpfs2024}, and NAS \cite{nas2024}) along with automated performance monitoring and optimization.

\subsection{Natural Language Interaction}
\label{sec:interface:natural-language}

\begin{dialogue}
    \user{Filter out too short text samples. The dataset path is input\_data.jsonl}
    \assistant[darkgray!70]{<thought>Filter out text samples that are too short based on the provided dataset path.</thought>\\
    <function>run\_text\_length\_filter</function>\\
    <dataset\_path>input\_data.jsonl</dataset\_path>}
    \assistant{The analyzed statistics are displayed in the image below: \raisebox{-\height/5}{\twemoji[height=\twemojiDefaultHeight]{bar chart}}}
    \assistant{Based on the analysis, please enter the minimum and maximum threshold values for filtering.}
    \user{80 120}
    \system[gray]{Execute function run\_text\_length\_filter\newline
    [ARGUMENTS] dataset\_path: input\_data.jsonl\newline
    [STATUS] SUCCESS\newline
    [RESULT] Filtered dataset path: output.jsonl}
    \assistant[gray]{<thought>The filtering has been successfully executed, and the filtered dataset is now at the specified path.</thought>\\
    <function>finish</function>\\
    <response>Your request to filter out too short text samples has been completed. The processed dataset is available at output.jsonl.</response>}
    \assistant{Your request to filter out too short text samples has been completed. The processed dataset is available at output.jsonl.}
\end{dialogue}

Querying and understanding over 100 diverse OPs have become challenging. Thankfully, advances in foundation models have transformed human-computer interactions. Research shows the synergy of reasoning and acting (ReAct) \cite{Yao-arxiv-2022-ReAct} in language models enhances task-solving. Using RESTful API services, we developed model-driven agents based on ReAct, enabling intuitive data processing through natural language. We adopted AgentScope \cite{gao2024agentscope}, a multi-agent platform, for low-code integration, using prompt-optimized ReAct agents to align OP functionalities with our RESTful APIs. Built-in function preprocessing, response parsing, and fault-tolerance streamline development. \oursysII thus allows flexible instruction input, interpreting user intents to execute appropriate data processing functions.

An example is shown above for interactive text length filtering using this agent, where users describe tasks with simple and vague language. Specifically, the user requests ``filter out too short text samples from \textitt{input\_data.jsonl}''. The agent's ReAct prompt helps analyze and reason user intent, signaled by the \textitt{<thought>} tag. The \textitt{<function>} tag notes the chosen function, while \textitt{<dataset\_path>} represents parameter mapping. Initially, a statistical analysis of text lengths is visualized, and the agent asks for filtering thresholds. After user input, filtering runs with parameters via RESTful API. Upon completion, the agent confirms the processed dataset path as \textitt{output.jsonl}, marks the task finished, and informs the user. Gray dialogs reveal the agent's reasoning and actions, providing insights into the workflow. This example illustrates step-by-step user understanding, function selection, parameter population, and task execution, emphasizing transparency and automation of \oursysII with agentic processing.


\section{Core Runtime Implementation Details}
\label{app:core_runtime}

\subsection{Execution Mode Configuration}
\label{app:exec_modes}

The core runtime of \oursysII is implemented in Python, featuring a flexible data processing pipeline characterized by two execution modes: a standalone mode for convenient single-node execution, based on the Hugging Face Dataset \cite{lhoest-etal-2021-datasets}, and a distributed mode that offers scalability across multiple nodes, leveraging the Ray Data \cite{ray} and MaxFrame-DataFrame \cite{maxcompute2024}. These frameworks offer diverse computational engines, each with distinct capabilities and suitable use cases, but they also come with heterogeneous programming interfaces and complex implementation details.

In \oursysII, we exploit the strengths of these diverse computational engines and their associated programming classes by abstracting a top-level \texttt{\oursys-Dataset} class and devising a standardized data representation schema (detailed in the next subsection).
The primary design principle employed is the \textit{Facade Pattern}. The abstracted class provides comprehensive and unified development interfaces, facilitating seamless use across standalone, distributed, and cloud cluster environments, meanwhile optimized to support extensive multimodal data processing.

\begin{listing}[!h]
\setlength{\belowcaptionskip}{0pt}
\begin{minted}
[
frame=lines,
framesep=2mm,
baselinestretch=1.2,
fontsize=\scriptsize,
linenos,
breaklines=true, 
bgcolor=lightgray!10, 
rulecolor=black, 
numbersep=5pt, 
]{python}
# Differernt input sources, such as from local files, dataset hubs from Hugging Face and Modelscope, and special types like arXiv, wiki, CommonCrawl, etc.  
data = DJDataset.load(src_file)  

data = data.process(op1)  # run a single operator
data = data.process(op1).process(op2)  # run multiple operators continuously
data = data.process([op1, op2])  # run a list of operators

data.export(tgt_file)  # tgt_file can be either local path or remote path
\end{minted}
\caption{Some basic interfaces of \texttt{\oursys-Dataset}.}%
\label{listing:dj_dataset_interface}
\end{listing}

Moreover, following the principle of \textit{Template Method Pattern}, the \texttt{\oursys-Dataset} class establishes templated workflows for OP executing on datasets and provides several basic and unified interfaces, catering to users with diverse programming preferences, as demonstrated in Listing~\ref{listing:dj_dataset_interface}. The complexities of underlying computational engines and cumbersome runtime issues related to instantiated dataset objects are abstracted away from these interfaces, ensuring transparency and ease of use for end-users and developers alike.

New functionalities can be added by merely implementing new OP classes without altering the internal logic of \texttt{\oursys-Dataset}. The standalone and distributed execution modes automatically recognize and switch configurations based on runtime settings. Coupled with unified \textit{function signatures}, the data processing can be consolidated into a batched processing mode, not only standardizing the \textit{data interface} for OP processing but also facilitating the implementation of a robust sample-level fault tolerance mechanism. The internal adaptation features are elaborated in Appendix~\ref{app:fault_tolerance}.

\subsection{Data Schema Implementation}
\label{app:schema_impl}

\begin{listing}[!h]
\setlength{\belowcaptionskip}{0pt}
\begin{minted}
[
frame=lines,
framesep=2mm,
baselinestretch=1.2,
fontsize=\scriptsize,
linenos,
breaklines=true, 
bgcolor=lightgray!10, 
rulecolor=black, 
numbersep=5pt, 
]
{json}
{
  // >>> core contents: texts, dialogs, ...
  // for general pretraining
  "text": "<__dj__image> desc of image 1 <|__dj__eoc|> desc of image 2 <__dj__image> <|__dj__eoc|>",
  // for post-tuning
  "query": "<query2>",
  "response": "<response2>",
  "history": [["<query1>", "<response1>"]],
  // <<< core contents

  // >>> extra data contents: multimodal data paths, ...
  "images": [
    "path/to/the/image1",
    "path/to/the/image2"
  ],
  // <<< extra data contents

  // >>> meta infos and stats, which could be primitive or produced by Data-Juicer
  "meta": {
    "src": "customized",
    "version": "0.1"
  },
  "stats": {
    "lang": "en",
    "image_widths": [224, 336]
  },
  // <<< meta infos and stats
}
\end{minted}
\caption{Illustration of data schema of \oursysII.}
\label{listing:mm_format}
\end{listing}

In the context of foundation models and post-tuning tasks, datasets are complex, originating from various sources and modalities. Target tasks are diverse, requiring different data formats and organizational structures. Integrating diverse data processing into a unified pipeline is challenging. \oursysII adopts an extensible and intermediate data representation schema that intrinsically supports multimodal data types and flexible data field customization.

An example of a data sample is illustrated in Listing~\ref{listing:mm_format}. In the data schema, each dataset encapsulates column-stored samples represented by a non-recursive dictionary whose level-1 fields include three categories: (1) the core contents within customizable fields, including ``text'' that is usually used in the pretraining task, and ``query'', ``response'', ``history'' fields represent dialogs in post-tuning tasks, which are directly related to the pretraining or post-tuning procedures in the downstream usage of the dataset; (2) the extra data contents contain the path lists of multimodal data for multimodal datasets; (3) the ``meta'' information concerning this sample that stems from either its raw data or \oursys's Mappers for tagging, and the ``stats'' information calculated by \oursys's Filters (all Filter OPs consist of a \texttt{compute\_stats()} method, utilized for subsequent analysis and conditional filtering).

For multimodal datasets, the content is primarily centered on the textual modality within the ``text'' field, capable of optionally including several text chunks split by customizable end-of-chunk tokens, which are \texttt{<|\_\_dj\_\_eoc|>} in default. Each chunk serves as a semantic unit, where the associated multimodal data within the same chunk relates to the same topic, thereby aligning with one another. Multimodal data (excluding text) are denoted by ordered customized special tokens in the text (e.g., \texttt{<\_\_dj\_\_image>}), storing entities at designated files, accessible through indexed file paths within the modality fields of samples to facilitate object sharing and elimination of redundant computing. For instance, in the example shown in Listing \ref{listing:mm_format}, the first chunk in the ``text'' field split by the \texttt{<\_\_dj\_\_eoc>} token contains the first \texttt{<\_\_dj\_\_image>} token and a description to this image, which corresponds to the first image stored in the ``images'' field.

This token-centric representation elegantly aligns multimodal data while preserving positional information, friendly to current prevalent learning paradigms for foundation models employing next-token prediction tasks \cite{bai2024survey,wang2024emu3}. And it's compatible with both simple cross-modal pairing datasets (e.g., image-text pairs \cite{chen2015microsoft} and video-text pairs \cite{xu2023youku}), as well as complex interleaved multimodal datasets (e.g., MMC4 \cite{zhu2023multimodal}), owing to its chunkable and special token design.

For post-tuning datasets, we provide several core fields in the core contents to represent the dialog datasets in our intermediate format, which is naturally aligned with Alpaca and Query-Response formats in the ModelScope-SWIFT repository \cite{zhao2024swiftascalablelightweightinfrastructure} for smooth training of 400+ foundation models. As for other widely-used formats in well-known repositories like LLaMA-Factory \cite{zheng2024llamafactory}, a suite of dataset conversion tools was developed to efficiently transform various popular multimodal datasets in them to and from the \oursys schema, enabling users to process extensive existing models and datasets within \oursys. These tools also serve as demonstrations for extending support to other uncovered datasets.

\subsection{Operator Factory \& Taxonomy}
\subsubsection{OP Design Principles}
\label{app:op_taxonomy}
Operators are fundamental units responsible for executing processing functionalities, such as enhancing the accuracy of descriptive text or filtering out images containing Not-Safe-for-Work (NSFW) content. As illustrated by the middle orange box in Fig.~\ref{fig:overview}, \oursys defines five previous atomic OP classes: Formatter, Filter, Mapper, Deduplicator, and Selector; alongside four new types of compositional OPs: Grouper, Aggregator, FusedOP, and ScriptOP. The first five OP classes handle dataset format conversion, sample filtering, modification, deduplication, and selection, respectively. Following the \textit{Strategy Pattern}, they are designed to encapsulate diverse algorithms that can be dynamically used to process the data. Each OP has a clearly defined role and can be interchanged or modified without impacting other system parts.

\begin{listing}[!ht]
\setlength{\belowcaptionskip}{0pt}
\begin{minted}[
    frame=lines,
    framesep=2mm,
    baselinestretch=1.2,
    fontsize=\scriptsize, 
    linenos,
    breaklines=true, 
    bgcolor=lightgray!10, 
    rulecolor=black, 
    numbersep=5pt, 
]{python}
# Default approach processes the whole dataset sequentially
data.process(op1).process(op2) # or data.process([op1, op2])

# Explict fine-grained processing at the batch level
data.process(FusedOP([op1, op2], bs=1000))
# Internally
for data_batch in data.next(batch_size):
    data_batch.process(op1).process(op2)
\end{minted}
\caption{Illustration of FusedOP in \oursysII.}%
\label{listing:dj_fused_op}
\end{listing}

Moreover, following the \textit{Decorator Pattern}, compositional OPs are provided to enhance existing functionality without modifying the prevailing OP structure while dynamically adding data processing behavior to objects. 
The FusedOP enables explicitly grouping multiple atomic OPs to process data in a fine-grained manner within the same data batch, opposed to the default sequential processing across datasets as demonstrated by Listing \ref{listing:dj_fused_op}.
The Grouper takes a \texttt{\oursys-Dataset} as input and groups data samples into batches, which can then be input into the Aggregator for subsequent aggregation. For example, we can employ \textitt{extract\_entity\_attribute\_mapper} and \textitt{ entity\_attribute\_aggregator} following a \textitt{key\_value\_grouper} for meta-information extraction from textual input.
Meanwhile, the ScriptOP allows users to incorporate existing Python files or execute code snippets, utilizing customized functionalities encoded within scripts like \texttt{helper\_func.py}. This includes leveraging existing \oursys command tools or integrating short Python scripts (e.g., lambda functions). 

Together, these nine OP types provide robust expressive capabilities for end-to-end data processing solutions that can be embedded within a single YAML configuration file (as depicted in Fig.~\ref{fig:recipe_cfg}).

\subsubsection{Unified Coordination of Logical Operations}
\label{app:fusedop_example}
In \oursysI, the logical operations of different OPs were coordinated within various executors (either standalone or distributed) rather than being bound to the OPs themselves. Disentangled from the executor's scheduling interface, it becomes challenging to determine the execution logic of different OPs, making the development and extension of individual OPs less intuitive and lacking self-explanatory qualities. In \oursysII, several design principles are utilized to address this issue, including the \textit{Abstract Factory Pattern, Template Method Pattern, and Single Responsibility Principle}.

Specifically, a top-level OP factory class is abstracted above the aforementioned fundamental OP classes. In this class, functionalities common to all OPs are extracted, such as preprocessing of instantiated parameters, support for serialization, and configuration of OP-aware runtime parallelism.
Besides, a unified \texttt{run()} method is implemented, maintaining a consistent interface for integration and API calls. Furthermore, parallelism in multi-processing and multi-GPU is automatically configured and decoupled from specific OPs, ensuring transparency for end users and developers, as introduced subsequently in Appendix~\ref{app:op-adapter}. Lower-level OP classes define their own templated execution logic behind the run invocation. Taking the Filter class as an example, its core logic first engages the \texttt{compute\_stats()} method to obtain statistical information based on specific metrics and then invokes its \texttt{process()} method to determine sample filtering based on thresholds.

On the one hand, this simplifies users' understanding of OP types. Users can instantiate any OP and invoke it with a unified parameter signature using \texttt{op.run()}, thereby reducing the learning curve. The templated execution flow of various OP types is self-contained within base classes, eliminating dependencies on external executors to oversee invocation logic. On the other hand, by templating execution logic within base classes, developers can readily modify, extend, or implement new OPs. For instance, a developer aiming to customize an existing Filter does not need to rewrite a new class entirely but can inherit from a related existing class and override specific methods such as \texttt{compute\_stats()} or others as required.

Regarding the naming and implementation of specific leaf OP classes, we adhere to extracting functionalities that are not tightly coupled with the OPs into common utility classes wherever feasible. This approach enables each OP to focus on its specific modalities and functionalities, facilitating a reduction in code complexity and enhancing clarity in understanding individual OP classes. Compared to previous implementations, the revised OP classes demonstrate easier inspection, integration, and testing. Users can utilize these robust OPs and seamlessly integrate them into their own tools or systems flexibly, both in source code or exposed RESTful API.

\subsection{Control Panel Implementation}
\label{app:job_detail}

With the fundamental \texttt{\oursys-Dataset} and \texttt{\oursys-Operators} primitives established, we offer a series of control panel modules (the red box in Fig.~\ref{fig:overview}) to organically combine them and accomplish data processing tasks. \texttt{Executor} encapsulates a series of standardized execution processes tailored for different standalone and distributed engines, leveraging modules such as \texttt{Config} and \texttt{Monitor} to accomplish end-to-end system configuration, data loading, analysis, processing iteration, data checkpointing, and more. Template workflows manage the complete data development lifecycle, including feedback, data storage, logging, and performance and operational monitoring.

To avoid the substantial costs from trial and error for data and model development in foundation model scenarios, we further develop a Sandbox suite in \oursysII for \textit{data-model co-development}, serving as a specialized intermediate layer connecting data processing jobs to numerous open-source infrastructures of model training and evaluation.
The suite offers template workflows that extend beyond dataset-only development by incorporating cost-effective model training and evaluation signals, and quantitatively studying interplays among data, model, and compute. 
Users can easily conduct small, quick, and comparative experiments to find insights and superior data recipes, which can then be scaled to larger models and datasets, thereby optimizing the return on investment in data-model co-development. Additional details and empirically validated support can be found in \cite{chen2024djsandbox}.


\section{Optimization Details on \oursys-Dataset}
\label{app:dataset-optimization}
\subsection{Engine-agnostic Processing}
\label{sec:engine-agnostic}

In \oursysI, the data processing pipeline is implemented using two distinctive execution modes: a standalone mode tailored for single-node operations and a distributed mode designed for scalability over multiple nodes. These modes exploit different dataset classes, each with a unique set of functionalities and interfaces.
The default standalone execution mode employs the Hugging Face \texttt{Dataset} class \cite{lhoest-etal-2021-datasets}, which provides a rich array of encapsulated functions, such as \texttt{dataset.map()} and \texttt{dataset.filter()}. It's also equipped with configurable batch processing capabilities essential for various computational needs.
In contrast, the distributed execution mode leverages the \texttt{Ray Dataset} class \cite{ray}, which scales effectively across multiple nodes. 

Despite both dataset classes using a storage format based on Apache Arrow \cite{apache-arrow}, they exhibit significant differences in the behaviors and interfaces exposed. For instance, the Ray Dataset delineates individual and batched sample processing using separate methods: \texttt{map} and \texttt{map\_batches}. It also allows to specify GPU counts for optimized scheduling. Meanwhile, the Hugging Face Dataset excels in supporting a broad spectrum of data modalities, such as image and audio.
Moreover, the Hugging Face Dataset is typically applied in read-heavy data processing scenarios, such as in-memory tokenization and tensor reshaping, which are crucial for training deep learning models. For processing tasks in the context of foundation models, especially those involving synthesis operations, write-heavy procedures warrant attention. Here, the Ray Dataset provides flexible data exportation techniques advantageous to such tasks.

As highlighted in Sec. \ref{sec:dj_dataset}, we introduce a top-level \texttt{\oursys-Dataset} class in \oursysII, along with common functions to bridge the variety of interfaces and implementation specifics across diverse computational engines. Besides the support of Hugging Face Dataset and Ray Dataset, our \texttt{\oursys-Dataset} is also seamlessly extendable to support distributed computing within Alibaba Cloud ecosystems, thanks to the compatibility between ours and MaxFrame-DataFrame classes with intermediate in-memory formats like Pandas \cite{reback2020pandas} and NumPy \cite{harris2020array} or external-memory formats like Parquet. 
Thus, we develop unified wrappers that extract the core execution functions of different \texttt{\oursys-Operators} into User-Defined Functions for the MaxCompute engine. 
Compared to Spark, MaxCompute is compatible in both syntax and runtime on Alibaba Cloud nodes and can be considered a commercially optimized version. An internal empirical comparison shows that MaxCompute SQL achieves 50\% better performance than native Spark SQL.

\textbf{The Deduplicators within \oursysII.}
Fuzzy deduplication is complex, involving a mixture of operations such as map, filter, group by, aggregate, and join \cite{BRODER2000630-minhash}. The support and performance of these operations vary across different engines, especially in large-scale scenarios.
To demonstrate the engine-agnostic feature of \oursysII, we use our \textitt{minhash\_deduplicator} as an example, which supports the aforementioned three different engines. Users only need to specify algorithm-specific parameters such as \texttt{jaccard\_threshold} and \texttt{num\_permutations}. 
These parameters remain consistent regardless of the engine used, while engine-specific details and optimizations are completely transparent to the user.
For instance, we utilize Ray Actors to implement our Ray-based deduplicator, starting with a load-balanced, distributed union-find algorithm~\cite{bts} and introducing a hash-based aggregation process to enhance memory utilization and efficiency. This method avoids fragmented unions caused by Ray's native \texttt{groupby} operation, which is computationally expensive in typical LSH implementations with traditional big-data engines~\cite{hf-dedup}.
As a result, we achieve a 3.3x speedup over our vanilla Ray version.

\subsection{Fault Tolerance}
\label{app:fault_tolerance}

In practical processing scenarios, datasets often contain schema-incompatible or corrupted data elements, such as improperly formatted JSON objects or damaged images that cannot be read. This issue becomes increasingly important with large-scale datasets, as processing tasks may extend over several hours or even days. In \oursysI, corruption of a single sample would halt the entire processing task, resulting in a waste of computational resources and the loss of already processed samples.

\begin{figure}[h!]
    \centering
    \includegraphics[width=0.98\linewidth]{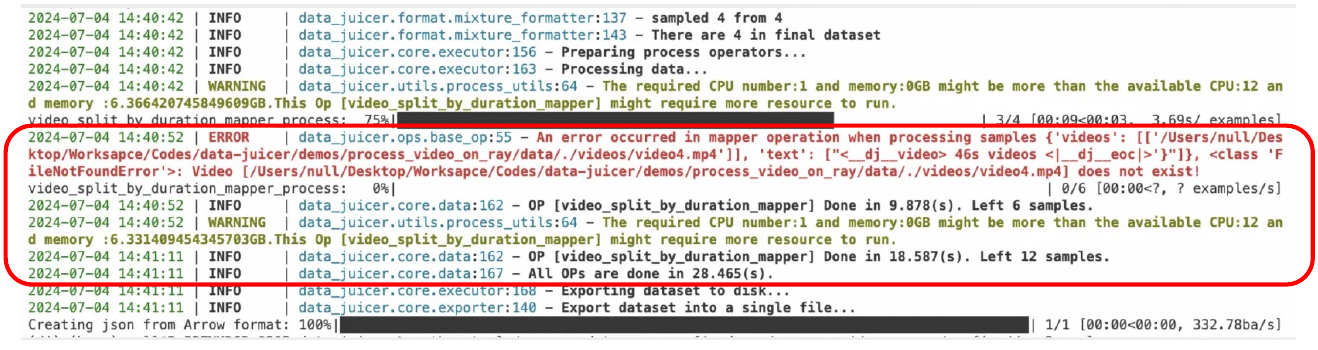}
    \caption{Illustration of fault tolerance for ``bad'' data.}
    \label{fig:demo_fault_tolerance}
\end{figure}

To address this issue, \oursysII introduces a sample-level fault tolerance mechanism designed to enhance processing reliability by providing a worst-case guarantee. A unified exception manager is implemented to automatically capture runtime errors with customizable handlers during the processing of each OP. By default, dataset processing operations are performed at the data-batch level for a general handler. As a result, the system can easily bypass problematic samples by skipping the affected batch (as demonstrated in Fig.~\ref{fig:demo_fault_tolerance}), while actively tracking and reporting these cases for subsequent debugging and correction. This ensures a seamless user experience and minimizes retry costs in scenarios involving large-scale data processing. Users have the flexibility to either discard these samples or mark them for future reprocessing.

Of particular note is the fact that implementing this mechanism in \oursysI presented challenges due to the rigid schema consistency mandates imposed by the underlying Hugging Face and Ray dataset classes. The newly refactored batching interfaces facilitate the unified construction of compatible empty samples based on schemas accessed during exception handling, thereby enabling seamless integration with valid dataset entries.

\subsection{Streaming I/O and Subset Splitting}
\label{sec:data_io_split}
Memory constraints frequently emerge as a bottleneck in data processing tasks associated with foundation models. Memory demands can be substantial, and its precise usage is difficult to predict beforehand. Influential factors include the actual storage demands of objects such as text, image, and audio within individual dataset samples, as well as the storage demands of auxiliary or newly generated objects at runtime. These can stem from varying OP specifications, model sizes, synthesis data volumes, intermediate variables, the number of concurrent processes, and specific computational engines. 

To effectively adapt to a diverse range of data processing scenarios with varying data volumes and available computational resources, we introduce streaming loading and data pre-splitting capabilities in \oursysII. They collectively facilitate improved computational resource utilization, and provide potential for a flexible programming space of hybrid stream and batch processing.

Firstly, \oursysII offers a streaming loading interface, addressing the current lack of native support in the Arrow framework underlying Hugging Face and Ray Datasets for streaming JSONL data. As many foundation model datasets use JSONL, we developed an in-house patch to alleviate Out-of-Memory (OOM) issues.

Secondly, we develop a user-friendly script that automatically pre-splits the original dataset based on two observations: (1) The size limit of the underlying Apache-Arrow block, and (2) The inherent automatic block-splitting strategy in Ray. 
When there are tens of thousands of nodes but with only a few dataset files, Ray would split the dataset files according to the available resources and distribute the blocks of the dataset to all nodes, which brings a huge network communication cost and decreases the CPU utilization of each node.
Thus, we split the original dataset into smaller 128MB files in advance automatically according to the dataset size and the number of distributed nodes, trying to adapt the features of Arrow and Ray for better performance. 
This approach reduces location and reprocessing costs associated with fault tolerance and helps mitigate network exchange overheads, especially beneficial in contexts involving large-scale multimodal data, as well as in scenarios that require handling global objects of Ray Actor in distributed modes.

\section{Implementation of Adaptation for \oursys-Operator}
\label{app:op-adapter}
This section explores the internal adaptations developed for \texttt{\oursys-Operator}, which are crucial for optimizing resource allocation and user experience without requiring users to understand hardware specifics or implementation details. We implement several automatic adaptation features for resource management, aiming at balancing resource constraints and operational efficiency within \oursysII. These strategies include automatic OP reordering at the recipe level (Appendix \ref{sec:op_order}), automatic parallelism at the OP level (Appendix \ref{sec:op_wise_parallel}), and automatic data insight mining to assess the impact of each OP on data samples, considering both upstream and downstream OPs (Appendix \ref{app:sec:op-insight}).
These features are encapsulated in a dedicated \texttt{Adapter} class, which uses a \texttt{probe\_small\_batch()} method to systematically probe and analyze essential information by applying individual OPs on randomly sampled data in runtime, with default sample size as \texttt{min(1000, remaining\_data\_size)}.

\subsection{Workloads-aware OP Reordering}
\label{sec:op_order}
At the recipe level, we introduce a new probe-based OP reordering strategy. In \oursysI, an OP fusion optimization was proposed to eliminate redundant computations for the Filter OPs, which involved three key steps: detection and grouping of fusible OPs, OP fusion, and OP reordering. 
The reordering strategy aimed to position more time-consuming OPs at the end of the group to process fewer samples to save time, as some were filtered out by preceding OPs. It is assumed that the fused OPs are the most time-consuming, and only the fused OPs are moved to the end of each group.

However, the assumption is not always correct and the prior reordering strategy omit the unfused OPs, rendering it greedy and suboptimal, especially when applied to diverse datasets and data recipes characterized by varied data distribution and OP orchestration. In \oursysII, we advance the reordering strategy to an adaptive approach, which is workload-aware and can be applied automatically to unfused OPs as well.

Specifically, before processing the full dataset, \oursysII utilizes the probe functionality of \texttt{Adapter} to obtain estimated processing speeds for individual OPs relevant to specific input datasets. When processing the entire dataset begins, OPs in each group (including the unfused ones) are reordered based on the probed speeds and the commutativity of the Filter OPs.
For the fused OP, assuming that there are $n$ fusible OPs in it and their probed speeds are $v_i$ where $i \in \{1, 2, \cdots, n\}$, the estimated speed of the fused OP can be calculated as:
\begin{equation}
    v_{fused} = \frac{N}{T_{total}} = \frac{N}{\sum_{i=1}^n \frac{N}{v_i}} = \frac{1}{\sum_{i=1}^n \frac{1}{v_i}},
\end{equation}
where $N$ is the total number of samples to be processed by the fused OP and $T_{total}$ is the total time cost of it.
Then, faster OPs are prioritized, while slower ones are deferred to later stages. This probe-based approach identifies the globally optimal reordering solution for each OP group, outperforming the suboptimal strategies of the previous version, as empirically validated in Appendix \ref{exp:op-fusion}.

\subsection{Automatic Operator-wise Parallelism}
\label{sec:op_wise_parallel}
In data processing for foundation models, it is crucial to recognize that different OPs require vastly different computational resources. Model-based OPs often need several gigabytes of GPU memory, while simple rule-based OPs may only need minimal CPU processing. Therefore, using a uniform parallelism granularity across all OPs in a data pipeline can cause OOM issues for some and resource underutilization for others.
To address this challenge, we introduce several automatic mechanisms for OP-wise parallelism. 

\begin{itemize}[leftmargin=*]
    \item 
    Model-based OPs that integrate substantial models require considerable computational time, potentially spanning hundreds of hours on CPUs for large datasets. To mitigate this, \oursysII expedites these OPs by automatically leveraging CUDA and GPUs when available. Given the diverse memory requirements of large models, we utilize the \texttt{Adapter} component to conduct quick VRAM benchmarking prior to full-size dataset processing. This information is systematically assigned to the \texttt{mem\_required} parameter of the respective OPs. 
    During the execution of these OPs on datasets, \oursysII continuously assesses the available GPU memory of the execution environment to dynamically and adaptively determine the optimal parallelism strategy. 
    To further alleviate GPU memory demand, we also integrate quantization libraries such as vLLM \cite{kwon2023efficient} to enable efficient inference of leveraged models.
    
    \item 
    For non-model-based OPs, attributes such as \texttt{cpu\_required} and \texttt{mem\_required} are crucial. In \oursysII, users can specify a global parallelism level, typically aligning with the available processor count. Meanwhile, we calculate the adaptive number of processors to finely optimize the entire processing pipeline, aiming to maximize resource utilization (90\% by default) as much as possible. For this purpose, the \texttt{Adapter} is also employed dynamically and instrumental in determining precise \texttt{cpu\_required} and \texttt{mem\_required} values at runtime.    
    
    \item 
    For general-purpose and I/O-intensive OPs, \oursysII enables batched processing, using a robust default \texttt{batch\_size} parameter guided by performance saturation trends shown in Appendix \ref{sec:batch_size}. Batched processing reduces I/O overhead, boosts efficiency per processor, and enhances overall speed. Additionally, we introduce hierarchical parallelism for OPs involving multimodal data I/O, such as \textitt{image\_aspect\_ratio\_filter}. 
    These OPs utilize multi-process and GPU parallelism, along with multiple threads, to handle data batches more efficiently, taking the concurrent opportunities between I/O and computation latencies.
    
\end{itemize} 

\subsection{Insight Mining Example}
\label{app:sec:op-insight}

\begin{figure}[h!]
    \centering
    \includegraphics[width=0.98\linewidth]{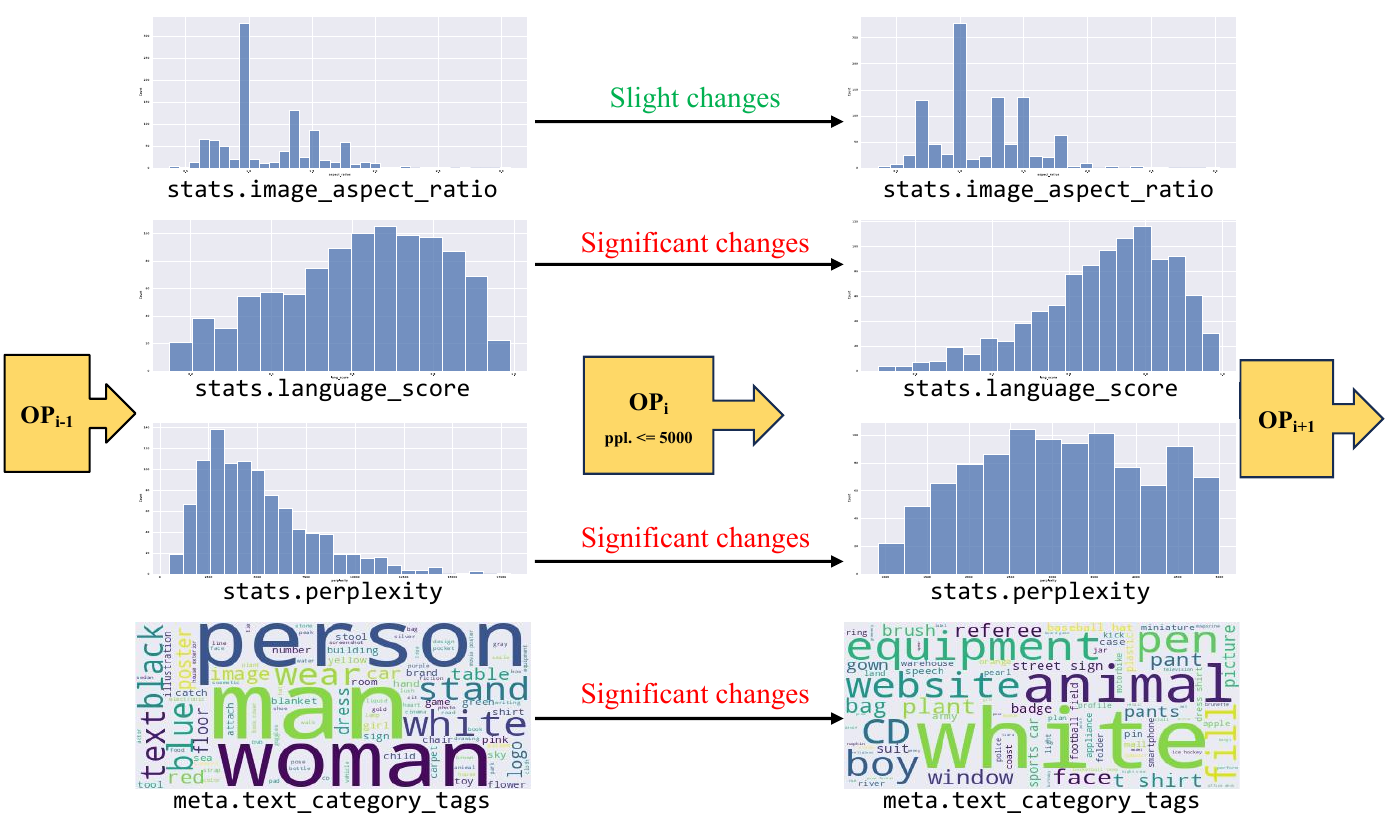}
    \caption{Illustration of OP-wise insight mining.}
    \label{app:fig:op-insight}
\end{figure}

Here we provide an example of OP-wise insight mining in Fig. \ref{app:fig:op-insight}.

\section{Future Directions in Pipeline Optimization for Exascale Data}
\label{app:pipeline-opt}

As data processing demands for foundation models scale towards the exabyte level, merely adding more computational resources becomes economically and technically unsustainable. Our future work thus focuses on a next-generation pipeline optimization framework that functions akin to a query optimizer for deep learning data pipelines. This framework is designed to intelligently rewrite and execute data-processing graphs to minimize fundamental bottlenecks in I/O, memory usage, and computation. Key architectural directions include:

\textbf{Advanced Operator Fusion.} Building upon the existing filter fusion, this strategy involves analyzing the pipeline's Abstract Syntax Tree (PipeAST) to identify and merge consecutive, computationally compatible operators. For example, a sequence of resize, crop, and normalize operations on an image could be fused into a single, highly optimized kernel. This significantly reduces intermediate data materialization and memory traffic, streamlining per-sample transformations.

\textbf{Pipeline-Aware Column Pruning.} For wide-schema datasets, such as Parquet files from autonomous driving logs containing hundreds of sensor columns, loading all data is highly inefficient. We envision a system that performs static analysis on the entire pipeline's data dependencies to determine the precise subset of columns required for the complete workflow. This enables selective loading of only the necessary data from storage, drastically reducing I/O and memory footprint from the very first read.

\textbf{Predicate Pushdown to the Storage Layer.} This powerful optimization involves pushing inexpensive filtering predicates as close to the data source as possible. For instance, a filter on image metadata (e.g., resolution > 1024x1024 or creation\_date > 2023) can be applied during the initial file listing or manifest scanning phase. This ``pre-rejects'' samples before their potentially large data payloads (e.g., multi-megabyte images) are downloaded or deserialized, preventing immense wastage of network bandwidth and compute resources.

\textbf{Dynamic and Adaptive Resource Allocation.} Transcending static resource configurations, this direction focuses on a runtime that monitors real-time system metrics (CPU utilization, GPU VRAM, memory bandwidth). By leveraging advanced features in modern execution engines (e.g., Ray's compiled execution graphs), the system can dynamically adjust operator-level parallelism and resource assignments. For example, it could temporarily allocate more GPUs to a bottleneck model-based filter, ensuring sustained optimal throughput across heterogeneous hardware and fluctuating workloads.

\section{Additional Experiment Details and Results}
\label{app:additional-exp-details-results}

\subsection{Implementation Details of Scalability Experiments}
\label{app:exp-setup-details}
We start with a base dataset comprising 560k image-text pairs (about totally 5.6 million textual tokens) from the pertaining dataset of LLaVA \cite{liu2023llava}. 
We expand this dataset by factors of \{2, 4, 20, 100, 500, 2500, 12500, 125000\}, resulting in nine datasets scaling to 70B data samples.
These datasets are categorized into three scales: small (1x, 2x, 4x), medium (4x, 20x, 100x), and large (100x, 500x, 2500x, 12500x, 125000x). For different scales, we prepare both multimodal and text-only data processing recipes, each containing 5 OPs.
We run the recipes on these datasets using various computing engines (Standalone, Ray, MaxCompute) with different Alibaba Cloud resources, including ECS instances, PAI-DLC, and MaxCompute, spanning 1 to 100 nodes and 64 to 12,800 CPU cores. 
To ensure a fair comparison, we use the same worker node configuration, such as CPU frequency, across different computing engines.

\subsection{Ablation Study on Runtime Adaptations}
\label{sec:exp_efficiency_runtime}

\subsubsection{Automatic workloads-aware OP reordering}
\label{exp:op-fusion}
We evaluate the performance improvements due to our probe-based OP reordering on two recipes with different numbers of OPs. The simple recipe contains 5 OPs, with 2 fusible OPs, while the complex recipe includes 13 OPs, with 5 fusible OPs. We run these recipes on the base dataset used before (560k image-text pairs), comparing processing times for all OPs, fusible OPs, and other OPs.

\begin{figure}[htb]
    \centering
    \subfloat{\includegraphics[width=0.49\linewidth]{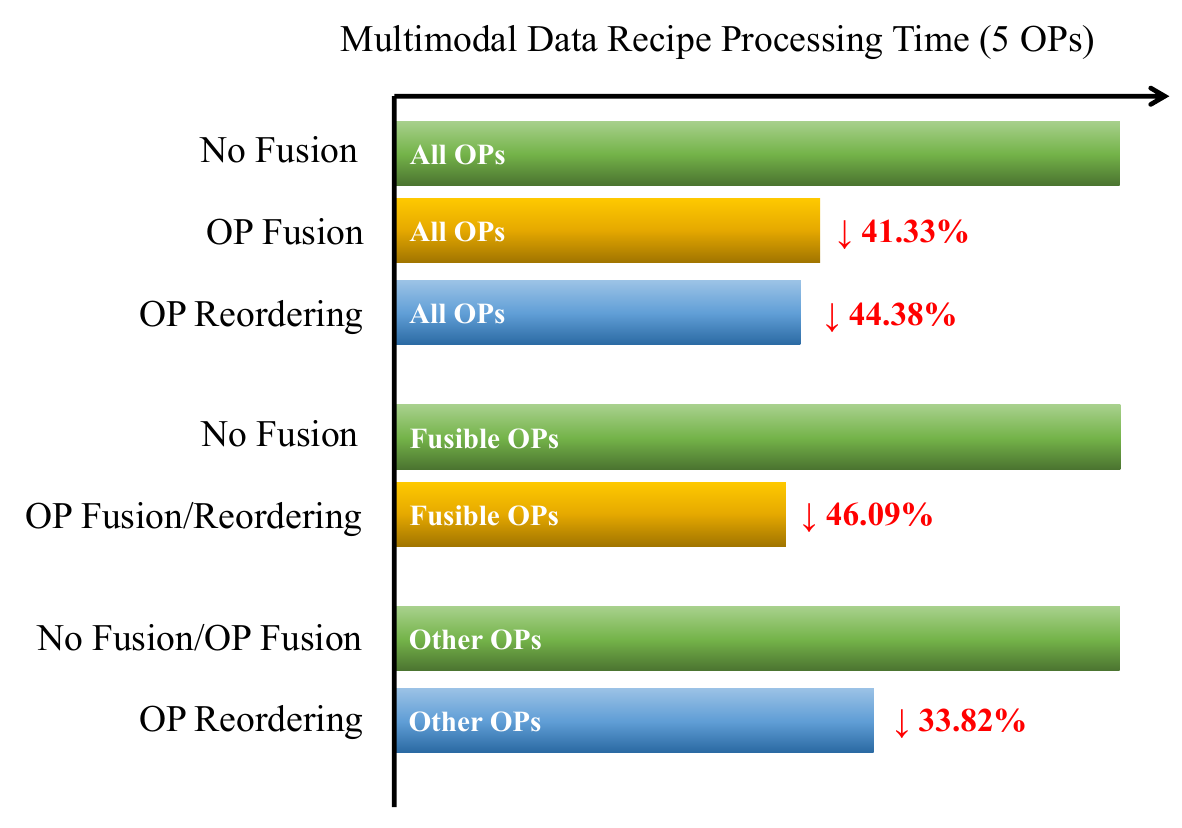}}
    \subfloat{\includegraphics[width=0.49\linewidth]{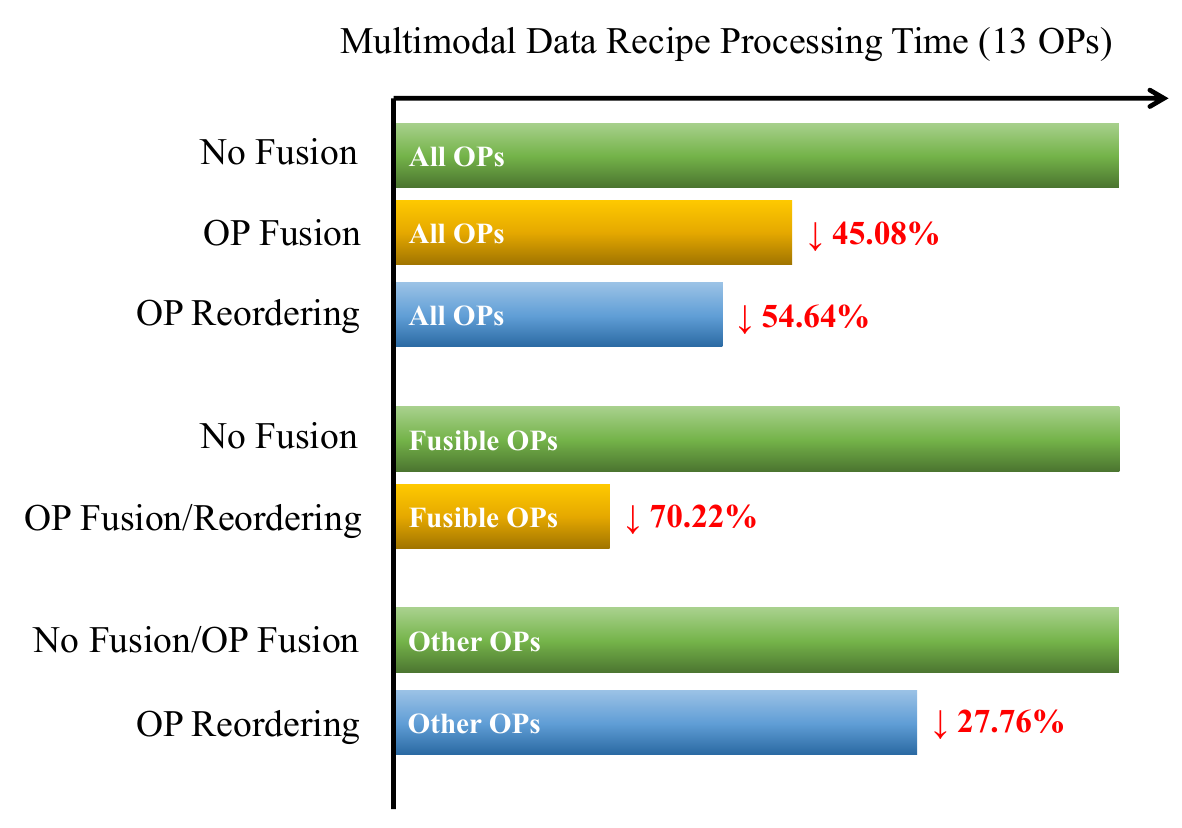}}
    \caption{Processing time for simple and complex recipes. 
    }
    \label{fig:probe_exps}
\end{figure}

Fig.~\ref{fig:probe_exps} illustrates the results. 
Generally, OP fusion is effective for accelerating data processing, and automatic OP reordering offers further improvements based on OP fusion. 
Notably, OP reordering is more effective in complex data recipes (the right sub-figure), saving more time, especially for fusible OPs (46.09\% v.s. 70.22\%).

\subsubsection{Automatic GPU resource allocation}
Efficient model inference on GPUs is critical for the processing speeds of model-based OPs. We select four image-related OPs, including an image-only OP and three image-text cross-modality OPs, integrating various models that require different GPU memories. Using a test dataset of 10k image-text pairs, we process these OPs on a machine with 64 processors and 8 Nvidia A100 GPUs, and compare the times with and without GPU support in \oursysII. Table \ref{tab:cuda_exp} shows these comparative results.

\begin{table}[h!]
\centering
\small
\caption{Multimodal OPs processing time comparison between CPU cores and GPUs on 10k samples. ``np'' denotes the number of processors.}
\label{tab:cuda_exp}

\begin{tblr}{
  width = \linewidth,
  colspec = {Q[520]Q[100]Q[40]Q[150]Q[150]},
  cell{1}{1} = {r=2}{},
  cell{1}{2} = {r=2}{},
  cell{1}{3} = {r=2}{},
  cell{1}{4} = {c=2}{},
  hline{1,7} = {-}{0.08em},
  hline{3} = {-}{},
  hline{2} = {4-5}{},
}
Multimodal OPs                  & VRAM                  & np & Processing Time &                     \\
                                &                       &    & CPU             & GPU                 \\
\footnotesize{\textitt{image\_captioning\_mapper}}       & \textasciitilde{}16GB & 32 & 8668.87s        & 305.44 s            \\
\footnotesize{\textitt{image\_diffusion\_mapper}}        & \textasciitilde{}8GB  & 64 & 100 h            & \textasciitilde{}1 h \\
\footnotesize{\textitt{image\_text\_similarity\_filter}} & \textasciitilde{}1.5GB & 64 & 73.04 s         & 35.84 s             \\
\footnotesize{\textitt{image\_nsfw\_filter}}            & \textasciitilde{}1GB   & 64 & 102.59 s        & 39.74 s             
\end{tblr}
\end{table}


As the results show, using GPUs saves at least 50\% of processing time for all selected model-based OPs, especially for large, slow models like BLIP-2 \cite{li2023blip2} used in \textitt{image\_captioning\_mapper} and SDXL model \cite{Rombach_2022_CVPR} in \textitt{image\_diffusion\_mapper}. Compared to the CPU version, GPU usage and adaptive resource allocation are much more efficient and necessary. Due to the GPU memory limit of around 80GB, we can only allocate at most 4 models of \textitt{image\_captioning\_mapper} OP on a single GPU, so the number of processors for this OP is automatically reduced to 32, preventing OOM errors.

\subsubsection{Batched data processing}
\label{sec:batch_size}
\begin{figure}[htb]
    \centering
    \subfloat[Different batch sizes and recipes. \label{fig:hie_exps:batch}]{\includegraphics[width=0.485\linewidth]{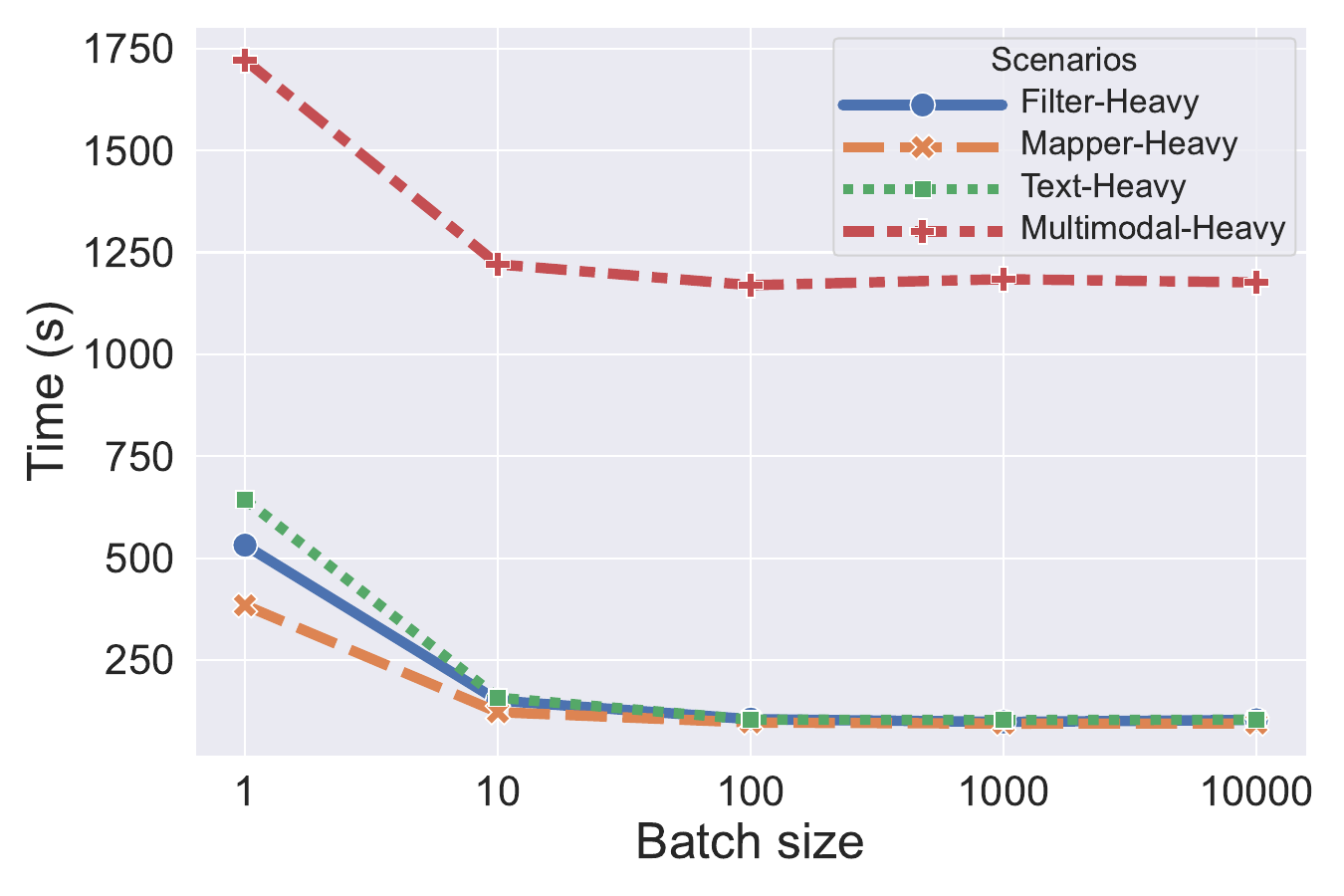} \vspace{-0.15in}}    
    \hspace{0.01\linewidth}
    \subfloat[Different parallelism levels.\label{fig:hie_exps:hierar}]
    {\includegraphics[width=0.485\linewidth]{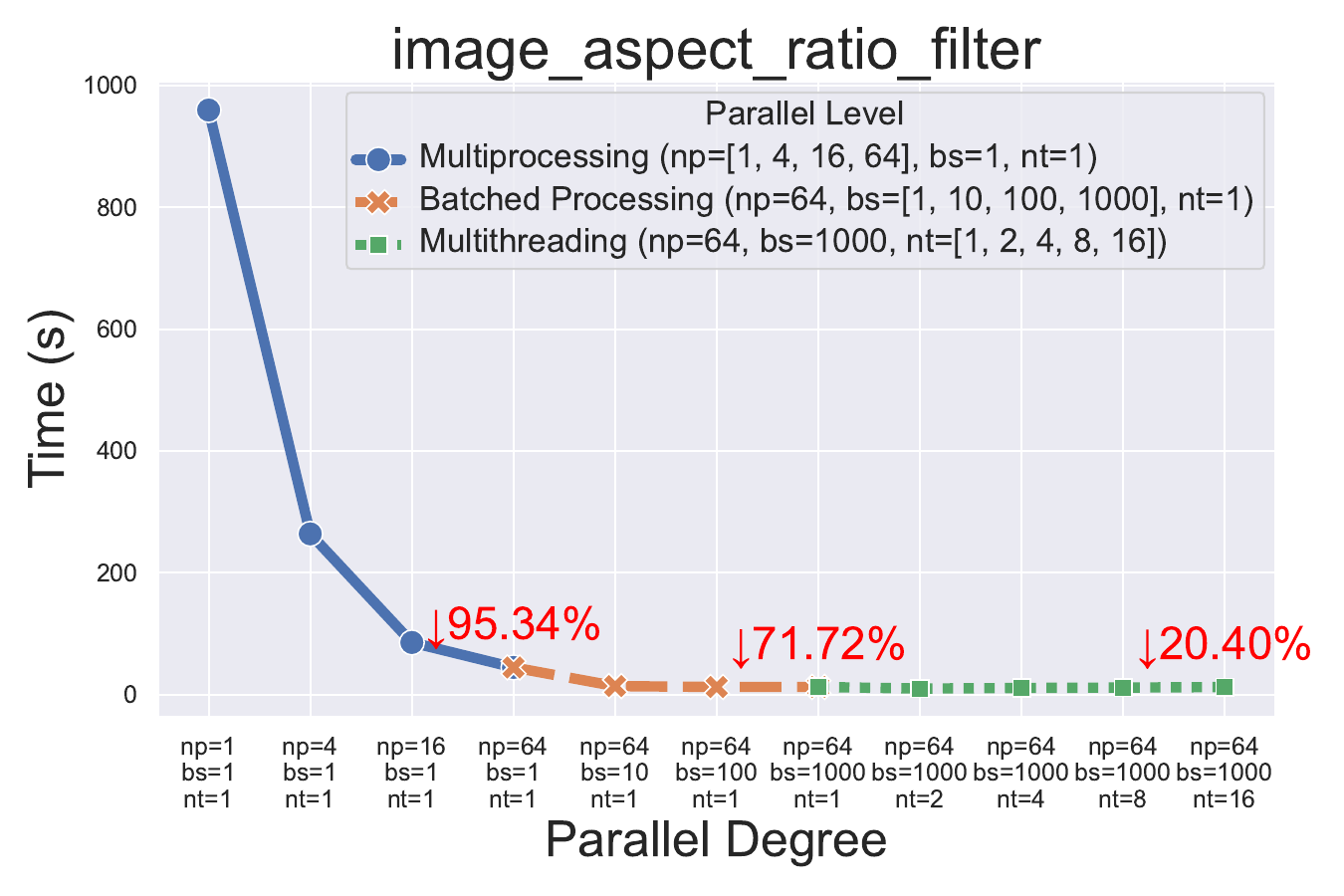}\vspace{-0.15in}}
    \caption{Data processing efficiency of OP-wise hierarchical parallelism. ``bs'' denotes the batch size, and ``nt'' dennotes the number of threads.}
    \label{fig:hie_exps}
\end{figure}

We compare the efficiency of single-sample processing in the previous \oursysI and batched processing in \oursysII. We use the previous 560k multimodal dataset, processed by 16 processors. Each processor handles about 30k samples. We select four recipes representing different scenarios: Filter-Heavy, Mapper-Heavy, Text-Heavy, and Multimodal-Heavy. Each recipe contains four OPs of the specified ``heavy'' type and one of another type. We run these recipes with five different batch sizes and summarize the results in Fig.~\ref{fig:hie_exps:batch}.
From these experiments, we conclude: (1) Batched processing is always more efficient. Larger batch sizes consistently speed up data processing in all scenarios, reducing processing time by 84\%. (2) Efficiency gains from larger batch sizes plateau beyond a certain threshold. Specifically, batch sizes of 100 or more yield similar benefits. (3) A batch size of 1000 is recommended, generally showing the most efficient processing in our experiments. Consequently, 1000 is selected as the default batch size in \oursysII.

\subsubsection{Automatic OP-wise hierarchical parallelism}
To verify the efficiency of the automatic OP-wise hierarchical parallel strategy, we select the multimodal OP \textitt{image\_aspect\_ratio\_filter} as an example. We conduct experiments with various processor counts, batch sizes, and thread numbers, covering three parallel levels: multiprocessing, batched processing, and multithreading. This OP processes approximately 560k image-text pairs. Fig.~\ref{fig:hie_exps:hierar} demonstrates the time consumption and efficiency gains.

As depicted, significant speed improvements are consistently achieved across all parallel levels. Beyond the multiprocessing and batched processing strategies examined earlier, multithreading further reduces processing time. This OP benefits from multithreading due to its intensive I/O procedures (reading images), which balance I/O and CPU utilization. Most other multimodal OPs also require heavy data read/write operations, thus benefiting from the multithreading strategy.

\section{High-Resolution Figures in the Paper}

\begin{figure}[htb]
    \centering
    \includegraphics[width=0.76\linewidth]{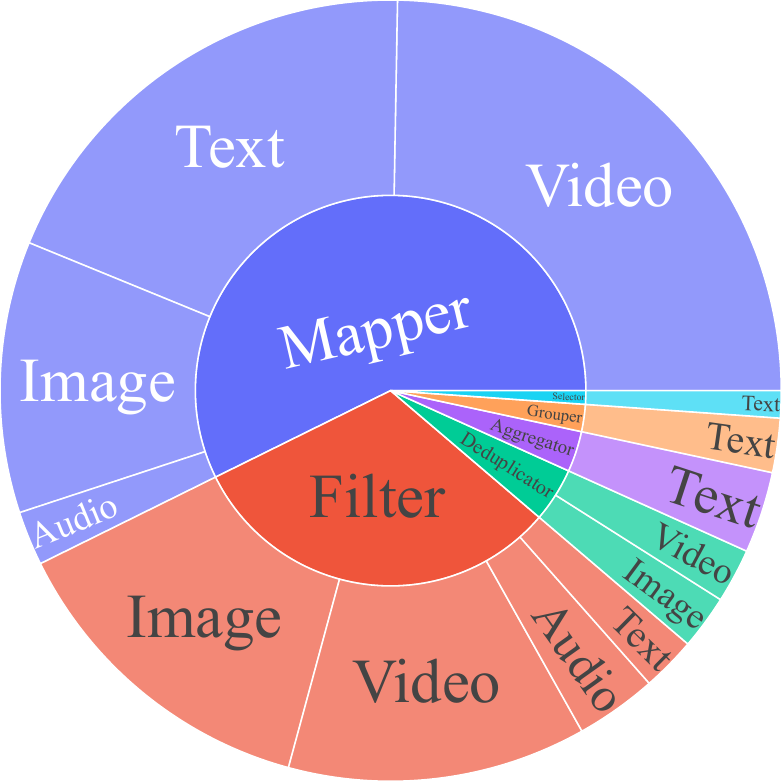}
    \caption{High-resolution version of Figure \ref{fig:op_num_dist:op}. The smallest sector in the inner circle is ``Selector''.}
    \label{fig:op_num_dist:op_hd}
\end{figure}

\begin{figure}[htb]
    \centering
    \includegraphics[width=0.76\linewidth]{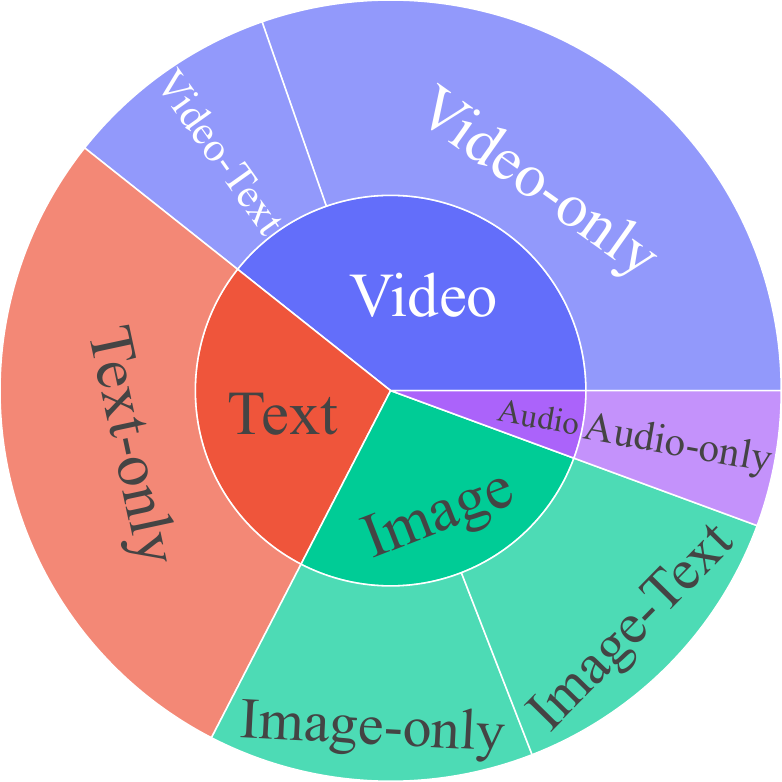}
    \caption{High-resolution version of Figure \ref{fig:op_num_dist:modality}.}
    \label{fig:op_num_dist:modality_hd}
\end{figure}

\begin{figure}[htb]
    \centering
    \includegraphics[width=0.76\linewidth]{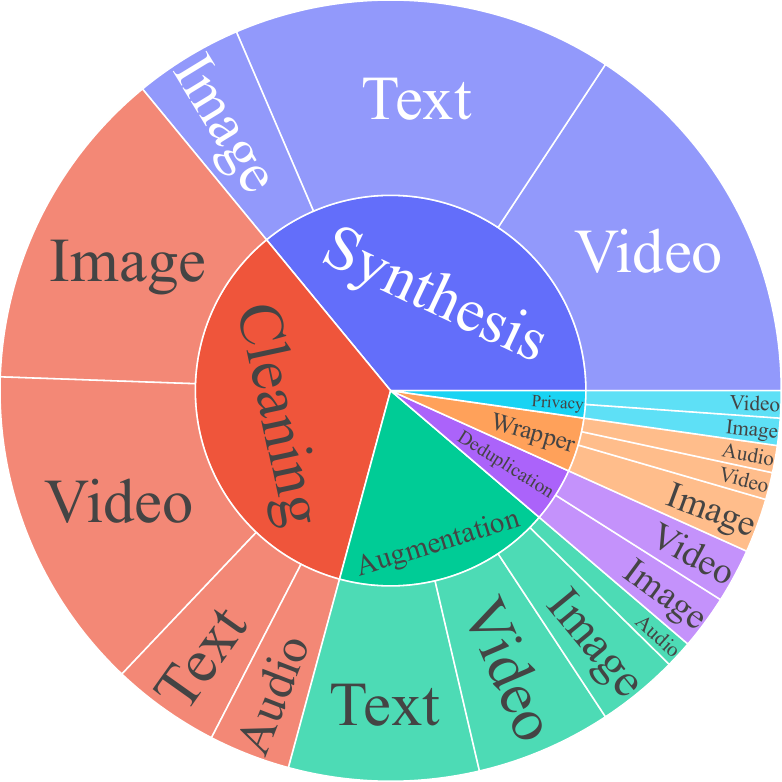}
    \caption{High-resolution version of Figure \ref{fig:op_num_dist:func}.}
    \label{fig:op_num_dist:func_hd}
\end{figure}

\begin{figure}[htb]
    \centering
    \includegraphics[width=0.76\linewidth]{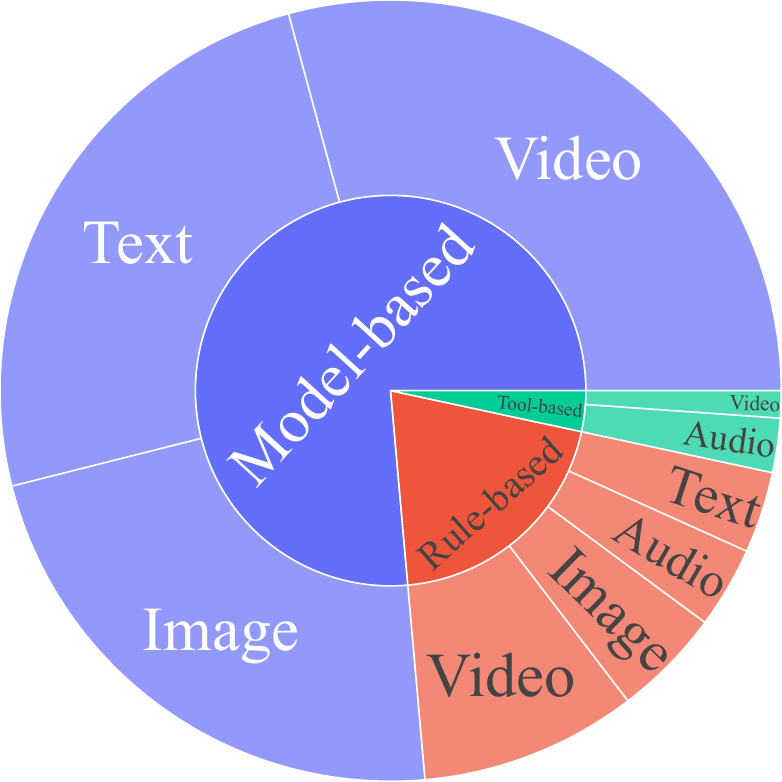}
    \caption{High-resolution version of Figure \ref{fig:op_num_dist:impl}.}
    \label{fig:op_num_dist:impl_hd}
\end{figure}

\begin{figure}[htb]
    \centering
    \includegraphics[width=0.76\linewidth]{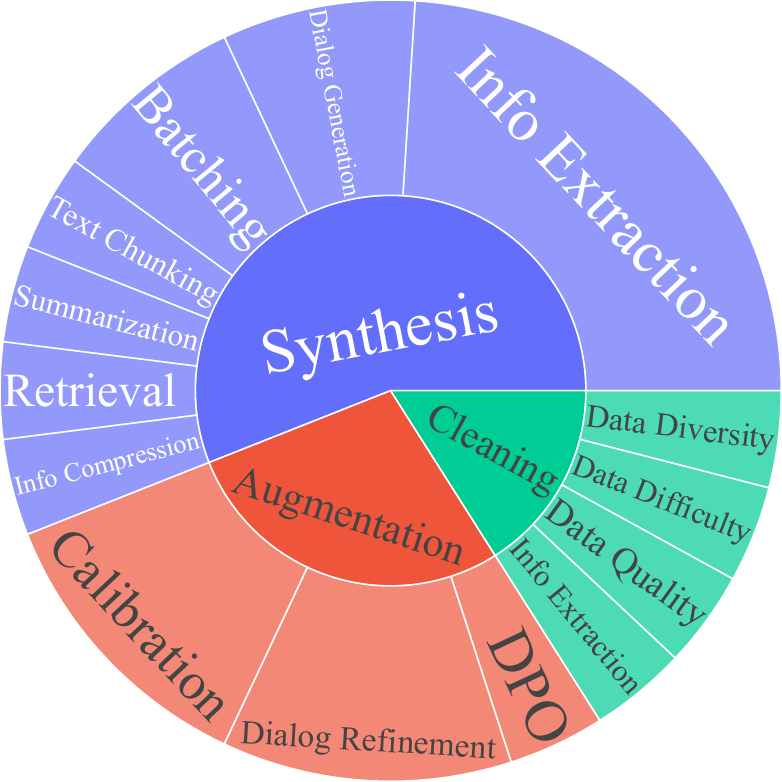}
    \caption{High-resolution version of Figure \ref{fig:fg_func_type}.}
    \label{fig:fg_func_type_hd}
\end{figure}

\begin{figure}[htb]
    \centering
    \includegraphics[width=\linewidth]{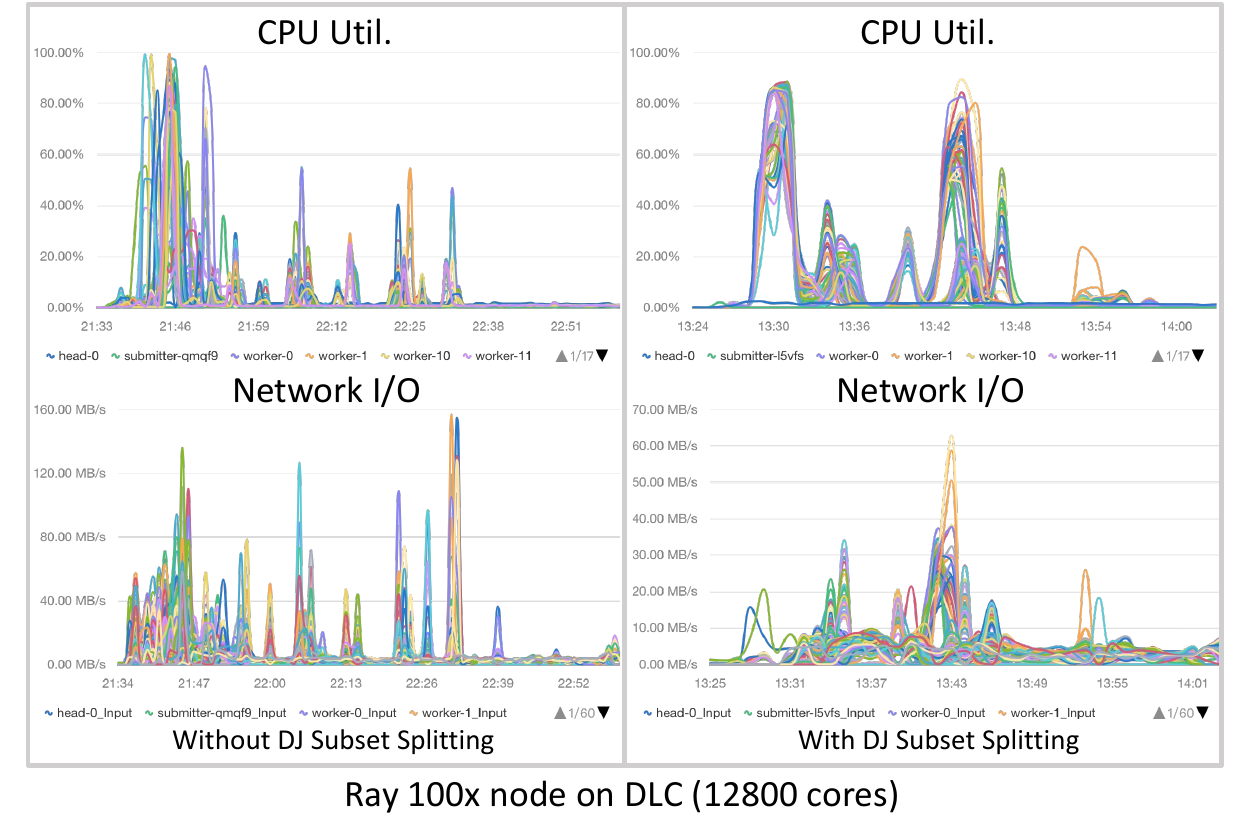}
    \caption{High-resolution version of Figure \ref{fig:cloud_exps:text_large_resource}.}
    \label{fig:cloud_exps:text_large_resource_hd}
\end{figure}

\clearpage

\newpage
\section*{NeurIPS Paper Checklist}

\begin{enumerate}

\item {\bf Claims}
    \item[] Question: Do the main claims made in the abstract and introduction accurately reflect the paper's contributions and scope?
    \item[] Answer: \answerYes{} 
    \item[] Justification: The claims in the abstract and introduction summarize the system designing, architecture, experimental results, and contributions of our paper.

\item {\bf Limitations}
    \item[] Question: Does the paper discuss the limitations of the work performed by the authors?
    \item[] Answer: \answerYes{} 
    \item[] Justification: The discussion on limitations of our work is in Sec. \ref{sec:conclusion}.

\item {\bf Theory assumptions and proofs}
    \item[] Question: For each theoretical result, does the paper provide the full set of assumptions and a complete (and correct) proof?
    \item[] Answer: \answerNA{} 
    \item[] Justification: This paper does not include theoretical results.

    \item {\bf Experimental result reproducibility}
    \item[] Question: Does the paper fully disclose all the information needed to reproduce the main experimental results of the paper to the extent that it affects the main claims and/or conclusions of the paper (regardless of whether the code and data are provided or not)?
    \item[] Answer: \answerYes{} 
    \item[] Justification: Appendix \ref{app:exp-setup-details} shows the implementation details and settings of our experiments.

\item {\bf Open access to data and code}
    \item[] Question: Does the paper provide open access to the data and code, with sufficient instructions to faithfully reproduce the main experimental results, as described in supplemental material?
    \item[] Answer: \answerYes{} 
    \item[] Justification: Code is open-sourced at \url{https://github.com/modelscope/data-juicer}, and no new data is released. Experiments are conducted based on public datasets such as LLaVA \cite{liu2023llava}, Redpajama \cite{together2023redpajama}, and Common Crawl \cite{commoncrawl}.

\item {\bf Experimental setting/details}
    \item[] Question: Does the paper specify all the training and test details (e.g., data splits, hyperparameters, how they were chosen, type of optimizer, etc.) necessary to understand the results?
    \item[] Answer: \answerYes{} 
    \item[] Justification: The basic setting of experiments is introduced in Sec. \ref{sec:exp_efficiency_dist:settting}, and the full details about the experiments are specified in Appendix \ref{app:additional-exp-details-results}, including dataset scales, data processing recipes, computing engines and resources, and so on.

\item {\bf Experiment statistical significance}
    \item[] Question: Does the paper report error bars suitably and correctly defined or other appropriate information about the statistical significance of the experiments?
    \item[] Answer: \answerNo{} 
    \item[] Justification: As mentioned in Appendix \ref{app:exp-setup-details}, due to the considerable costs involved 100 machines and 12,800 CPU cores, we only run these experiments once. Besides, all experiments relied on deterministic algorithms, architectures, and implementations.

\item {\bf Experiments compute resources}
    \item[] Question: For each experiment, does the paper provide sufficient information on the computer resources (type of compute workers, memory, time of execution) needed to reproduce the experiments?
    \item[] Answer: \answerYes{} 
    \item[] Justification: The full details about the experiments specified in Appendix \ref{app:additional-exp-details-results} already include the information on the computer resources, as well as each ablation study in the following subsections.
    
\item {\bf Code of ethics}
    \item[] Question: Does the research conducted in the paper conform, in every respect, with the NeurIPS Code of Ethics \url{https://neurips.cc/public/EthicsGuidelines}?
    \item[] Answer: \answerYes{} 
    \item[] Justification: Nothing in this paper violates the NeurIPS Code of Ethics.

\item {\bf Broader impacts}
    \item[] Question: Does the paper discuss both potential positive societal impacts and negative societal impacts of the work performed?
    \item[] Answer: \answerYes{} 
    \item[] Justification: The discussion on societal impacts is in Sec. \ref{sec:conclusion}.
    
\item {\bf Safeguards}
    \item[] Question: Does the paper describe safeguards that have been put in place for responsible release of data or models that have a high risk for misuse (e.g., pretrained language models, image generators, or scraped datasets)?
    \item[] Answer: \answerNA{} 
    \item[] Justification: This paper does not release any new data or models.

\item {\bf Licenses for existing assets}
    \item[] Question: Are the creators or original owners of assets (e.g., code, data, models), used in the paper, properly credited and are the license and terms of use explicitly mentioned and properly respected?
    \item[] Answer: \answerYes{} 
    \item[] Justification: This paper credits and cites all the public datasets used in the experiments, and the license and terms of use are properly respected. See Appendix \ref{app:additional-exp-details-results} and Sec. \ref{exp:large-scale}.

\item {\bf New assets}
    \item[] Question: Are new assets introduced in the paper well documented and is the documentation provided alongside the assets?
    \item[] Answer: \answerYes{} 
    \item[] Justification: The code of \oursysII is open-sourced at \url{https://github.com/modelscope/data-juicer} with detailed documentation.

\item {\bf Crowdsourcing and research with human subjects}
    \item[] Question: For crowdsourcing experiments and research with human subjects, does the paper include the full text of instructions given to participants and screenshots, if applicable, as well as details about compensation (if any)? 
    \item[] Answer: \answerNA{} 
    \item[] Justification: This paper does not involve crowdsourcing nor research with human subjects.

\item {\bf Institutional review board (IRB) approvals or equivalent for research with human subjects}
    \item[] Question: Does the paper describe potential risks incurred by study participants, whether such risks were disclosed to the subjects, and whether Institutional Review Board (IRB) approvals (or an equivalent approval/review based on the requirements of your country or institution) were obtained?
    \item[] Answer: \answerNA{} 
    \item[] Justification: This paper does not involve crowdsourcing nor research with human subjects.

\item {\bf Declaration of LLM usage}
    \item[] Question: Does the paper describe the usage of LLMs if it is an important, original, or non-standard component of the core methods in this research? Note that if the LLM is used only for writing, editing, or formatting purposes and does not impact the core methodology, scientific rigorousness, or originality of the research, declaration is not required.
    \item[] Answer: \answerYes{} 
    \item[] Justification: As the title of our paper and Sec. \ref{sec:mm_abilities} mentioned, the core contribution of \oursysII is data processing \textbf{with} foundation models and LLMs. They are the core characters to provide data processing capabilities in many operators.

\end{enumerate}

\end{document}